\documentclass[11pt]{article}	% notitlepage= no new page after title (default article)
\usepackage[english]{babel}
\usepackage[latin1]{inputenc}
\usepackage[a4paper,top=2.5cm,bottom=3.5cm,left=2.5cm,right=2.5cm]{geometry}
\usepackage{amsmath,amssymb,amsfonts,amsthm,mathrsfs}
\usepackage{mathtools}
\usepackage{bbm}			% additional mathemtical fonts
\usepackage[subnum]{cases}	%multi-case equations with separate number for each
\makeatletter
\newcommand*{\distas}[1]{\mathbin{\overset{#1}{\kern\z@\sim}}}	
\makeatother
\newcommand*{\norm}[1]{\left\lVert#1\right\rVert}		% norm
\DeclareMathOperator*{\argmin}{arg\,min}	% define \argmin \argmax

\DeclareMathOperator*{\vect}{vec}		% define vectorization operator
\newcommand*\vecc[1]{\vect\left(#1\right)}
\DeclareMathOperator*{\diagg}{diag}		% define \diag
\newcommand*\diag[1]{\diagg\left(#1\right)}
\allowdisplaybreaks			% breaks "align" in multiple pages
\usepackage{soul,xcolor,graphicx}
\usepackage{rotating} %sidewaysfigure, sidewaystable
\usepackage{subfigure}
\usepackage{caption}
\usepackage{tikz}
\usetikzlibrary{shapes,arrows,bayesnet}	% flow charts
\usepackage{tkz-berge}
\usepackage[shortlabels]{enumitem}	% change enumerate items
\usepackage{float}
\usepackage{array}
\usepackage{multirow}
\usepackage{longtable}
\newcolumntype{L}[1]{>{\raggedright\let\newline\\\arraybackslash\hspace{0pt}}m{#1}}
\newcolumntype{C}[1]{>{\centering\let\newline\\\arraybackslash\hspace{0pt}}m{#1}}
\newcolumntype{R}[1]{>{\raggedleft\let\newline\\\arraybackslash\hspace{0pt}}m{#1}}
\usepackage{booktabs}			
\usepackage{tabularx}			%tabulars with equal-width cols
\usepackage{epsfig}
\usepackage[bottom]{footmisc}	%put footnotes always at bottom
\usepackage{mdframed}
\usepackage{algorithm}
\usepackage{algpseudocode}
\usepackage{natbib} % square,comma,numbers,round,authoryear
\usepackage{hyperref} % pagebackref=false, linktoc=all, colorlinks=true, citecolor=red
\hypersetup{bookmarks=true, unicode=true, urlcolor=blue, linkcolor=blue, filecolor=blue, colorlinks=true, citecolor=red, final=true}
\newtheoremstyle{break}		% Theorem (and all dependent classes) body on new line
  {\topsep}{\topsep}%
  {\itshape}{}%
  {\bfseries}{}%
  {\newline}{}%
\theoremstyle{break}
\newtheorem{lemma}{Lemma}[section]

\newtheorem{definition}{Definition}[section]
\newtheorem{example}{Example}[section]
\newtheorem{remark}{Remark}[section]
\usepackage[titletoc,toc,title]{appendix}	 % \begin{appendices}
\usepackage{authblk}				% affiliation of authors

\title{\LARGE{\textbf{Nonparametric forecasting of multivariate probability density functions}}}
\author[1,2,3,4]{Dominique Gu\'{e}gan\thanks{e-mail: \href{mailto: dominique.guegan@univ-paris1.fr}{dominique.guegan@univ-paris1.fr}}}
\author[1,2,3,4]{Matteo Iacopini\thanks{e-mail: \href{mailto: matteo.iacopini@unive.it}{matteo.iacopini@unive.it}, corresponding author.}}
\affil[1]{Universit\'{e} Paris I - Panth\'{e}on-Sorbonne}
\affil[2]{Centre d'\'{E}conomie de la Sorbonne (CES)}
\affil[3]{LabEx ReFi (Financial Regulation), Paris}
\affil[4]{Ca' Foscari University of Venice}
\date{\today}
\begin{document}

\maketitle
\thispagestyle{empty}		% removes number from title page
\begin{abstract}
The study of dependence between random variables is the core of theoretical and applied statistics. Static and dynamic copula models are useful for describing the dependence structure, which is fully encrypted in the copula probability density function. However, these models are not always able to describe the temporal change of the dependence patterns, which is a key characteristic of financial data.

We propose a novel nonparametric framework for modelling a time series of copula probability density functions, which allows to forecast the entire function without the need of post-processing procedures to grant positiveness and unit integral. We exploit a suitable isometry that allows to transfer the analysis in a subset of the space of square integrable functions, where we build on nonparametric functional data analysis techniques to perform the analysis.

The framework does not assume the densities to belong to any parametric family and it can be successfully applied also to general multivariate probability density functions with bounded or unbounded support. Finally, a noteworthy field of application pertains the study of time varying networks represented through vine copula models.

We apply the proposed methodology for estimating and forecasting the time varying dependence structure between the S\&P500 and NASDAQ indices.
\end{abstract}

\newpage
\section{Introduction}\label{sec:introduction}
One of the most relevant research fields in theoretical and applied statistics is devoted to the study of the dependence between random variables. In finance, the analysis of the dependence patterns is a challenging problem and its understanding serves several purposes: control of risk clustering, credit, market and systemic risk measurement, pricing and hedging of credit sensitive instruments (such as collateralized debt obligations or CDOs) and credit portfolio management. The analysis of the relationships between economic and financial variables is crucial for the identification of causality relations (e.g., see~\cite{Granger88Causality},~\cite{White10Granger_Structural_Causality}).
From a statistical perspective, the main purpose is the development of models able to describe and forecast the joint dynamic behaviour of financial variables. Moreover, these models may provide an effective support for the financial regulator (for example, in predicting and counteracting an increase of the systemic risk).

Firstly developed by~\cite{Sklar59book}, copula functions have attracted signifcant attention over the last decade, particularly within the financial and econometric communities, as a flexible instrument for modelling the joint distribution of random variables (see~\cite{Joe97Copula_book},~\cite{Nelsen13Copula_book} and~\cite{Durante15Copula_Measure_book} for an introduction and a compelling review).
Let $(X_1,\ldots,X_d)$ be a random vector with continuous marginal cumulative distribution functions (cdf) $F_i(\cdot)$ and probability density function (pdf) $f_i(\cdot)$. The random vector $(U_1,\ldots,U_d) = (F_1(X_1),\ldots,F_d(X_d))$, obtained by application of the probability integral transform, has uniform marginals. The copula of $(X_1,\ldots,X_d)$ is defined as the joint cumulative distribution function $C:[0,1]^d \rightarrow [0,1]$ of $(U_1,\ldots,U_d)$, that is $C(u_1,\ldots,u_d) = \mathbb{P}(U_1\leq u_1,\ldots, U_d \leq u_d)$. Moreover, denoting $F(\cdot)$ the joint cumulative distribution of $(X_1,\ldots,X_d)$ and by $f(\cdot)$ its probability density function, Sklar's theorem (\cite{Sklar59book}) states that there exists a unique copula $C(\cdot)$ with probability density function $c:[0,1]^d \rightarrow \mathbb{R}_+$ such that $F(x_1,\ldots,x_d) = C(F_1(x_1),\ldots,F_d(x_d))$ and $f(x_1,\ldots,x_d) = c(F_1(x_1),\ldots,F_d(x_d)) \prod_{i=1}^d f_i(x_i)$.

The use of a copulas permits to separately deal with the marginal distributions and the dependence structure among a set of random variables, thus providing a high degree of flexibility in modelling the corresponding joint distribution. The literature on quantitative finance and financial econometrics has widely recognized the importance of this instrument, as documented by the review of~\cite{Patton12CopulaFinance_review} and the textbooks by~\cite{Cherubini04Copula_finance} and~\cite{Cherubini11DynamicCopula_finance}.

A standard practice in financial econometrics, motivated by the fact that multivariate data (e.g., returns of a portfolio of assets) have non-Gaussian marginal distributon, consists in assuming a heavy-tailed distribution for the marginals (or to estimate them nonparametrically) and to join them with a parametric copula function, which fully describes the dependence structure (\cite{Deheuvels78CaracterisationLoisExtremes}, \cite{Deheuvels79Empirical_Copula}) through its parameters.
This approach allows a parsimonious description of the dependence between two variables by means of the few parameters of a copula function.
% Estimation is generally conducted by means of a two-step procedure which consists in estimating the parameters of the marginal distributions and then the parameters of the copula function (inference for the margins, see~\cite{Chen06Estimation_semiparam_copula},~\cite{Cherubini04Copula_finance}).

This method has some undesirable shortcomings. First, each parametric copula family is designed to describe only a specific dependence pattern (for example, see Table~\ref{tab:tail_dependence_copula_families}), which makes the selection of the family a crucial aspect of every modelling strategy. To this aim, we recall the definition of upper (lower) tail dependence from \cite{Cherubini04Copula_finance}. This concept is used to describe situations where high (low) values of a variables are likely to be observed together with high (low) values of the other. In terms of the copula pdf, this means that the probability is concentrated to the top-right (bottom-left) corner. Formally, given two random variables $X\sim G_X$ and $Y\sim G_Y$ with bivariate copula $C(\cdot)$, the upper and lower tail dependence coefficients (upper TDC and lower TDC, respectively) are given by:
\begin{align}
\label{eq:TDC_u}
\lambda_U & = \lim_{u\rightarrow 1^-} \mathbb{P}(G_X(X) > u | G_Y(Y) > u) = \lim_{u\rightarrow 1^-} \frac{1-2u+C(u,u)}{1-u} \, , \\
\label{eq:TDC_l}
\lambda_L & = \lim_{u\rightarrow 0^+} \mathbb{P}(G_X(X) < u | G_Y(Y) < u) = \lim_{u\rightarrow 0^+} \frac{C(u,u)}{u} \, ,
\end{align}
which are asymptotically equivalent to:
\begin{align}
\label{eq:TDC_u_l_log}
\lambda_U = 2- \lim_{u\rightarrow 1^-} \frac{\log(C(u,u))}{\log(u)} \, , \qquad \qquad \lambda_L = 2- \lim_{u\rightarrow 0^+} \frac{\log(1-2u+C(u,u))}{\log(1-u)} \, .
\end{align}
The copula $C(\cdot)$ is said to have upper (lower) tail dependence when $\lambda_U \neq 0$ ($\lambda_L \neq 0$). Table~\ref{tab:tail_dependence_copula_families} reports the tail dependence coefficients for some commonly used copula families (see \cite{Cherubini04Copula_finance}). Only some of them (e.g., Gaussian, $t$-student and Fr\'{e}chet, for some values of their parameters) have simultaneously upper and lower tail dependence: this happens only for some values of the parameter of the copula and, in any case, the tail dependence coefficients are equal, thus implying that the tail dependence is symmetric.

\begin{table}[!t]
\centering
\begin{tabular}{| l | c | c |}
\toprule
\textbf{Copula family} & \textbf{Upper} $\lambda_U$ & \textbf{Lower} $\lambda_L$ \\
\midrule
Gaussian$(\rho)$			& 1 if $\rho =1$ 			& 1 if $\rho =1$ \\
$t$-student$(\nu,\rho)$ 	& $\lambda > 0$ if $\rho >1$	& $\lambda > 0$ if $\rho >1$ \\
Gumbel$(\theta)$   		& $2-2^{1/\theta}$ & 0 \\
Clayton$(\theta)$  		& 0 & $2^{-1/\theta}$ \\
Frank$(\theta)$      	& 0 & 0 \\
Fr\'{e}chet$(p,q)$ 		& $q$ & $q$ \\
\bottomrule
\end{tabular}
\caption{Upper and lower tail dependence for some copula families. In brackets the parameters of the copula family.}
\label{tab:tail_dependence_copula_families}
\end{table}

Second, when the copula parameter is assumed to be fixed, these constructions are able to identify only the overall, static relations and fail to account for any kind of spatial or temporal change.
This constraint is particularly restrictive in time series analysis of financial data, as pointed out by~\cite{Fermanian04Pitfalls_copula_models}. In fact, very often the relations between financial variables are non linear and change over time.

%%%% Dynamic copulas
To address this shortcoming,~\cite{Patton06TimeVarying_ConditionalCopula_II} and~\cite{Patton06TimeVarying_ConditionalCopula_I} introduced dynamic copula models by assuming that the parameters of the copula function are driven by an autoregressive process. Instead, \cite{Fermanian12TimeVarying_Copula} allowed the parameters to depend on past realizations of the observables.
These seminal works, have opened a new avenue to research (see~\cite{Manner12Survey_TimeVarying_Copula} for a review) and has brought outstanding improvements to the econometrician's toolbox.
For example,~\cite{So14VineGARCH_dynamicCopula} and~\cite{Jondeau06CopulaGARCH_Dynamic} incorporate dynamics into a copula-GARCH model improving its forecasting performance, \cite{Dias04DynamicCopula_HighFreq} and~\cite{Goorbergh05DynamicCopula_OptionPricing} exploited dynamic copulas in modelling high-frequency data and option pricing, respectively, whereas~\cite{Oh17TimeVar_SystRisk_DynamicCopula} has recently applied this methodology to the study of systemic risk. Other relevant empirical contributions exploiting this construction include~\cite{Almeida16TimeVarying_Copula_D_Vine}, \cite{Bartram07TimeVarying_dependence}, \cite{Weiss13VineCopula_timeDependence}, \cite{Hu10TimeVarying_dependence}, \cite{Hafner10Estimation_semiparam_dynamic_copula}, \cite{Hafner12Dynamic_copula}, \cite{Guidolin06NonlinearDynamicDependence_jointDistribution}.

%%%% Problem with dynamic copulas
Despite the greater flexibility, dynamic copulas may fail to account for the characteristics of the dependence structure among financial data. Since each copula family is constructed for describing a specific dependence pattern, the change of its parameters may not be enough to capture other types of dependence.

The recent paper by \cite{Guegan10DynamicCopula_test} found empirical evidence supporting this theory. They developed a strategy for testing the null of a static copula against a dynamic copula, then upon rejection they tested for the change of the copula family over different temporal windows. The main results is that the dependence structure between the S\&P500 and NASDAQ indices experienced a great variability over time, thus stressing the need for a dynamic model; nonetheless the null hypothesis of equal copula function family was rejected for several windows. This suggests that, in this dataset, the evolution of the dynamic copula parameter is insufficient to account for the variation of the dependence structure and different copula families should be used for different temporal windows.

\bigskip

%%%% FDA time series in finance
To overcome these limitations, we propose a methodology that has the whole function as the object of interest, instead of a finite-dimensional parameter vector. We do this by exploiting some results developed in the literature on functional data analysis, which we briefly introduce in the following.

Starting from the seminal work of \cite{Bosq00LinearProcesses_FunctionSpaces}, functional data analysis (see~\cite{Ramsay05Functional_data_analysis_book} and \cite{FerratyVieu06Nonparam_FDA_Book} for a thorough treatment) has found applications also in finance and financial econometrics (see \cite{Hormann12functionalTS}, \cite{Kokoszka12dependent_Functional_data} and \cite{Kidzinski15Functional_TimeSeries} for an introduction to the topic).
Different models have been proposed for time series of functional data: for example, \cite{Sen15functionalTS} assumed stationarity and estimated a VARMA functional process for electricity spot prices, conversely \cite{Liebl10NonStat_Functional_app} used the same data but proposed a method for dealing with non-stationary data and applied it in~\cite{Liebl13FunctionalTS_app} for forecasting. Within the same stream of research, \cite{Horvath10Test_Stability_Functional_AR} and \cite{Horvath14Test_Stationarity_Functional_AR} designed a testing procedure for detecting non-stationarity.
\cite{Aue15forecast_stat_functionalTS} and \cite{Kargin08Curve_Forecasting_Functional_AR}, instead, used time series functional for improving on the forecasting performance of multivariate on forecasting.
More recently,~\cite{Kidzinski16PCA_functional_periodicTS} and~\cite{Klepsch17functional_ARMA} extended the theory of univariate ARMA models to the functional framework, by introducing also seasonal effects. Finally,~\cite{Petris13BayesFTSA} and~\cite{Canale16BNP_Func_TS} developed an inferential procedure following the Bayesian paradigm, following a parametric and non-parametric approach, respectively.

%%%% Functional TS models
The literature on functional time series modelling can be partitioned into two main classes, according to the methodology developed.
The parametric framework, firstly introduced by~\cite{Bosq00LinearProcesses_FunctionSpaces}, hinges on the linear functional autoregressive model (FAR) which can be considered an infinite-dimensional analogue of vector autoregressive (VAR) processes, widely used in times series analysis.
By contrast, the non-parametric approach (see~\cite{FerratyVieu06Nonparam_FDA_Book} and~\cite{Ramsay05Functional_data_analysis_book} for an overview) relies on functional principal component analysis (fPCA). See Appendix~\ref{sec:apdx_fPCA} for a short introduction.

%%%% Problem with pdfs
Unfortunately, none of the previously mentioned approaches is suited for dealing with constrained functions, such as probability density functions (pdfs), which must be positive (on their support) and have unit integral. A statistical model for the analysis of a time series of pdfs should include a mechanism for dealing with these constraints. In the literature, three main approaches have been proposed to address this issue.
One possibility consists in ignoring the constraints and treating the pdfs as an unrestricted functions, then after the estimation and forecasting steps, the output is re-normalized in order to get a probability density function. \cite{Sen15forecastPDF} adopted this approach for studying a time series of pdfs of financial data from the S\&P500 and the Bombay Stock Exchange.

More appealing alternatives do not need to post-process the output and allow to perform the analysis taking into account the constraints. The seminal works by~\cite{Egozcue06Hilbert_Density}, \cite{Boogaart10Bayes_Linear}, \cite{Boogaart14Bayes_Hilbert} and \cite{Egozcue15Bayes_space_Change_Measure} introduced the notion of Bayes space, that is a Hilbert space of probability density functions. They borrowed from compositional data analysis and Aitchison's geometry (see~\cite{Aitchison86CompositionalData}) and interpreted probability density functions as infinite-dimensional compositional vectors. They replaced the pointwise sum and multiplication by the operations of perturbation and powering which, for $f(\cdot),g(\cdot) \in \mathbb{D}(I)$, $I\subset \mathbb{R}^n$, and $\alpha \in \mathbb{R}$, are defined by:
\begin{equation}
f(\mathbf{x}) \oplus g(\mathbf{x}) = \frac{f(\mathbf{x}) g(\mathbf{x})}{\int_I f(\mathbf{x}) g(\mathbf{x}) \: \mu(\mathrm{d}\mathbf{x})} \, , \qquad \alpha \odot f(\mathbf{x}) = \frac{f(\mathbf{x})^\alpha}{\int_I f(\mathbf{x})^\alpha \: \mu(\mathrm{d}\mathbf{x})} \, .
\label{eq:oplus_odot}
\end{equation}
Instead, the analogue of subtraction is given by $f(\cdot) \ominus g(\cdot) = f(\cdot) \oplus [-1 \odot g(\cdot)]$. They showed that the tuple $(\mathbb{D}(I),\oplus,\odot)$ is a space and that the subset $\mathbb{D}^*(I) \subset \mathbb{D}(I)$ of probability density functions whose logarithm is square integrable is a Hilbert space. These remarkable results permit to conduct the analysis directly in $\mathbb{D}^*(I)$, provided that it is possible to re-define the necessary statistical models by means of the new operations $\oplus,\odot$.
\cite{Boogaart14Bayes_Hilbert} proved that $\mathbb{D}^*(I)$ is isomorphic to the space $\mathcal{L}_2^*(I)$ of functions on $I$ with square integrable logarithm (written $\mathbb{D}^*(I) \cong_{\textnormal{clr}} \mathcal{L}_2^*(I)$) via the centred log-ratio isometry defined as follows (see Section~\ref{sec:notation} for the notation).

\begin{definition}[Centred log-ratio]\label{def:clr}
Let $\nu$ be a measure on $\mathbb{R}^n$ and $f:I \rightarrow \mathbb{R}_+$ be a probability density function supported on a set $I \subset \mathbb{R}^n$ of finite $\nu$-measure, that is $\nu(I) < \infty$ and $\nu(I) \neq 0$. The centred log-ratio (clr) transformation is an invertible map is defined as:
\begin{equation}
\textnormal{clr}(f)(\mathbf{x}) = g(\mathbf{x}) = \log(f)(\mathbf{x}) - \frac{1}{\nu(I)} \int_I \log(f)(\mathbf{y}) \: \nu(\mathrm{d}\mathbf{y}) \, ,
\label{eq:clr}
\end{equation}
with inverse given by:
\begin{equation}
\textnormal{clr}^{-1}(g)(\mathbf{x}) = f(\mathbf{x}) = \frac{\exp(g)(\mathbf{x})}{\int_I \exp(g)(\mathbf{y}) \: \nu(\mathrm{d}\mathbf{y})} \, .
\label{eq:clr_inverse}
\end{equation}
\end{definition}

Consequently, by definition~\ref{def:clr} it follows that the clr transform of a pdf supported on $I$ has to satisfy the following constraint (which we will call zero integral constraint in the rest of this paper):
\begin{equation}
\int_I \text{clr}(f)(\mathbf{x}) \: \mu(\mathrm{d}\mathbf{x}) = \int_I \log(f)(\mathbf{x}) \: \mu(\mathrm{d}\mathbf{x}) -\int_I \frac{1}{\mu(I)} \left[ \int_I \log(f)(\mathbf{y}) \: \mu(\mathrm{d}\mathbf{y}) \right] \: \mu(\mathrm{d}\mathbf{x}) = 0 \, .
\label{eq:integral_constraint}
\end{equation}
The spaces $\mathbb{D}(I),\mathbb{D}^*(I),\mathcal{L}_1(I),\mathcal{L}_2(I),\mathcal{L}_2^*(I)$ are defined with respect to a reference measure $\nu$ and contain equivalence classes of functions which are proportional each other, that is we implicitly defined $\mathbb{D}(I) = \mathbb{D}_\nu(I)$, $\mathbb{D}^*(I)=\mathbb{D}_\nu^*(I)$, $\mathcal{L}_{1,\nu}(I)=\mathcal{L}_{1,\nu}(I)$, $\mathcal{L}_2(I)=\mathcal{L}_{2,\nu}(I)$, $\mathcal{L}_2^*(I)=\mathcal{L}_{2,\nu}^*(I)$.
In this paper we consider to be the Lebesgue reference measure, i.e. $\nu=\mu$. In order to single out a specific element it is necessary to normalize the reference measure. This can be easily done if the set $I$ is $\nu$-finite, whereas if $\nu(I) = \infty$, normalization can be done using the centring procedure (see \cite{Boogaart14Bayes_Hilbert}).
Moreover, the following relations hold: $\mathbb{D}^*(I) \subset \mathbb{D}(I) \subset \mathcal{L}_1(I)$ and $\mathbb{D}^*(I) \cong_{\textnormal{clr}} \mathcal{L}_2^*(I) \subset \mathcal{L}_2(I) \subset \mathcal{L}_1(I)$.
%\begin{equation}
%\begin{matrix}
%\mathbb{D}^*(I) & \subset & \mathbb{D}(I) & \subset & \mathcal{L}_1(I) \\[5pt]
%\mathbb{D}^*(I) & \cong_{\textnormal{clr}} & \mathcal{L}_2^*(I) & \subset & \mathcal{L}_2(I) & \subset & \mathcal{L}_1(I) \\
%\end{matrix}
%\label{eq:relation_spaces}
%\end{equation}

It is possible to use the clr transform to project a pdf into the Hilbert space $\mathcal{L}_2^*(I)$ (provided that its logarithm is square integrable), which is a space embedded with the operations of pointwise addition and multiplication. We can perform the statistical analysis in this space and then project the output back into $\mathbb{D}^*(I)$ by the inverse map $\text{clr}^{-1}(\cdot)$.
This strategy via the centred log-ratio map has been proposed by~\cite{Hron16densitySFPCA} for performing fPCA on univariate pdfs with compact support. \cite{Canale16Hilbert_Density_Forecast} developed a different isometric, bijective function which maps constrained functions into a pre-Hilbert space, then estimated a FAR model on this space and transformed back the result into the original space. Though general, this framework is not explicitly designed for dealing with pdfs, but only bounded and monotonic functions, thus preventing from its use in the current setting.
The same idea of transforming pdfs into $\mathcal{L}_2(I)$ via an invertible map has been followed by~\cite{Petersen16densityFPCA}, who defined two different transformations satisfying this property: the log hazard and the log quantile density transformations, respectively. Despite their strong theoretical properties, both maps have the shortcoming of not having an equivalent transformation applicable in the multivariate case, which makes them unsuited for the analysis of multivariate probability density functions.

%Forecasting probability density functions is more complicated than forecasting unrestricted functional time series, because of the non-negativity and unit integral constraints. The main implication is that the set of $d$-dimensional probability density functions is not closed with respect to the operations of pointwise addition and scalar multiplication. Consequently the linear models proposed by~\cite{Bosq00LinearProcesses_FunctionSpaces}, such as the FAR($p$), cannot be used in this framework.

\bigskip

%%%%%%% CONTRIBUTION %%%%%%%
The empirical finding by \cite{Guegan10DynamicCopula_test} represents the key stylized fact motivating our work.
Given that a dynamic copula model may not be sufficiently flexible to describe the time varying dependence between financial variables, we contribute to this active field of research by proposing a different statistical framework for forecasting multivariate probability density functions.
To address the issues related with modelling pdfs, we extend the procedure of~\cite{Hron16densitySFPCA} who build on the previous work by~\cite{Boogaart10Bayes_Linear,Boogaart14Bayes_Hilbert}. The idea is to map the space of probability density functions to the space of integrable functions through an isometry, perform the analysis in this space (which has nicer properties), then use the inverse mapping to get the solution in the original space.
Our contribution is also related to the studies of \cite{Liebl13FunctionalTS_app} and \cite{Hays12FunctionalPCA_TimeSeriesForecast}, who developed dynamic models for forecasting functional time series of electricity prices on the basis of fPCA.
However, our focus is on the modelling of probability density functions, which call for the adoption of more complex tools than that of unrestricted functions. Finally, we contribute to the literature on dynamic dependence modelling in finance by providing a tool able to forecast the temporal evolution of the dependence pattern between the S\&P500 and NASDAQ indices.

We propose a nonparametric framework for forecasting multivariate probability density functions by extending existing models for the analysis of cross sectional, univariate probability density functions with compact support. We focus on bivariate copula pdfs because of their great importance in finance, however the proposed methodology is flexible and general, thus permitting to deal with the prediction of general pdfs with bounded or, under some conditions, unbounded support.
Thanks to the fact that a copula pdf encompasses all information on the dependence structure, we can interpret our approach as a general framework for modelling the (temporally evolving) dependence patterns between random variables.

The reminder of the paper is as follows. In Section~\ref{sec:preliminaries} we introduce the notation as well as the fundamental concepts that will be used throughout the paper. Section~\ref{sec:methodology} presents the details of the proposed baseline methodology, whereas Section~\ref{sec:extensions} provides insights on potential issues and extensions. Section~\ref{sec:application} provides an overview of the financial dataset used and presents the results of the empirical analysis. Finally, Section~\ref{sec:conclusions} concludes and describes some extensions of the current work and lines of future research.

\section{Preliminaries}\label{sec:preliminaries}
In this section we describe the proposed methodology after having introduced the main notation that will be used throughout the rest of the paper.

\subsection{Notation}\label{sec:notation}
Throughout the paper, if not differently specified, greek letters denote unknown quantities to be estimated, whereas latin letters any other variable. We denote scalars with lower-case letters, vectors with boldface lower-case letters and matrices with boldface upper-case letters. We use the shorthand $f(\cdot)$ for denoting a function, regardless of the number of arguments it takes, moreover we denote the composition of functions by $(g \circ f)(\cdot) = g(f(\cdot)) = g(f)(\cdot)$. The inner product between two functions $f(\cdot),g(\cdot)$ supported on $I\subseteq \mathbb{R}^n$ is defined in the standard way by $\langle f(\cdot), g(\cdot) \rangle = \int_I f(\mathbf{x}) g(\mathbf{x}) \: \mathrm{d}\mathbf{x}$.
The integer part of the scalar $x$ is denoted $\lfloor x \rfloor$.

We use the notation $\mathbf{A} = [\mathbf{A}_1,\ldots,\mathbf{A}_T]$ to denote a collection of $T$ matrices $\mathbf{A}_t$ of equal size $N\times M$.
The symbol $\mathbf{I}_k$ is used for the identity matrix of size $k\times k$, whereas $\mathbf{0}_k$ for the $k\times 1$ column vector of zeros. Moreover, empty spaces in the matrices stand for zero entries. Let $\mathbf{L}_N$ be the matrix representation of the first difference operator $L$, that is the $N\times (N+1)$ matrix which post-multiplied by a $(N+1)$-dimensional vector $\mathbf{a}$ yields a vector of size $N$, $\mathbf{La}$, whose entries are the first differences of the elements of $\mathbf{a}$.
The Moore-Penrose pseudo-inverse of the $N\times M$ matrix $\mathbf{A}$ is denoted by $\mathbf{A}^\dag$.
If $\mathbf{A}$ is positive definite, we define by $\mathbf{A}^{1/2}$ its (unique) principal square root.

\bigskip

In Section~\ref{sec:methodology} we will refer to the spaces of functions described as follows.
We define $\mathbb{F}_+(I)$ to be the space of non-negative, integrable functions on $I \subseteq \mathbb{R}^n$, whose general element is the function $f:I\rightarrow\mathbb{R}_+$, and we let $\mathbb{F}_0(I)$ be the space of functions on $I$ with zero integral. $\mathbb{D}(I)$ denotes the set of probability density functions with support $I$ and we define $\mathbb{D}^*(I)$ to be the space of probability density functions with support $I$ whose logarithm is square integrable.
We denote by $\mu(\cdot)$ the Lebesgue measure on $\mathbb{R}^n$, for $n \geq 1$. In the case $n=1$ we also use the shorthand notation $\mathrm{d}x = \mu(\mathrm{d}x)$, whereas for $n > 1$ we define $\mathrm{d}\mathbf{x} = \mu(\mathrm{d}\mathbf{x})$. Consequently, if $I = [a,b]$ then $\mu(I) = b-a$. All integrability definitions are made using the Lebesgue measure as reference measure, if not differently specified. Let $\mathcal{L}_p(I)$ be the space of $p$-integrable functions supported on $I$ and let $\mathcal{L}_p^*(I)$ be the space of functions on $I$ whose logarithm is $p$-integrable.
The $n$-dimensional unit simplex is defined as $\mathcal{S}^n = \lbrace \mathbf{x}\in\mathbb{R}^n : x_i>0, \: i=1,\ldots,n, \text{ and } \sum_{i=1}^n x_i = 1 \rbrace$, whereas $\mathcal{S}_0 = \lbrace \mathbf{x}\in\mathbb{R}^n : \sum_{i=1}^n x_i = 0 \rbrace$ is the subspace of $n$-dimensional vectors whose elements have zero sum.
We define $\lbrace \mathbf{e}_1,\ldots,\mathbf{e}_N \rbrace$ be the canonical basis of the space of $N \times N$ matrices.
For two spaces $X,Y$ we use the notation $X \cong_{f} Y$ to indicate that they are isomorphic through the isometric isomorphism $f:X\rightarrow Y$.

Let $\mathbf{x} = (x_1,\ldots,x_n)'$ be a vector of observations. In Section~\ref{sec:procedure_copula_estimation} we denote the empirical (marginal) cumulative distribution function of $\mathbf{x}$ by $F^n(\mathbf{x}) = (F^n(x_1),\ldots,F^n(x_n))'$. Moreover, define the rank transformation of $\mathbf{x}$ to be the function that maps each element $x_i$ of $\mathbf{x}$ to:
\begin{equation}
R_i = \sum_{j=1}^n \mathbbm{1}(x_j \leq x_i) \, .
\label{eq:rank_transformation_def}
\end{equation}
Denote with $\mathbf{u} = (u_1,\ldots,u_n)'$ with $u_i \in [0,1]$, $i=1,\ldots,n$, the vector of pseudo-observations associated to the observations $\mathbf{x}$, used for the estimation of the empirical copula in Section~\ref{sec:procedure_copula_estimation}. Each pseudo-observation is defined as:
\begin{equation}
u_i = \frac{1}{n} R_i \, . 
\label{eq:pseudo-observations_def}
\end{equation}

\bigskip

In Section~\ref{sec:procedure_modified_fPCA} and Appendix~\ref{sec:apdx_fPCA}, we will use the following notation in performing functional principal component analysis (fPCA). We define a sample of observed functions by $\mathbf{f}(\cdot) = (f_1(\cdot),\ldots,f_T(\cdot))'$, with $f_t: I \rightarrow \mathbb{R}$, for some domain $I \subseteq \mathbb{R}^n$. Moreover, we let $\mathbf{\breve{f}}(\cdot) = (\breve{f}_1(\cdot),\ldots,\breve{f}_T(\cdot))'$ denote the approximation of the observed functions obtained as an outcome of the fPCA. The principal component functions are denoted by $\boldsymbol{\xi}(\cdot) = (\xi_1(\cdot),\ldots,\xi_J(\cdot))'$ and the scores associated to the function $\breve{f}_t(\cdot)$ are $\boldsymbol{\beta}_t = (\beta_{t,1},\ldots,\beta_{t,J})'$, moreover let $\mathbf{B} = (\boldsymbol{\beta}_1,\ldots,\boldsymbol{\beta}_T)$. The corresponding estimated quantities are denoted by $\boldsymbol{\widehat{\xi}}(\cdot)$, $\boldsymbol{\widehat{\beta}}_t$ and $\mathbf{\widehat{B}}$.

\bigskip

In Section~\ref{sec:procedure_modified_fPCA} we also use spline functions, for which we adopt the following notation. We denote by $\mathfrak{I}_n$ a $n$-dimensional index set whose elements are the $n$-tuple $(i_1,\ldots,i_n)$ with entries $i_j \in [1,I_j]$, where $I_j$ is a positive integer for each $j=1,\ldots,n$.
We denote with $\boldsymbol{\lambda}$ the vector with entries $\lambda_0 < \ldots < \lambda_{g+1}$ representing the points of the knot sequence used for the spline functions.
For a $m$-order spline, define the extended knot sequence as the vector $\boldsymbol{\bar{\lambda}}$ of length $2m+g+2$ whose entries satisfy the relations:
\begin{equation*}
\underbrace{\lambda_{-m} = \ldots = \lambda_{-1}}_{m} = \underbrace{\lambda_0 < \ldots < \lambda_{g+1}}_{g+2} = \underbrace{\lambda_{g+1+1} = \ldots = \lambda_{g+m+1}}_{m} \, .
\end{equation*}
An extended knot sequence for a bivariate spline is defined by $\boldsymbol{\bar{\lambda}}^{x,y} = \boldsymbol{\bar{\lambda}}^x \otimes \boldsymbol{\bar{\lambda}}^y$, where $\boldsymbol{\bar{\lambda}}^x, \boldsymbol{\bar{\lambda}}^y$ are the extended knot sequences along each axis and the generic entry is the couple $\bar{\lambda}_{i,j}^{x,y} = (\bar{\lambda}_i^x, \bar{\lambda}_j^y)$.
$B_{i}^k(x)$ denotes the univariate (basis) B-spline function of degree $m-1$, with knot sequence indexed by $i$ and $D_x^\ell[f](x)$, $\ell \leq m-1$, is the partial derivative operator of order $\ell$, applied to the function $f$ with respect to the variable $x$.
For univariate splines of degree $m$, let $\mathbf{b} = (b_{-m},\ldots,b_g)'$ be the $(g+m+1)$ coefficient vector, whereas for bivariate splines of the same degree we define the $(m+g+1) \times (m+g+1)$ coefficient matrix by $\mathbf{\bar{B}}$.
Moreover, let $\mathbf{C}^{m+1}(\mathbf{x}^n)$ be the $n\times(g+m+1)$ collocation matrix of B-spline functions evaluated at the observation points $\mathbf{x}^n = (x_1,\ldots,x_n)'$:
\begin{equation}
\mathbf{C}^{m+1}(\mathbf{x}^n) = \begin{bmatrix}
B_{-m}^{m+1}(x_1) & \ldots & B_g^{m+1}(x_1) \\
\vdots & \ddots & \vdots \\
B_{-m}^{m+1}(x_n) & \ldots & B_g^{m+1}(x_n) \\
\end{bmatrix}
\end{equation}
Following \cite[ch.10]{DeBoor01Spline_book}, a univariate spline function of degree $k$ and the corresponding partial derivative of order $\ell$ are given by:
\begin{align}
\label{eq:univariate_spline_def}
s_m(x) & = \sum_{i=-m}^g b_i B_i^{m+1}(x) \, , \\
\label{eq:univariate_spline_derivative_def}
D_x^\ell[s_m](x) = s_m^{(\ell)}(x) & = \sum_{i=-m}^g b_i^\ell B_i^{m}(x) \, ,
\end{align}
Given an extended knot sequence $\boldsymbol{\bar{\lambda}}$ and evaluation points $\mathbf{x}^n = (x_1,\ldots,x_n)'$, are given by:
\begin{align}
\label{eq:univariate_spline}
s_k(\mathbf{x}^n) & = \sum_{i=-k}^g b_i B_i^{k+1}(\mathbf{x}^n) = \mathbf{C}^{k+1}(\mathbf{x}^n) \mathbf{b} \, , \\
\label{eq:univariate_spline_derivative}
D_x^\ell[s_m](\mathbf{x}^n) = s_k^{(\ell)}(\mathbf{x}^n) & = \mathbf{C}^{m+1-\ell}(\mathbf{x}^n) \mathbf{b}^{(\ell)} = \mathbf{C}^{m+1-\ell}(\mathbf{x}^n) \mathbf{S}_\ell \mathbf{b} \, ,
\end{align}
where $\mathbf{C}^{m+1}(\mathbf{x})$ is a matrix of B-splines evaluated at the points $\mathbf{x}$, $\mathbf{b}$ is the coefficient vector and $\mathbf{S}_\ell$ is a matrix transforming the coefficient vectors of splines of degrees $m$ to those of their derivatives of degree $\ell$. The direct link between the coefficients of a spline and its derivative are due to the property that the derivative of a spline is another spline of lower degree (see~\cite{DeBoor01Spline_book}), that is:
\begin{equation}
s_{m+1}(x) = \int s_m(x) \: \mathrm{d}x \, .
\end{equation}
Similarly, we define the $d$-dimensional tensor product spline function by the tensor product between univariate splines (see \cite{Schumaker07Spline_book}):
\begin{equation}
s_m(x_1,\ldots,x_d) = \sum_{i_1} \cdots \sum_{i_d} b_{i_1,\ldots,i_d} B_{i_1}^m(x_1) \cdots B_{i_d}^m(x_d) = \sum_{i_1,\ldots,i_d} b_{i_1,\ldots,i_d} B_{i_1}^m(x_1) \cdots B_{i_d}^m(x_d) \, ,
\label{eq:multivariate_spline}
\end{equation}
with $b_{i_1,\ldots,i_d} \in \mathbb{R}$ for $(i_1,\ldots,i_d) \in \mathfrak{I}_d$. Notice that the coefficients $b_{i_1,\ldots,i_d}$, with $(i_1,\ldots,i_d) \in \mathfrak{I}_d$, can be represented as a vector of length $\prod_{j=1}^d I_j$ or, equivalently, as a $d$-order array (or tensor) with dimensions $I_1 \times \ldots \times I_d$.
The partial derivatives of the multivariate spline in eq.~\eqref{eq:multivariate_spline} are given by \cite{Schumaker07Spline_book} (and can be easily computed via Algorithm 5.11 in~\cite{Schumaker07Spline_book}):
\begin{equation}
D_{x_1}^{\ell_1} \cdots D_{x_d}^{\ell_d} [s_k](x_1,\ldots,x_d) = \sum_{i_1,\ldots,i_d} b_{i_1,\ldots,i_d}^{\ell_1,\ldots,\ell_d} B_{i_1}^{m-\ell_1}(x_1) \cdots B_{i_d}^{m-\ell_d}(x_d) \, .
\label{eq:multivariate_spline_derivative}
\end{equation}
Finally, in Section~\ref{sec:procedure_prediction} we use the notation $\mathbf{\widetilde{B}} = (\boldsymbol{\widetilde{\beta}}_{T+1}',\ldots,\boldsymbol{\widetilde{\beta}}_{T+H})'$, where $\boldsymbol{\widetilde{\beta}}_{T+h}$ is the forecast for the vector $\boldsymbol{\widehat{\beta}}_T$ at horizon $h=1,\ldots,H$. The corresponding forecast for the fPCA approximate functions are denoted by $\mathbf{\widetilde{f}}_{T+H}(\cdot) = (\widetilde{f}_{T+1}(\cdot),\ldots,\widetilde{f}_{T+H}(\cdot))'$ whereas $\mathbf{\widetilde{c}}_{T+H}(\cdot) = (\widetilde{c}_{T+1}(\cdot),\ldots,\widetilde{c}_{T+H}(\cdot))'$ denotes the forecast of the copula probability density functions.

\subsection{Related literature}
Given an observed bivariate time series of relevant economic or financial variables $(\mathbf{x},\mathbf{y}) = \lbrace x_{t,i},y_{t,i}) \rbrace_{ti}$, with $i=1,\ldots,N$ for each $t=1,\ldots,T$, with unknown time varying copula probability density function $c_t(\cdot)$, the purpose of this methodology is to obtain a $h$-step ahead forecast the copula pdf $\widetilde{c}_{T+h}(\cdot)$, $h=1,\ldots,H$. In order to achieve this result, we will borrow some concepts from various streams of literature. The core of the methodology is grounded on functional data analysis (FDA), in particular the technique of functional principal component analysis (fPCA), and on the centred log-ratio isometry between $\mathbb{D}^*(I)$ and $\mathcal{L}_2^*(I)$. Furthermore, we exploit several concepts from the literature on nonparametric estimation of copula functions and, finally, we use standard techniques for multivariate time series analysis.

\bigskip

The use of functional autoregressive processes (FAR) proposed by \cite{Bosq00LinearProcesses_FunctionSpaces} is prevented by the constraints holding on pdfs and the fact that $\mathbb{D}(I)$ is not closed under pointwise addition and multiplication. In situations like this, post-processing techniques are necessary for mapping the output of a given procedure into the desired space. However, this procedure is suboptimal as there are guarantees that all the information is preserved by this mapping. By contrast, an efficient  forecasting model for probability density functions should yield a consistent output, that is the predicted function must be pdfs.

In second instance, as functions are infinite-dimensional objects, the original forecasting problem would require to work with infinite-dimensional spaces. Though natural, this brings in a significant degree of complexities that a similar problem in finite-dimensional spaces (i.e., Euclidean spaces). Clearly, a direct matching between an infinite-dimensional problem and a finite-dimensional one, does not exist. Moreover, na\"ive techniques for moving from an infinite-dimensional problem into one a finite one via discretization of the functions could be too rough and lose too much information. Nonetheless, under suitable assumptions it is possible to approximate the infinite-dimensional functional forecasting problem by a simpler one in which the parameters of interest are finite-dimensional vectors. Moreover, under certain conditions this approximation is optimal (according to a weighted least squares criterion).

\bigskip

In order to avoid post-processing and rough approximations, in Section~\ref{sec:procedure_modified_fPCA} we exploit the centred log-ratio isometry between the spaces $\mathbb{D}^*(I)$, $\mathcal{L}_2^*(I)$ and define a factor model for approximating the clr-transformed densities (that is, $\text{clr}(c_t)(\cdot) \approx \breve{f}_t(\cdot)$):
\begin{equation}
\breve{f}_t(\cdot) = \boldsymbol{\beta}_t' \boldsymbol{\xi}(\cdot) = \sum_{j=1}^J \beta_{t,j} \xi_j(\cdot) \, ,
\label{eq:factor_model}
\end{equation}
where $\xi_j(\cdot)$ are the principal component functions (or factors) and the coefficients $\beta_{t,j} \in \mathbb{R}$ are the principal component scores, both estimated by means of the fucntional principal component analysis. The factor model defines an approximation of the functions $\text{clr}(c_t)(\cdot)$ by means of a finite linear combination of common, time-invariant factors with component specific time-varying scores. The optimality criterion given by the quadratic distance $||\text{clr}(c_t)(\cdot) - \breve{f}_t(\cdot)||_2$ is minimized by the choice of the principal component functions that maximize the explained variability of the series $\text{clr}(c_t)(\cdot)$, $t=1,\ldots,T$ (see~\cite{Ramsay05Functional_data_analysis_book}).

\bigskip

Functional data analysis is a growing field of research and the existing results dealing with probability density functions are scarce.
A possible interpretation of fPCA, in analogy with multivariate PCA, identifies the principal component functions with the eigenfunctions of their covariance operator of the observed functions. Following this interpretation, \cite{Hormann15Dynamic_fPCA} provided a remarkable extension of fPCA to time series functional data. They worked on the frequency domain using the techniques of \cite{Brillinger01TimeSeries_book} for estimating the dynamic principal component functions, which account for the temporal dependence among functional observations. Unfortunately, their results are not straightforwardly extendible to pdfs.

%The literature on functional data analysis exploiting multivariate K-L decomposition is rather scarce: a numerical implementation is provided by~\cite{Wang08KL_expansion_thesis} under particular assumptions on the shape of the covariance function, whereas~\cite{Boumaza15Multivariate_fPCA} focus on the interpretation of principal component scores and functions. An extension of the K-L decomposition to multiple series of functions has been prposed by~\cite{Boumaza15Multivariate_fPCA} and \cite{Chen17MultiDimensional_fPCA}. In the latter papaer, the authors developed an efficient algorithm for solving a $d$-dimensional fPCA problem by exploiting parallel computation via GPUs. Their approach is based on the discretization of the eigenproblem stemming from fPCA on a rectangular grid, thus reducing the computations to matrix products. To cope with the high dimensionality of the resulting matrices and speed up the computation, they exploited random projection approximations (\cite{Halko11RandomProjection_Matrix_Approx}).

In fact, when dealing with pdfs, the estimation of the principal component functions $\boldsymbol{\xi}(\cdot)$ poses some issues which call for the development of specific procedures.
\cite{Egozcue06Hilbert_Density} has proved the analogy between probability density functions and compositional vectors, whcih are vectors belonging to the $n$-dimensional unit simplex $\mathcal{S}^n$ representing fractions or proportions and contitue the conrerstone of compositional data analysis (see~\cite{Aitchison86CompositionalData}). \cite{Egozcue06Hilbert_Density} interpreted pdfs as infinite-dimensional compositional vectors and translated into the functional domain the main results of compositional data analysis: this includes the definition of the operations of perturbation and powering (analogue of addition and scalar multiplication), $\oplus,\odot$, that make $(\mathbb{D}(I),\oplus,\odot)$ a space.
\cite{Boogaart10Bayes_Linear} and \cite{Boogaart14Bayes_Hilbert} proved that $(\mathbb{D}(I),\oplus,\odot)$ is indeed a Hilbert space and showed that the centred log-ratio widely used in compositional data analysis is an isometry (i.e. an isometric isomorphism) between the spaces $\mathbb{D}^*(I),\mathcal{L}_2^*(I)$.

These results opened new possibilities to the functional analysis of pdfs. While it is possible to exploit the operations $\oplus,\odot$ for performing statistical analyses directly on $\mathbb{D}(I)$, the need for re-definition of standard techniques by $\oplus,\odot$ has lead the researchers to prefer the use of isometries. For the sake of working out fPCA of (transformed) univariate pdfs in $\mathcal{L}_2(I)$, \cite{Petersen16densityFPCA} proposed two isometries (the log-hazard and the log-quantile transforms) between $\mathbb{D}(K)$ and $\mathbb{D}(K)$, for $K$ a compact subset of $\mathbb{R}$, whereas \cite{Hron16densitySFPCA} exploited the clr map. Other contributions in this area include \cite{Salazar15Predict_Density_Compositional}, who proposed a forecasting model for univariate pdfs, and \cite{Menafoglio14Kriging_pdf_compositional}, who studied the problem of interpolation via the kriging method for probability density functions.
The common strategy consists in three steps: the transformation of the pdfs into a suitable Hilbert subspace of $\mathcal{L}_2(I)$, where the statistical model is defined and the analysis is undertaken. Finally, the use of the inverse of the isometry for mapping the result back into $\mathbb{D}(I)$.
Finally, the contribution of \cite{Machalova16clr_spline} (see also~\cite{Machalova02opt_interp_smooth_spline}, \cite{Machalova02opt_interp_leastsq_smooth_pline}) is based on the interpretation of fPCA as an eigenproblem, which the authors solved proposing a solution within the class of spline functions (see Appendix~\ref{sec:apdx_fPCA} for more details on fPCA and related solution methods). This allowed for the inclusion of the zero integral constraint as a constraint on the coefficients of the basis spline functions.

\bigskip

In this paper we follow~\cite{Boogaart14Bayes_Hilbert} and use the centred log-ratio transform to map pdfs into the Hilbert space $\mathcal{L}_2^*(I)$. Then, we extend to the multivariate framework the strategy developed by \cite{Machalova16clr_spline} for dealing with the integral constraint for univariate pdfs with compact support, thus obtaining a way to account for the constraint eq.~\eqref{eq:integral_constraint} in the estimation of the principal component factors and scores.

The most appealing feature of the factor model in eq.~\eqref{eq:factor_model} we specify is that all the information about the temporal dependence between the functions is carried by the scores, which form a vector-valued time series. Therefore, a forecast for the approximated function $\widetilde{f}_{T+h}(\cdot)$ at horizon $h\geq 1$, can be obtained by plugging-in a forecast for the scores, computed by well-known methods (e.g., VAR models). Then we get a forecast for the pdf, $\widetilde{c}_{T+h}(\cdot)$, by simply applying the inverse centred log-ratio map in eq.~\eqref{eq:clr_inverse}.

Our strategy shares some similarities with \cite{Liebl13FunctionalTS_app} and \cite{Hron16densitySFPCA}, but the methodologies differ in some key aspects. First and most important, we are interested in forecasting pdfs, which complicates the analysis with respect to the unrestricted case of \cite{Liebl13FunctionalTS_app}. Moreover, we extend the analysis of \cite{Hron16densitySFPCA} to the bivariate case (though the methodology generalizes easily to multidimensionality). Finally, we provide some remarks about how to deal with the case of densities with unbounded support.

\section{Methodology}\label{sec:methodology}
We propose a strategy for estimating the factor model in eq.~\eqref{eq:factor_model}, then forecasting the clr-transformed functions $\widetilde{f}_{T+h}(\cdot)$ and the corresponding pdfs $\widetilde{c}_{T+h}(\cdot)$, $h=1,\ldots,H$. The methodology focuses on the forecast of bivariate copula probability density functions, however, the method is general and can be applied without structural changes to general multivariate pdfs with bounded support as well as to pdfs with unbounded support that satisfy an additional constraint (see Section~\ref{sec:extensions}). The modelling framework can be summarized as follows:
\begin{itemize}
\item in Section~\ref{sec:procedure_copula_estimation} we partition the raw dataset in sub-samples corresponding to different periods $t$, then for each of them we estimate the copula probability density function (or, in the generally case, the multivariate pdf).\\
When dealing with copula pdfs, we use the nonparametric density estimator proposed by \cite{Chen99EstimationPDF_BetaKernel} for avoiding the boundary bias (for general multivariate pdfs we suggest standard product kernel estimators).

\item next, in Section~\ref{sec:procedure_modified_fPCA} we estimate the factor model in eq.~\eqref{eq:factor_model} by a modified version of the functional principal component analysis. In this section we combine the centred log-ratio transform and spline functions for estimating the principal component functions and the scores ($\boldsymbol{\widehat{\xi}}(\cdot)$ and $\mathbf{\widehat{B}}$, respectively) such that the resulting functions $\breve{f}_t(\cdot)$ (approximating $\text{clr}(c_t)(\cdot)$) satisfy the restrictions of probability density functions.\\
We generalize the strategy proposed by \cite{Machalova16clr_spline} and its application in \cite{Hron16densitySFPCA} to the multivariate (and potentially unbounded) case.

\item finally, in Section~\ref{sec:procedure_prediction} we estimate a VAR($p$) process for the time series of scores previously estimated and forecast the scores $h$ steps ahead, $h=1,\ldots,H$. Then, we get the forecast of the approximated function $\breve{f}_{T+h}(\cdot)$ and by applying the inverse centred log-ratio transform we obtain a predicted copula probability density function (or the multivariate pdf) $\widetilde{c}_{T+h}(\cdot)$.\\
The forecasting strategy extends \cite{Liebl13FunctionalTS_app} from univariate unconstrained functions to multivariate pdfs.
\end{itemize}

Algorithm~\ref{alg:procedure} synthetically represents the proposed strategy. Each block is described in detail in the following subsections.

\begin{algorithm}[h!t]
\caption{Methodology}
\label{alg:procedure}
\begin{algorithmic}[1]
\Function{CopulaEstim}{$\mathbf{x},\mathbf{y}$}
	\State a) split data $\lbrace x_{t'},y_{t'} \rbrace_{t'=1}^{TN}$ into $T$ sub-samples $\lbrace x_{t,i},y_{t,i} \rbrace_{i=1}^N$ \Comment{data for 2-steps}
	\For{$t=1,\ldots,T$}
		\State b) compute pseudo-obs $\lbrace u_{t,i},v_{t,i} \rbrace_{i=1}^N$ from $\lbrace x_{t,i},y_{t,i} \rbrace_{i=1}^N$
		\State c) estimate copula $\widehat{c}_t(u,v)$  \Comment{Beta kernel}
		\State d) compute clr transform of copula values $\text{clr}(\widehat{c}_t)(u_{t,i},v_{t,i})$  \Comment{clr}
	\EndFor
	
	\State \Return $(\mathbf{U},\mathbf{V},\mathbf{C}) = \lbrace u_{t,i}, v_{t,i}, \text{clr}(\widehat{c}_t)(u_{t,i},v_{t,i}) \rbrace_{t,i}$
\EndFunction
\Statex
\Function{mod\_fPCA}{$\mathbf{U},\mathbf{V},\mathbf{C},\boldsymbol{\bar{\lambda}}^{u,v}$}
	\For{$t=1,\ldots,T$}
		\State a) $(\mathbf{d}_t,\boldsymbol{\phi}) \leftarrow$ solve constrained optimal smoothing problem $(\mathbf{U},\mathbf{V},\mathbf{C},\boldsymbol{\bar{\lambda}}^{x,y})$  %\Comment{constrained splines}
		%\State    $\quad$ get coefficient matrix $\mathbf{D}$
	\EndFor
	\State b) $(\boldsymbol{\widehat{\xi}},\mathbf{\widehat{B}}) \leftarrow$ solve eigenproblem $(\mathbf{D},\boldsymbol{\phi})$
	%\State    $\quad$ get estimated principal components $\boldsymbol{\widehat{\xi}}$ and scores $\boldsymbol{\widehat{\beta}}_t$

	\State \Return $\mathbf{\widehat{B}} = (\boldsymbol{\widehat{\beta}}_1,\ldots,\boldsymbol{\widehat{\beta}}_T)$
\EndFunction
\Statex
\Function{Prediction}{$\mathbf{\widehat{B}}$}
	\State a) estimate VAR($p$) for $\lbrace \boldsymbol{\widehat{\beta}}_t \rbrace_{t}$ %\Comment{autoregressive process for scores}
	\State b) forecast scores $\mathbf{\widetilde{B}} = (\boldsymbol{\widetilde{\beta}}_{T+1},\ldots,\boldsymbol{\widetilde{\beta}}_{T+H})$
	\State c) forecast transformed pdfs $\mathbf{\widetilde{f}}_{T+H}(\cdot) = (\widetilde{f}_{T+1}(\cdot),\ldots,\widetilde{f}_{T+H}(\cdot))'$
	\State d) forecast pdfs $\mathbf{\widetilde{c}}_{T+H}(\cdot) = (\widetilde{c}_{T+1}(\cdot),\ldots,\widetilde{c}_{T+H}(\cdot))'$    \Comment{inverse clr}
	\State \Return $\mathbf{\widetilde{c}}_{T+H}(\cdot)$
\EndFunction
\end{algorithmic}
\end{algorithm}

\subsection{Step 1 - Copula estimation}\label{sec:procedure_copula_estimation}
After the introduction of the empirical copula (\cite{Deheuvels78CaracterisationLoisExtremes}, \cite{Deheuvels79Empirical_Copula}), which is a nonparametric estimatior for the copula cumulative distribution function, several non-paramteric techniques for the estimation of a copula pdf and cdf have been proposed. We follow~\cite{Chen99EstimationPDF_BetaKernel} and \cite{Charpentier07Estimation_CopulaPDF_Kernel} and estimate the copula pdfs from raw data via a product Beta kernel estimator. Among the main advantages of this approach we remark the greater flexibility with respect to parametric methods, the smoothness of the estimated function (as opposed to the empirical copula) and the absence of boundary bias.

Consider a sample of observations $\lbrace x_{t'},y_{t'} \rbrace_{t'}$ of size $T'$ (for instance, with daily frequency). First of all, we fix the reference period $t$ (i.e., year, quarter) and split the raw sample accordingly into $T$ sub-samples of size $N$ (to be interpreted, for instance, as $T$ years of $N$ daily observations), $\lbrace x_{t,i},y_{t,i} \rbrace_{i,t}$ of size $N$, for $t=1,\ldots,T$.
The reference period coincides with the frequency of the functional time series we want to analyse, whereas the intra-period observations are interpreted as noisy measurements of the discretized continuous function of interest $c_t(\cdot)$. Consequently, we are going to use the $N$ data points in each period $t$ to estimate the function $c_t(\cdot)$, then we use the resulting functional time series for performing forecasts of the probability density through a modified fPCA algorithm.

We exploit the intra-period information (i.e. $N$ observations, for fixed period $t$) for estimating a copula pdf for each period, $\widehat{c}_t(\cdot)$. Recall that a copula probability density has uniformly distributed marginals representing the marginal cumulative distributions. As the latter are unknown, it is necessary to estimate them as first step. In practice, we compute the pseudo-observations (see~\cite{Nelsen13Copula_book}, \cite{Cherubini04Copula_finance}) defined as follows:
\begin{equation}
(u_{t,i},v_{t,i}) = \left( F_x^N(x_{t,i}),F_y^N(y_{t,i}) \right) \, .
\label{eq:pseudo_observations}
\end{equation}
The marginals $F_x^N(x_{t,i}),F_y^N(y_{t,i})$ can be estimated via the empirical cumulative distribution function or through the rank transformed data.

\bigskip

We choose the second method, as it is computationally faster and provides distributions closer to the uniform. The rank transformation generates pseudo-observations according to (similarly for $y$):
\begin{equation}
R_{t,i}^x = \sum_{j=1}^N \mathbbm{1}(x_{t,j} \leq x_{t,i}) \, ,  \qquad   u_{t,i} = \frac{1}{N} R_{t,i}^x \, .
\label{eq:rank_transformed_data}
\end{equation}
Given the pseudo observations, we estimate the copula probability density function by using a nonparametric kernel density estimator obtained as the product of univariate Beta kernels. Given a sample $\lbrace x_t \rbrace_{t=1}^T$, the Beta kernel density estimator (\cite{Chen99EstimationPDF_BetaKernel}, \cite{Charpentier07Estimation_CopulaPDF_Kernel}) is defined as:
\begin{equation}
\widehat{f_h}(x) = \frac{1}{T} \sum_{t=1}^T \mathcal{K}_\beta \left( x_t; 1+\frac{x}{h}, 1+\frac{1-x}{h} \right) \, ,
\label{eq:beta_kernel_estimator}
\end{equation}
where $\mathcal{K}_\beta(\cdot;a,b)$ is the pdf of a Beta distribution with parameters $(a,b)$ and $h$ is the bandwidth.
Alternative nonparametric methods for the estimation of a probability density function, such as the kernel estimator of \cite{Fermanian03Nonparam_estim_copula_TS}, do not fit well the current framework because of the inadequateness of the methods to deal with the compact support. This causes a boundary bias problem if an unbounded kernel (such as the Gaussian) is used or a lack of smoothness, if the derivatives of the empirical copula distribution are chosen. Both shortcomings are instead solved by the product Beta kernel estimator. The price to pay is the lack of adequate rules of thumb for the specification of the bandwidth, which must be tuned case-by-case. 
The estimated smooth functions $(\widehat{c}_1(\cdot),\ldots,\widehat{c}_T(\cdot))$ are used to compute the values of the copula function at specific couples of pseudo-observations, that is $\widehat{c}_t(u_{t,i},v_{t,j})$, for $i,j=1,\ldots,N$, $t=1,\ldots,T$. Finally, we apply the clr transform in eq.~\eqref{eq:clr} to obtain $\mathbf{C}_t = (\text{clr}(\widehat{c}_t)(u_{t,i},v_{t,j}))_{ij}$, for $i,j=1,\ldots,N$ and $t=1,\ldots,T$.
In compact notation, denote the matrices of pseudo-observations $\mathbf{U} = (\mathbf{u}_1,\ldots,\mathbf{u}_T)$, $\mathbf{V} = (\mathbf{v}_1,\ldots,\mathbf{v}_T)$ and the associated collection of matrices of clr-transformed copula pdf values $\mathbf{C} = [\mathbf{C}_1,\ldots,\mathbf{C}_T]$.
The series of matrices $\mathbf{C}$ is required for estimating the constrained spline functions in Section~\ref{sec:procedure_modified_fPCA}.

In the general case, when the interest lies on multivariate pdfs with unbounded support, we propose to estimate the density via product kernel estimators, with standard choices of the univariate kernels such as Gaussian or Epanechnikov.
%\end{remark}

\bigskip

\noindent Concerning the interpretation of the method, we make the following remarks:
\begin{itemize}
%\begin{remark}[Interpretation I]
\item
functions are infinite-dimensional objects, thus from a computational perspective it is impossible to deal with them directly, but a discretization step is in order. Functional data in a strict sense do no exist, instead available data can be defined as noisy observations of discretized functions. Each discretized functional data point, broadly speaking, consists of a pair of location and value of the function at location (where location has no particular meaning). For instance, for univariate functions the location is the point on the x-axis, whereas the value at the location is the corresponding value of the function (on the y-axis).
Therefore, in the current framework, we may think of the $N$ rank transformed  observations $\lbrace u_{t,i},v_{t,i} \rbrace_i$, for a given $t$, as a set of location points whereas $\widehat{c}_t(u_{t,i},v_{t,i})$ represents the value of the copula function at those locations (i.e., a discretized version of the underlying smooth function).
We remark that this is a standard interpretation in functional data analysis (e.g., see~\cite{Ramsay05Functional_data_analysis_book}) and is unrelated to the procedure developed here.
%\end{remark}

%\begin{remark}[Interpretation II]
\item
From a financial point of view, the copula pdfs are a flexible instrument providing all the information about the dependence between the marginal series $\mathbf{x}_t = \lbrace x_{t,i} \rbrace_i$ and $\mathbf{y}_t = \lbrace y_{t,i} \rbrace_i$ at time $t$. They are remarkably richer than a single scalar parameter: in addition to the extreme cases of independence (corresponding to a product copula) and perfect dependence (diagonal copula), they permit to study several particular forms of dependence, such as tail dependence (i.e. the probability of comovements of the variables in the upper/lower tail of the distribution, see~\cite{Joe97Copula_book}, \cite{Nelsen13Copula_book}).\\
From this perspective, the availability of a (estimated) time series of copula pdfs permits to have information on different forms of dependence across several periods. Instead of limiting to a descriptive analysis on the variation of (finite-dimensional) synthetic statistics built from each function $c_t(\cdot)$, we aim at characterizing how the whole dependence pattern evolves over time.
%\end{remark}
\end{itemize}

\subsection{Step 2 - Modified fPCA}\label{sec:procedure_modified_fPCA}
Starting from $\mathbf{C}$, the time series of clr-transformed pdfs values estimated in Section~\ref{sec:procedure_copula_estimation}, our goal in this section is to estimate the factor model in eq.~\eqref{eq:factor_model} using the tools from functional principal component analysis (fPCA). In words, we estimate the function $\breve{f}_t(\cdot)$ that approximates the centred log-ratio transform of the pdf $c_t(\cdot)$, for $t=1,\ldots,T$. In this section we are considering bivariate copula pdf, whose support is compact $[0,1]^d$, with $d=2$. See Section~\ref{sec:extensions} for a discussion about the general frameworks when the pdfs have unbounded support or are multivariate with $d >2$.
The strategy does not impose any assumption except that the decay of the pdf at infinity must be such that its logarithm is square integrable.
Moreover, given that probability density functions represent a special case of constrained functions, the proposed methodology can be applied as well for forecasting multivariate square integrable functions.
%\begin{equation}
%\text{clr}(c_t)(\cdot) \approx \breve{f}_t(\cdot) = \boldsymbol{\beta}_t' \boldsymbol{\xi}(\cdot) = \sum_{j=1}^J \beta_{t,j} \xi_j(\cdot) \, ,
%\label{eq:factor_model}
%\end{equation}

The outcome of this step is a vector of (time invariant) estimated factors $\boldsymbol{\widehat{\xi}}(\cdot)$ and a vector-valued time series of scores $\mathbf{\widehat{B}} = (\boldsymbol{\widehat{\beta}}_1,\ldots,\boldsymbol{\widehat{\beta}}_T)$ which will be used in Section~\ref{sec:procedure_prediction} for building a forecast of the pdf $\widetilde{c}_{T+h}(\cdot)$, with $h=1,\ldots,H$.
Appendix~\ref{sec:apdx_fPCA} provides a summary of the results from functional data analysis used in this paper, we refer to \cite{Ramsay05Functional_data_analysis_book}, \cite{FerratyVieu06Nonparam_FDA_Book} for a more detailed presentation.

\bigskip

%Our main contribution in this section is twofold: first, we extend the strategy of \cite{Machalova16clr_spline} to the multivariate case; second, we extend \cite{Liebl13FunctionalTS_app} to deal with pdfs.

We present the outline of the strategy and the results, referring to Appendix~\ref{sec:apdx_computations} for detailed computations.
Ordinary fCPA is designed for the analysis of unconstrained functions, however in our framework the object of interest are pdfs, that is functions constrained to be positive on their support and to have unit integral. This calls for a modification of standard fPCA in order to account for the constraints without the need to post-process the output.
We propose a strategy for addressing this issue consisting in the exploitation of the centred log-ratio transform and spline functions.
The clr transform allows the analysis to be carried out in the space $\mathcal{L}_2^*(I)$, which is preferred over $\mathbb{D}^*(I)$ due to its nicer properties that make easier ordinary calculus. Then, we are left with the estimation of the factor model in eq.~\eqref{eq:factor_model}, which we interpret as an eigenproblem. A first approach consists in the discretization of the functions involved and the solution of the resulting multivariate problem: despite being intuitive, this approach easily breaks down as the dimension increases because of the number of points necessary for providing a good discrete grid.
Instead, we choose to express both the target function to be approximated by the factors and the factors themselves by a finite linear combination of pre-specified basis functions. This implicitly reduces the infinite-dimensional problem to an eigenproblem for the vecor coefficients of the basis expansion. Following \cite{Machalova16clr_spline}, we choose a B-spline basis as it allows to analytically solve the resulting eigenproblem taking into account the integral constraint in eq.~\eqref{eq:integral_constraint}.

More formally, we propose to estimate the factor model in eq.~\eqref{eq:factor_model} by interpreting the functions $\boldsymbol{\xi}(\cdot)$ as the eigenfunctions of the covariance operator of the functions $\mathbf{\breve{f}}(\cdot)$, denoted $G$. For each period $t=1,\ldots,T$ and $k=1,2,\ldots$, this yields the eigenproblem:
\begin{subnumcases}{}
\label{eq:eigenproblem}
\displaystyle\int G(\mathbf{x},\mathbf{y}) \xi_k(\mathbf{x}) \: \mathrm{d}\mathbf{x} = \beta_{t,k} \xi_k (\mathbf{y}) & \\[5pt]
\displaystyle\int \xi_k(\mathbf{x}) \xi_k(\mathbf{x}) \: \mathrm{d}\mathbf{x} = 1 &
\end{subnumcases}
subject to the additional constraints $\langle \xi_k, \xi_j \rangle = 0$, for $k\neq j$ and $\int \xi_j \: \mu(\mathrm{d}\mathbf{x}) = 0$ for $j=1,2,\ldots$.
Then, we look for a solution within the class of tensor product, bivariate spline functions (see \cite[ch.8]{Ramsay05Functional_data_analysis_book} for a review of alternative solution methods), which allows to include the zero integral constraint as a linear constraint on the coefficients of the basis spline functions, thanks to the relation between splines with their derivatives.

Since a spline function can be expressed as a linear combination of known basis B-splines (see Section~\ref{sec:notation} for the notation), we need to solve a finite dimensional optimization problem for the coefficient vector of the spline. The constrained optimal smoothing problem, for each period $t=1,\ldots,T$, is:
\begin{align}
\begin{cases}
\displaystyle\min_{s_m} \int_{a_1}^{b_1} \int_{a_2}^{b_2} \left[ s_m^{(\ell_1,\ell_2)}(u,v) \right]^2 \: \mathrm{d}v \: \mathrm{d}u + \alpha \left[ \sum_{i=1}^N \sum_{j=1}^N w_{i,j} \left( \text{clr}(\widehat{c}_t)(u_{t,i},v_{t,j})-s_m(u_{t,i},v_{t,j}) \right)^2 \right] \\[3pt]
\text{s.t. } \displaystyle\int_{a_1}^{b_1} \int_{a_2}^{b_2} s_m(u,v) \: \mathrm{d}v \: \mathrm{d}u = 0
\label{eq:constrained_smoothing_problem}
\end{cases}
\end{align}
where $s_m(\cdot,\cdot)$ is a spline of degree $m$, $\ell_1,\ell_2$ are the degree of the partial derivatives with respect to $u,v$, respectively, $\lbrace u_{t,i}, v_{t,j} \rbrace_{ij}$ with $i,j=1,\ldots,N$ , are the evaluation points and $\lbrace \text{clr}(\widehat{c}_t)(u_{t,i},v_{t,j}) \rbrace_{ij}$ is the corresponding value of the clr-transformed pdf.
Notice that $N$ is number of observations allocated to each period $t=1,\ldots,T$. $\lbrace w_{i,j} \rbrace_{ij}$ is a sequence of point-specific weights, whereas $\alpha$ is the global weight of the least squares component in the smoothing problem. Finally, the interval $(a_1,b_1) \times (a_2,b_2)$ is support of the original function and of the spline.
In the following we assume: $a_1=a_2 = 0$, $b_1=b_2 = 1$, $\ell_1 = \ell_2 = 2$, meaning that we look for a solution in the class of cubic splines on the interval $[0,1]^2$. Moreover, we consider an extended knot sequence given by the regular grid $\boldsymbol{\bar{\lambda}}^{u,v} = \boldsymbol{\bar{\lambda}}^u \otimes \boldsymbol{\bar{\lambda}}^v$, with:
\begin{subequations}
\begin{align}
\boldsymbol{\bar{\lambda}}^u & = (\lambda_{-m}^u,\lambda_{-m+1}^u,\ldots,\lambda_{g+m+1}^u)' \, , \\
\boldsymbol{\bar{\lambda}}^v & = (\lambda_{-m}^v,\lambda_{-m+1}^v,\ldots,\lambda_{g+m+1}^v)' \,
\end{align}
\end{subequations}
with:
\begin{subequations}
\begin{align}
 & \lambda_{-m}^u = \ldots = \lambda_0^u < \ldots < \lambda_{g+1}^u = \ldots = \lambda_{g+m+1}^u \, \\
 & \lambda_{-m}^v = \ldots = \lambda_0^v < \ldots < \lambda_{g+1}^v = \ldots = \lambda_{g+m+1}^v \, .
\end{align}
\end{subequations}
This is a square grid with the same knots along both directions (that is, the $x$-axis and the $y$-axis, respectively), however we may choose a different number of interpolation knots for each dimension. We have decided to use the same number of knots and the same location because we are interpolating a copula probability density function with support $[0,1]^2$.

\begin{lemma}
Define $\phi_k^{m+1}(\cdot)$, $k=1,\ldots,K$ the B-spline basis functions of order $m$. The optimal spline function solving the problem in eq.~\eqref{eq:constrained_smoothing_problem} is given by:
\begin{equation}
s_m(u,v) = \mathbf{C}^{m+1}(u,v) \mathbf{d} = \sum_{k=1}^K d_k \psi_k^{m+1}(u,v) \, .
\label{eq:spline_solution}
\end{equation}
See Appendix~\ref{sec:apdx_computations_opt_smooth} for the detailed computations.
\end{lemma}

The spline functions in eq.~\eqref{eq:spline_solution} represent an interpolated multivariate probability density function, with evaluation points $(u_{t,i},v_{t,i})_i$ and values $\text{clr}(\widehat{c}_t)(u_{t,i},v_{t,i})$, for $i=1,\ldots,N$.
By repeating this procedure for each sub-sample $(\mathbf{u}_t,\mathbf{v}_t)_t$, with $t=1,\ldots,T$, we end up with a series of $T$ multivariate spline functions satisfying the zero integral constraint.
With a slight abuse of notation, define $\breve{f}_t(\cdot) = s_m(\cdot)$ the spline in eq.~\eqref{eq:spline_solution} estimated using the sub-sample $(\mathbf{u}_t,\mathbf{v}_t)_t$, for each period $t=1,\ldots,T$. Therefore, we can write in compact notation:
\begin{equation}
\breve{f}_t(	\cdot) = \sum_{k=1}^K d_{t,k} \psi_k^{m+1}(\cdot) = \mathbf{d}_t' \boldsymbol{\psi}(\cdot) \, ,
\end{equation}
where $\mathbf{d}_t = (d_{t,1},\ldots,d_{t,K})'$ and $\boldsymbol{\psi}(\cdot) = (\psi_1^{m+1}(\cdot),\ldots,\psi_K^{m+1}(\cdot))'$.
It is now possible to solve the eigenproblem in eq.~\eqref{eq:eigenproblem} using the same B-spline functions $\boldsymbol{\psi}(\cdot)$ as a basis for the principal component functions $\xi_j(\cdot)$, $j=1,2,\ldots$:
\begin{equation}
\xi_j(\cdot) = \sum_{k=1}^K a_{j,k} \psi_k^{m+1}(\cdot) = \mathbf{a}_j' \boldsymbol{\psi}(\cdot) \, ,
\label{eq:eigenproblem_eigenfuntions}
\end{equation}
where $\mathbf{a}_j = (a_{j,1},\ldots,a_{j,K})'$.
From this basis expansion, the infinite-dimensional eigenproblem in eq.~\eqref{eq:eigenproblem} reduces to a finite-dimensional optimization problem for the coefficient vectors $\mathbf{a}_j$, for $j=1,\ldots,J$.
For selecting the number of principal components $J$, we sort the estimated eigenvalues in decreasing order and compute the proportion of total variability explained by $v_j = \rho_j / \sum_k \rho_k$, for $j=1,2,\ldots$. Then, we retain the first $J$ factors accounting for a given share $\bar{d}$ of the total variability, that is $J = \argmin_{j} \lbrace \sum_j v_j \geq \bar{d} \rbrace$.
The solution of this multivariate eigenproblem is obtained by first finding the optimal $\mathbf{u}_j$ satisfying (see Appendix~\ref{sec:apdx_computations_eigenproblem} for detailed computations):
\begin{equation}
T^{-1} \mathbf{M}^{1/2} \mathbf{D}' \mathbf{D} \mathbf{M}^{1/2} \mathbf{u}_j = \rho_j \mathbf{u}_j \, ,
\label{eq:eigenproblem_final}
\end{equation}
then transforming $\mathbf{\widehat{a}}_j = \mathbf{M}^{1/2} \mathbf{\widehat{u}}_j$, for $j=1,2,\ldots$.
The solution of eq.~\eqref{eq:eigenproblem_final} yields an estimate of the principal component functions by plugging $\mathbf{\widehat{a}}_j$ in eq.~\eqref{eq:eigenproblem_eigenfuntions}:
\begin{equation}
\widehat{\xi}_j(\cdot) = \mathbf{\widehat{a}}_j' \boldsymbol{\psi}(\cdot) \, .
\label{eq:eigenproblem_eigenfuntions_estimate}
\end{equation}
Since the eigenvectors are not uniquely identified, we follow \cite{Liebl13FunctionalTS_app}  and transform them by applying the VARIMAX orthonormal rotation (see \cite{Kaiser58Varimax_rotation}, \cite{Abdi03Factor_rotations}).
The eigenvalues provide an estimate for the scores $\widehat{\boldsymbol{\beta}}_t = (\widehat{\beta}_{t,1},\ldots,\widehat{\beta}_{t,J})'$, for each period $t=1,\ldots,T$. This coincide with (see~\cite{Ramsay05Functional_data_analysis_book}):
\begin{equation}
\widehat{\boldsymbol{\beta}}_t = \begin{bmatrix}
\langle \widehat{\xi}_1, \widehat{\xi}_1 \rangle & \ldots & \langle \widehat{\xi}_1, \widehat{\xi}_J \rangle \\      \vdots & \ddots & \vdots \\    
\langle \widehat{\xi}_J, \widehat{\xi}_1 \rangle & \ldots & \langle \widehat{\xi}_J, \widehat{\xi}_J \rangle \end{bmatrix}^{-1}
\begin{bmatrix}
\langle \breve{f}_t, \widehat{\xi}_1 \rangle \\ \vdots \\ \langle \breve{f}_t, \widehat{\xi}_J \rangle
\end{bmatrix} \, .
\label{eq:scores_estimate}
\end{equation}
As final output of this step we obtain the estimated time series of scores $\mathbf{\widehat{B}} = (\widehat{\boldsymbol{\beta}}_1,\ldots,\widehat{\boldsymbol{\beta}}_T)$.

\bigskip

%\begin{remark}[Convergence]
Each estimated eigenfunction can be seen as a continuous function of the clr-transformed functions, that is $\widehat{\xi}_j(\cdot) = g(\breve{f}_1(\cdot),\ldots,\breve{f}_T(\cdot))$. Hence, by the continuous mapping theorem, the estimator of the eigenfunction is consistent provided that the estimators for the clr-transformed functions are consistent too. Recall that each $\breve{f}_t$ corresponds to the centred log-ratio (continuous and smooth) transformation of a copula pdf, and it is estimated via a spline. It is known (see \cite{Schumaker07Spline_book}) that splines approximate arbitrarily well continuous smooth functions on a bounded interval.

Consequently, from the consistency of splines in approximating a smooth function (as is $\breve{f}_t(\cdot)$, $t=1,\ldots,T$ in our case) it descends the consistency of the estimator for each eigenfunction $\widehat{\xi}_j(\cdot)$ and, by another application of the continuous mapping theorem, the consistency of the estimator of the associated scores $\boldsymbol{\widehat{\beta}}_t$.
%\end{remark}

\subsection{Step 3 - Prediction}\label{sec:procedure_prediction}
In this last step, we aim at obtaining a $H$ steps ahead forecast $\widetilde{c}_{T+H}(\cdot)$ of the pdf $c_{t}(\cdot)$. The task is accomplished in three steps: first, we estimate a VAR($p$) process on the time series of estimated principal component scores from Section~\ref{sec:procedure_modified_fPCA}, $\lbrace \boldsymbol{\widehat{\beta}}_t \rbrace_{t=1}^T$, then we use the fitted values for obtaining a forecast of the scores $\boldsymbol{\widetilde{\beta}}_{T+h}$, $h=1,\ldots,H$. Next, for $h=1,\ldots,H$ we derive a forecast for the approximated function $\widetilde{f}_{T+h}(\cdot)$ by plugging-in eq.~\eqref{eq:factor_model} and finally we get the forecast of the pdf $\widetilde{c}_{T+h}(\cdot)$ by applying the inverse clr transform to $\widetilde{f}_{T+h}(\cdot)$.

The estimated scores from the Section~\ref{sec:procedure_modified_fPCA} for a vector-valued time series, where each vector has length $J$. We propose to model the time series through a VAR($p$), as follows:
\begin{equation}
\boldsymbol{\widehat{\beta}}_t = \boldsymbol{\phi}_{const} + \boldsymbol{\phi}_{trend} t + \sum_{l=1}^p \boldsymbol{\Phi}_l \boldsymbol{\widehat{\beta}}_{t-l} + \boldsymbol{\epsilon}_t \, ,   \qquad   \boldsymbol{\epsilon}_t \distas{iid} \mathcal{N}(\mathbf{0}, \sigma^2 \mathbf{I}_J) \, .
\label{eq:VAR_scores}
\end{equation}
Denoting the estimated coefficients by $(\widehat{\boldsymbol{\phi}}_{const},\widehat{\boldsymbol{\phi}}_{trend},\boldsymbol{\widehat{\Phi}}_1,\ldots,\boldsymbol{\widehat{\Phi}}_p)$, we perform forecasts for each $h=1,\ldots,H$ steps ahead in the usual way:
\begin{equation}
\boldsymbol{\widetilde{\beta}}_{T+h} = \widehat{\boldsymbol{\phi}}_{const} +\widehat{\boldsymbol{\phi}}_{trend} (T+h) + \sum_{l=1}^p \boldsymbol{\widehat{\Phi}}_l \boldsymbol{\widehat{\beta}}_{T-l} \, .
\label{eq:VAR_forecast}
\end{equation}
Then, we obtain the predicted clr-transformed function $\widetilde{f}_{T+h}(\cdot) \in \mathcal{L}_2^*(I)$ by substituting $\boldsymbol{\widetilde{\beta}}_{T+h}$ and the estimated principal components $\boldsymbol{\widehat{\xi}}(\cdot)$ into eq.~\eqref{eq:factor_model}, thus obtaining for $h=1,\ldots,H$:
\begin{equation}
\widetilde{f}_{T+h}(\cdot) = \boldsymbol{\widetilde{\beta}}_{T+h}' \boldsymbol{\widehat{\xi}}(\cdot) = \sum_{j=1}^J \widetilde{\beta}_{T+h,j} \widehat{\xi}_j(\cdot) \, .
\end{equation}
Finally, in order to compute the predicted probability density function $\widetilde{c}_{T+h}(\cdot) \in \mathbb{D}^*(I)$ we apply the inverse centred log-ratio transformation, for $h=1,\ldots,H$:
\begin{equation}
\widetilde{c}_{T+h}(\cdot) = \text{clr}^{-1}(\widetilde{f}_{T+h})(\cdot) = \frac{\exp \left\lbrace \widetilde{f}_{T+h}(\cdot) \right\rbrace}{\displaystyle\int \exp \left\lbrace \widetilde{f}_{T+1}(\cdot) \right\rbrace } \, . %\: \mathrm{d}u \: \mathrm{d}v
\end{equation}
The final outcome of the whole procedure is the set of forecasts of the multivariate pdf $\mathbf{\widetilde{c}}_{T+H}(\cdot) = (\widetilde{c}_{T+1}(\cdot),\ldots,\widetilde{c}_{T+H}(\cdot))'$.

\bigskip

%\begin{remark}[Dimensionality]\label{rmk:VAR_lasso}
The size of the VAR process in eq.~\eqref{eq:VAR_scores} corresponds to the number of principal components selected in the fPCA, $J$ and is generally small. Therefore, the dimensionality of the VAR does not hamper the estimation procedure even though the length $T$ of the time series is not really long. This is a consequence of the dimensionality reduction brought by the fPCA, interpreted as a factor model here.

Nonetheless, in higher-dimensional settings it may be still possible to estimate the coefficient matrix in eq.~\eqref{eq:VAR_scores} by adding a regularization term. The recent contributions, \cite{Nicholson16HighDim_VAR_forecast} and \cite{Nicholson17VARX_regularization} designed and implemented\footnote{Estimation can be carried out using the \texttt{R} (\href{https://cran.r-project.org}{https://cran.r-project.org}) package ``\texttt{BigVAR}'' (\href{https://cran.r-project.org/web/packages/BigVAR/index.html}{https://cran.r-project.org/web/packages/BigVAR/index.html}), see~\cite{Nicholson17VARX_regularization}.} several types of penalized regression for large VARX models (including the LASSO case) allowing up to $d=130$ marginal series.
%\end{remark}

As regards the numerical implementation of the procedure, the core of the proposed methodology relies on standard linear algebra operations, for which computationally efficient algorithms are available. Moreover, the dimensionality reduction brought by the fPCA has the additional advantage of reducing the size of the coefficient matrix of the VAR process do be estimated. Overall, the entire procedure represented in Algorithm~\ref{alg:procedure} is quite fast (see the details for the application in Section~\ref{sec:application}).

\section{Extensions}\label{sec:extensions}
Here we briefly discuss some possible extensions of the methodology discussed in Section~\ref{sec:methodology}.

\subsection{Unbounded support}
The results in \cite{Boogaart10Bayes_Linear}, \cite{Boogaart14Bayes_Hilbert} hold also for pdfs with unbounded support, provided that they are absolutely continuous with respect to a measure with finite total mass. This requirement is a direct consequence of the formula for the centred log-ratio in eq.~\eqref{eq:clr}, which involves at the denominator the total mass of the support.
In fact, the problem when dealing with pdfs defined on an unbounded region is that the Lebesgue measure of the whole domain is not finite, hence it would be necessary to choose a different, finite the reference measure of the spaces $\mathbb{D}(I),\mathbb{D}^*(I),\mathcal{L}_2(I),\mathcal{L}_2^*(I)$. If the new reference measure $\nu$ is absolutely continuous with respect to the Lebesgue measure, i.e. $\mathrm{d}\nu = g(\cdot) \: \mathrm{d}\mu$, then for $h(\cdot) \in \mathbb{D}_\nu(I)$ it holds $f(\cdot) = h(\cdot) g(\cdot) \in \mathbb{D}_\mu(I)$. Therefore in the particular case $\nu \ll \mu$ performing the analysis of the original pdf series $h_t(\cdot)$ under the reference measure $\nu$ is equivalent to perform the analysis of the modified series $h_t(\cdot) g(\cdot)$ under the Lebesgue measure.

\begin{example}[Alternative reference measure]
Let $I=\mathbb{R}^n$ and let $g = \mathrm{d}\mathbb{P}_\mathcal{N} / \mathrm{d}\mu$ to be the Radon-Nikodym derivative of the finite standard Gaussian measure $\mathbb{P}_\mathcal{N}$ with respect to the $n$-dimensional Lebesgue measure on $\mathbb{R}^n$ (thus, $g$ is the pdf of a standard normal distribution), the change of measure yields:
\begin{align*}
\int_{\mathbb{R}^n} f(\mathbf{x}) \: \mathrm{d}\mu = \int_{\mathbb{R}^n} f(\mathbf{x}) \: \frac{\mathrm{d}\mu}{\mathrm{d}\mathbb{P}_\mathcal{N}} \mathrm{d}\mathbb{P}_\mathcal{N} = \int_{\mathbb{R}^n} \frac{f(\mathbf{x})}{g(\mathbf{x})} \: \mathrm{d}\mathbb{P}_\mathcal{N} = \int_{\mathbb{R}^n} h(\mathbf{x}) \: \mathrm{d}\mathbb{P}_\mathcal{N} \, .
\end{align*}
If $\log(h)(\mathbf{x})$ is square integrable, then all the previous results can be applied, since $\mathbb{P}_\mathcal{N}(\mathbb{R}^n) = 1$. If instead $I=\mathbb{R}_+$ one may use the measure $\omega$ induced by a Gamma distribution, since $\omega(\mathbb{R}_+) = 1$.
\end{example}

\begin{example}[Clr with unbounded support]
Let $p_0 = \mathrm{d}\mathbb{P}_\mathcal{N} / \mathrm{d}\mu$ be the density of the standard Gaussian measure with respect to the Lebesgue measure on $\mathbb{R}$. Let $\nu$ be a measure and $p_\nu = \mathrm{d}\nu / \mathrm{d}\mu$ be its density with respect to the Lebesgue measure. Let $g = \mathrm{d}\nu / \mathrm{d}\mathbb{P}_\mathcal{N}$ be the density of $\nu$ with respect to the Gaussian measure.
Since $\mu(\mathbb{R}) = \infty$, the centred log-ratio for $g$ is not defined. However, by changing measure from $\mu$ to $\mathbb{P}_\mathcal{N}$ we obtain:
\begin{align}
\notag
\textnormal{clr}(g)(\cdot) & = \log \left( \dfrac{\mathrm{d}\nu}{\mathrm{d}\mathbb{P}_\mathcal{N}} \right)(\cdot) - \dfrac{1}{\mathbb{P}_\mathcal{N}(\mathbb{R})} \int_\mathbb{R} \log \left( \dfrac{\mathrm{d}\nu}{\mathrm{d}\mathbb{P}_\mathcal{N}} \right)(u) \: \mathrm{d}\mathbb{P}_\mathcal{N}(u) \\ \notag
 & = \log \left( \dfrac{\mathrm{d}\nu}{\mathrm{d}\mu} \dfrac{\mathrm{d}\mu}{\mathrm{d}\mathbb{P}_\mathcal{N}} \right)(\cdot) - \int_I \log \left( \dfrac{\mathrm{d}\nu}{\mathrm{d}\mu} \dfrac{\mathrm{d}\mu}{\mathrm{d}\mathbb{P}_\mathcal{N}} \right)(u) \:  \dfrac{\mathrm{d}\mathbb{P}_\mathcal{N}}{\mathrm{d}\mu} \: \mathrm{d}\mu(u) \\
 & = \log \left( \dfrac{p_\nu}{p_0} \right)(\cdot) -\int_I \log \left( \dfrac{p_\nu}{p_0} \right)(u) \: p_0(u) \: \mathrm{d}\mu(u) \, .
\end{align}
Notice that the integral on the last line is an expectation with respect to the probability measure $\mathbb{P}$, also Monte Carlo methods for numerical integration can be applied if the density $p_0$ can be easily sampled from, as is, for example, when $p_0$ is the pdf of a normal distribution.
\end{example}

Once a new reference measure has been chosen and the clr transform has been applied accordingly, the unbounded support is no more of concern for the methodology. In fact, the B-spline basis functions are defined also on unbounded regions and are computed for a given, finite knot sequence. The location of the knots would depend on the fatness of the tails of the densities, since fatter tails would require the knot sequence to be more scattered for having the resulting spline interpolating well the pdf. For example, a standard normal random variable has unbounded support, but almost the $95\%$ of the mass in the interval $[-2,2]$.

Consequently, the unboundedness of the support of the pdfs affect the spaces $\mathbb{D}_\nu^*(I),\mathcal{L}_{2,\nu}^*(I)$ to which the functions belong, but does not require a modification of the other parts of the procedure, since the basic constructions behind the result in eq.~\eqref{eq:spline_solution} are left unchanged.

\subsection{Multivariate case: $d > 2$}
The proposed methodology can be easily extended to deal with $d$-dimensional ($d>2$) probability density functions. The change would be involve the size of the sparse block diagonal matrices described in Appendix~\ref{sec:apdx_computations}.

The only concern that arises when $d > 2$ is the curse of dimensionality, as is typical in nonparametric statistics. In the proposed model this occurs through the need for an increasingly high number of observations $\lbrace x_{1,t,n}, \ldots, x_{d,t,n} \rbrace_n$ for each period $t$ in order to provide a good kernel estimation of the copula probability density function.
In addition, if the high dimension is associated to a high degree of complexity of the dependence structure, it may be necessary also to increase the number of principal components to keep, $J$. This in turn results in a higher dimensionality of the VAR model for the scores in Section~\ref{sec:procedure_prediction}. However, we do not expect this to be a significant obstacle, as compared to the previous issue which represents the true bottleneck to high-dimensional applications.

%%%%%%%%%%%%%%%%%%%%%%%%%%%%%%%%%%%%%%%%%%%%%%%%%%%%%%%%%%%%%%%%%%%%%%%%%%%%%%%%
\section{Simulation example}\label{sec:simulation}
\textbf{To be included}.

\section{Application}\label{sec:application}
The dataset is composed by daily observations of S\&P500 and NASDAQ indices from $1$st January $1980$ to $31$st December $2017$, for a total of $10,032$ observations over $38$ years.
We  make the following assumptions. We start by taking first differences of the two series in order to remove non-stationarity, then for each period $t=1,\ldots,T$ we assume to observe a sample $\lbrace x_{t,i},y_{t,i} \rbrace_{i=1}^N$, with $N=247$, of intra-period observations $(x_{t,i}, y_{t,i}) \in I = [0,1]^2$.
We compute the copula pseudo-observations $(u_{t,i}, v_{t,i}) = (F_x^N(x_{t,i}), F_y^N(y_{t,i}))$, $i=1,\ldots,N$, for each $t=1,\ldots,T$ via the rank of the observations.
Then, the empirical copula probability density function is estimated non-parametrically with the Beta kernel density estimator (\cite{Charpentier07Estimation_CopulaPDF_Kernel}, \cite{Chen99EstimationPDF_BetaKernel}), using a diagonal bandwidth. The choice of the bandwidth for Beta kernel estimators is tricky since no rules of thumb are available for its optimal choice. As a consequence, we performed several experiments with varying $h$ and found that the value $h=0.05$ provides a good balance between bias and variance. Smaller or greater values resulted in too wiggly or too wide contours, respectively, which we interpreted as indicators of a bad value of the bandwidth.

%\begin{table}[H]
%\centering
%\begin{tabular}{c c c c c}
%\toprule
%\textbf{Frequency} $t$ & $N$ & spline & $T$ & time series  \\ \midrule
%Year    & 248 & $(u_i^{(t)},v_i^{(t)}) \rightarrow c_t(\cdot,\cdot)$ &  38 & $c_t(\cdot,\cdot) \rightarrow b_{t,j}$ \\
%Quarter &  60 & & 152 & \\
%Month   &  18 & & 456 & \\
%\bottomrule
%\end{tabular}
%\end{table}

The motivation for the choice of this splitting of the sample into $T=38$ years is twofold. First, this allows us to estimate the function $c_t(\cdot)$, for each $t$, using up to  $N=248$ data points, while keeping a time series of estimated functions of length $T=38$, thus providing a good balance of the data between the intra-period and the temporal dimensions. Second, a intra-period window corresponding to one year permits to embed the seasonal effects (with period up to one year) in the estimated function, thus removing the need to account for them when analysing the time series $c_t(\cdot)$, $t=1,\ldots,T$.
We take first differences of the raw data (see Appendix~\ref{sec:apdx_application_plots} for additional plots), thus reducing the size of each sub-sample to $N=247$, in order to account  for non-stationarity of the original series.

We choose the following values of the parameters:
\begin{equation}
\begin{matrix*}
T=38 & N=247 & H = 10 & h=0.05    & \bar{d} = 0.92 \\
g=4  & m=3   & \ell=2 & \alpha=0.8 & \mathbf{W} = \mathbf{I}_{n^2}
\end{matrix*}
\end{equation}
After having estimated the copula pdfs $\widehat{c}_1(\cdot),\ldots,\widehat{c}_{38}(\cdot)$, we de-meaned them using the perturbation and powering operations defined in Section~\ref{sec:preliminaries}, obtaining $	\widehat{\widehat{c}}_t(\cdot) = \widehat{c}_t(\cdot) \ominus \bar{c}$, where $\bar{c} = 1/T \odot \bigoplus_{t=1}^T \widehat{c}_t(\cdot)$, which has been used as input for the step 2 of Algorithm~\ref{alg:procedure}.

The number of eigenfunctions to take has been estimated as described in Section~\ref{sec:procedure_modified_fPCA}, by $J= \argmin_j \lbrace \sum_j \widehat{\rho}_j \geq \bar{d} \rbrace$, yielding $J=4$.
Values of $\alpha$ lower (greater) than unity imply higher (lower) relative weight of the smoothing component with respect to the least squares in the constrained optimal smoothing problem in eq.~\eqref{eq:constrained_smoothing_problem}. We found that $\alpha = 0.8$ provides a good balance between the two.
As robustness check, we performed the analysis with different values of $\bar{d}$ (thus implying different number of eigenfunctions $J$) without significant changes.
%Alternative values of the number of points $n$ to be used in the constrained optimal smoothing problem lead to slightly different results, therefore we choose $n=60$ in order to reduce the computational cost of the algorithm.
We run Algorithm~\ref{alg:procedure} on an Apple MacBookPro with a 3.1GHz Intel Core i7 processor, RAM 16GB, using MATLAB r2017b without exploiting parallel calculus. This required around ten minutes of computation, with step 1 being the most computational intensive part.

The value of the BIC for several specifications of the VAR($p$) model in eq.~\eqref{eq:VAR_scores} for the time series of scores are reported in Table~\ref{tab:BIC} and suggest to choose a VAR($5$) model including a constant and a time trend. All the estimated VAR models are stable.

\begin{table}[!t]
\centering
\begin{tabular}{| c | c | c | c | c |}
\toprule
\textbf{Lags} & \textbf{Model A} & \textbf{Model B} & \textbf{Model C} & \textbf{Model D} \\
\midrule
1 & $1823.5$ & $1799.6$ & $1821.5$ & $1797.8$ \\
2 & $1831.7$ & $1804.0$ & $1827.0$ & $1799.3$ \\
3 & $1834.2$ & $1813.6$ & $1826.3$ & $1804.4$ \\
4 & $1820.7$ & $1804.2$ & $1801.2$ & $1769.1$ \\
%5 & $1702.3$ & $1677.0$ & $1672.3$ & $\mathbf{1613.7}$ \\
\bottomrule
\end{tabular}
\caption{BIC for different VAR specifications of the VAR($p$) model in eq.~\eqref{eq:VAR_scores}. Model A: no constant, no trend; model B: constant, no trend; model C: trend, no constant; model D: constant and trend. The best model according to BIC is in bold.}
\label{tab:BIC}
\end{table}

\begin{sidewaysfigure}
%\centering
\captionsetup{width=0.9\linewidth}
\includegraphics[trim=40mm 10mm  0mm 10mm, clip, height= 15.0cm, width= 25.0cm]{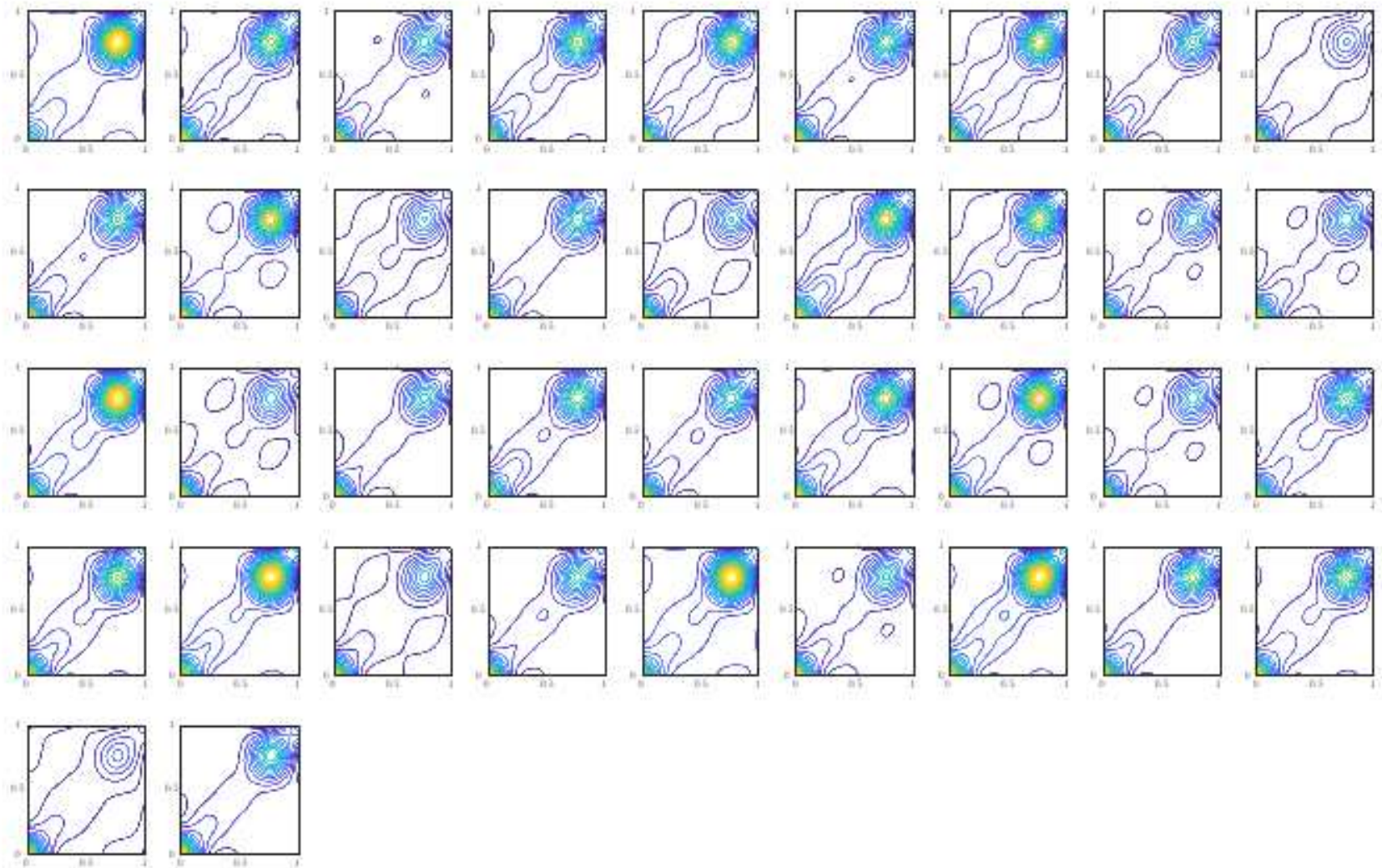}
\caption{\label{fig:ts_contour} Contour plot of time series of bivariate copula pdfs, approximated via fPCA, for each year $t=1,\ldots,T$, starting from $t=1$ in the top-left panel.}
\end{sidewaysfigure}

\begin{figure}[H]
\centering
\captionsetup{width=0.9\linewidth}
\includegraphics[trim=10mm 20mm  10mm 10mm, clip, height= 10.0cm, width= 15.0cm]{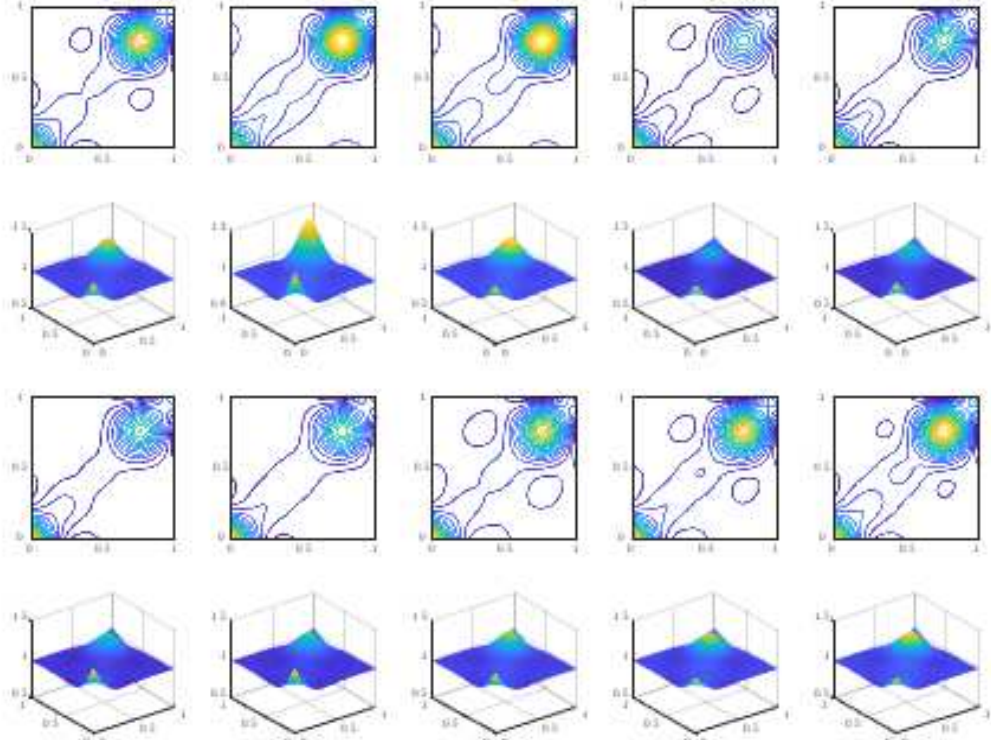}
\caption{\label{fig:forec_all} Contour plots (\textit{first} and \textit{third} row) and the corresponding 3D density plot (\textit{second} and \textit{fourth} row) of the forecasted bivariate copula pdfs, approximated via fPCA, for each horizon $h=1,\ldots,5$ (first and second rows) and $h=6,\ldots,10$ (third and fourth rows), starting from the top-left panel.}
\end{figure}

For comparing the results, the estimated copula pdfs $\widehat{c}_t(\cdot)$, $t=1,\ldots,T$ (respectively, the forecasted copula pdfs $\widetilde{c}_{T+h}(\cdot)$, $h=1,\ldots,H$) have been computed by applying the inverse clr map to the functions $\breve{f}_t(\cdot)$ ($\widetilde{f}_{T+h}(\cdot)$) estimated (forecasted) from the factor model in eq.~\eqref{eq:factor_model}, using $J$ eigenfunctions. Fig.~\ref{fig:ts_contour} shows the contour plot of the time series of the estimated bivariate copula pdfs $\widehat{c}_t(\cdot)$, $t=1,\ldots,T$, whereas Fig.~\ref{fig:forec_all} reports the contour and 3D density plots of the forecasted pdfs $\widetilde{c}_{T+h}(\cdot)$, $h=1,\ldots,H$. 
We found that:
\begin{itemize}
\item there is evidence of significant temporal changes of the estimated pdfs $\widehat{c}_t(\cdot)$. Periods (i.e. years) where the joint probability is concentrated around the top-right corner, meaning strong upper tail dependence, alternate with periods where the opposite happens, that is, periods with strong lower tail dependence. There are two main implications of this stylized fact:
\begin{itemize}
\item it signals that none of the copula families considered in Table~\ref{tab:tail_dependence_copula_families}, which are the most commonly used in econometrics, is able to account for the varying dependence over the whole time span of the sample, not even by letting the copula parameter vary over time. The reason is that all of them have either only one type of tail dependence (upper or lower), or both but in symmetric way.
Moreover, the same conclusion holds even if a dynamic copula model is specified by allowing the copula parameter to change over time.

\item it is consistent with the results of \cite{Guegan10DynamicCopula_test} discussed in Section~\ref{sec:introduction}, who found that a dynamic copula model for the whole sample is not satisfactory and that different parametric copula families should be used for modelling different temporal windows.
\end{itemize}

\item Fig.~\ref{fig:ts_forec_beta} in Appendix~\ref{sec:apdx_application_plots} shows the time series of the fPCA scores along with their forecasts (with $95\%$ confidence intervals). For all series we do not reject the null hypothesis of stationarity (using the ADF test).
Moreover, by comparing Fig.~\ref{fig:ts_contour} with Fig.~\ref{fig:ts_forec_beta} we find that smooth evolutions of the fPCA scores of the clr-transformed pdfs are able to generate significant changes of the pdf.

\item the forecasts of the bivariate copula pdf in Fig.~\ref{fig:forec_all} are smoothly varying over the forecasting horizon. The heterogeneity of the tail dependence observed in the sample seems to be replicated in the forecasts: for example, the upper tail dependence forecasted at horizon 2-3 is significantly reduced between horizons 4-7 and shows again at horizons 9-10. Consequently, we find that the proposed methodology is able to provide non-flat forecasts which can capture and describe the temporal evolution of the bivariate time series.
\end{itemize}

%\begin{sidewaysfigure}
%%\centering
%\captionsetup{width=0.9\linewidth}
%\includegraphics[trim=40mm 0mm  0mm 0mm, clip, height= 5.0cm, width= 25.0cm]{forec_all60_line.eps}
%\caption{\label{fig:forec_all} Contour plots (\textit{first} row) and 3D density plot (\textit{second} row) of the forecasted bivariate copula pdfs, approximated via fPCA, for each horizon $h=1,\ldots,H$, starting from $h=1$ on the left.}
%\end{sidewaysfigure}

\begin{figure}[!t]
\centering
\includegraphics[height= 4.0cm, width= 12.0cm]{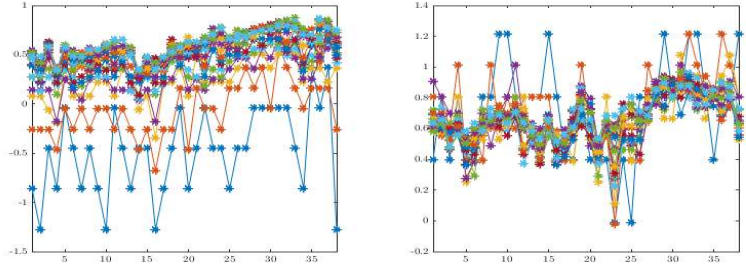}
\caption{Upper (\textit{left}) and lower (\textit{right}) tail dependence coefficients of the bivariate time series $(\mathbf{x}_t,\mathbf{y}_t)$, for $t=1,\ldots,38$ (\textit{x-axis}). Each curve corresponds to a different threshold $u=0.01,0.02,\ldots,0.20$.}
\label{fig:lUL_allT}
\end{figure}

\begin{figure}[!t]
\centering
\includegraphics[height= 4.0cm, width= 12.0cm]{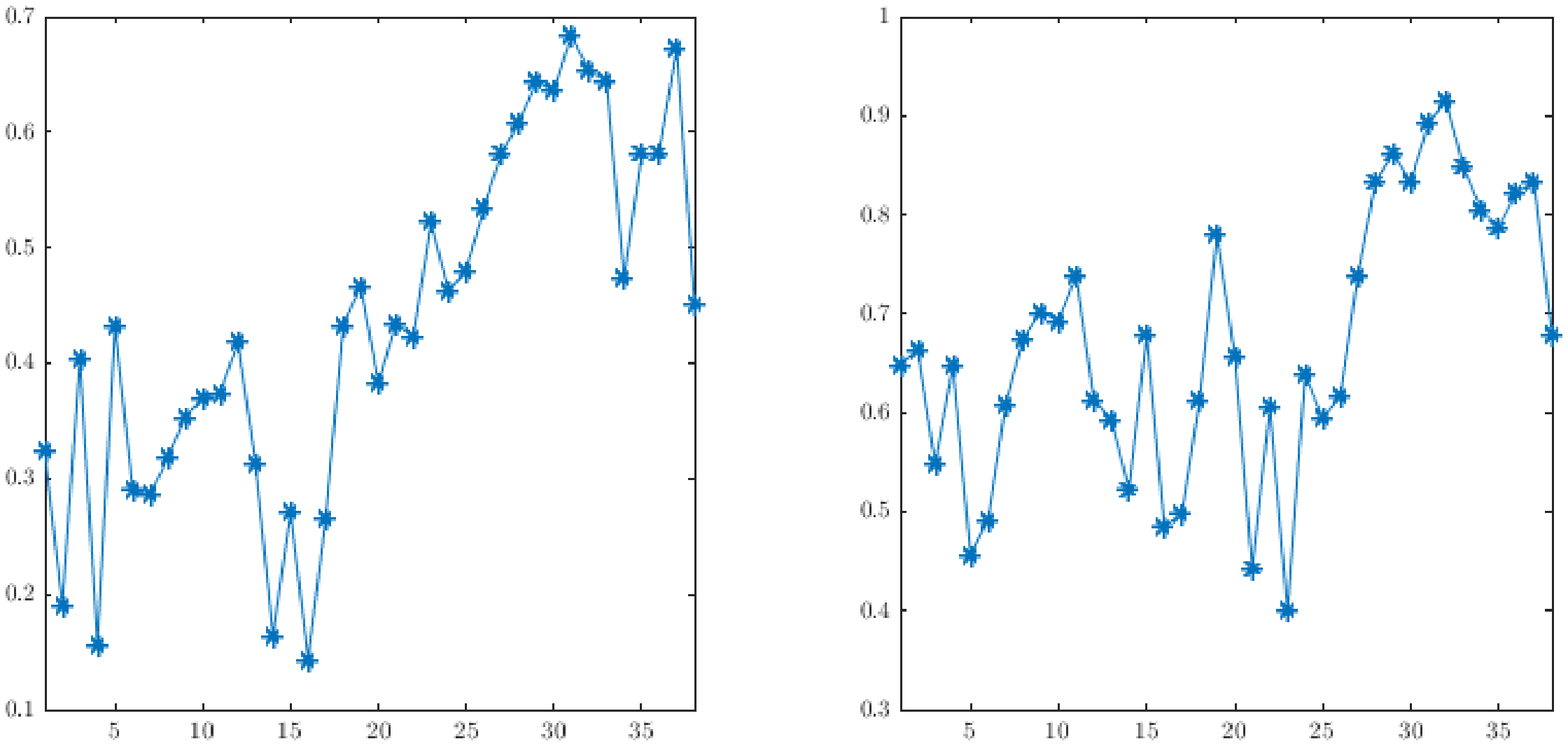}
\caption{Upper (\textit{left}) and lower (\textit{right}) tail dependence coefficients of the bivariate time series $(\mathbf{x}_t,\mathbf{y}_t)$, for $t=1,\ldots,38$ (\textit{x-axis}), threshold $u=0.10$.}
\label{fig:lUL_avgT}
\end{figure}

Several parametric and nonparametric estimators for the TDC $\lambda_U,\lambda_L$ defined in eq.~\eqref{eq:TDC_u}-\eqref{eq:TDC_u_l_log} have been proposed in the literature. Here we use the non-parametric estimator obtained from eq.~\eqref{eq:TDC_u_l_log}. Let $u \in [0,1]$ be an arbitrarily small threshold and let $\hat{C}_N(\cdot)$ be the empirical copula cumulative probability function, then the estimator is defined by (see \cite{Frahm05Estimator_TailDependence}):
\begin{align}
\label{eq:TDC_estimator}
\widehat{\lambda}_U = 2- \frac{\log(\hat{C}_N(1-u,1-u))}{\log(1-u)} \, , \qquad \qquad \widehat{\lambda}_L = 2- \frac{\log(1-2u+\hat{C}_N(u,u))}{\log(1-u)} \, .
\end{align}
Fig.~\ref{fig:lUL_allT} shows the estimated tail dependence coefficients for the sample observations, for each period $t=1,\ldots,38$, using a grid of $20$ equally spaced threshold values between $0.01$ and $0.20$. Instead, Fig.~\ref{fig:lUL_avgT} plots only the case for the median value of the threshold values, i.e. $u=0.10$.
We find significant variation of both the upper and lower tail dependence coefficients over time, which are always different from zero. In addition, the values of the upper TDC differ from those of the lower TDC, thus highlighting an asymmetric tail dependence. The threshold parameter seems to exert a minor role, as almost all the trajectories of both $\widehat{\lambda}_U$, $\widehat{\lambda}_L$ remain quite close to each other, except for few values of $u$, thus indicating robustness of the results (with respect to $u$). Moreover, the range of variation of the lower TDC is slightly higher than that of the upper TDC, in line with the previous findings in the financial econometrics literature (see \cite{Cherubini04Copula_finance}).

\begin{figure}[!t]
\centering
\includegraphics[height= 4.0cm, width= 12.0cm]{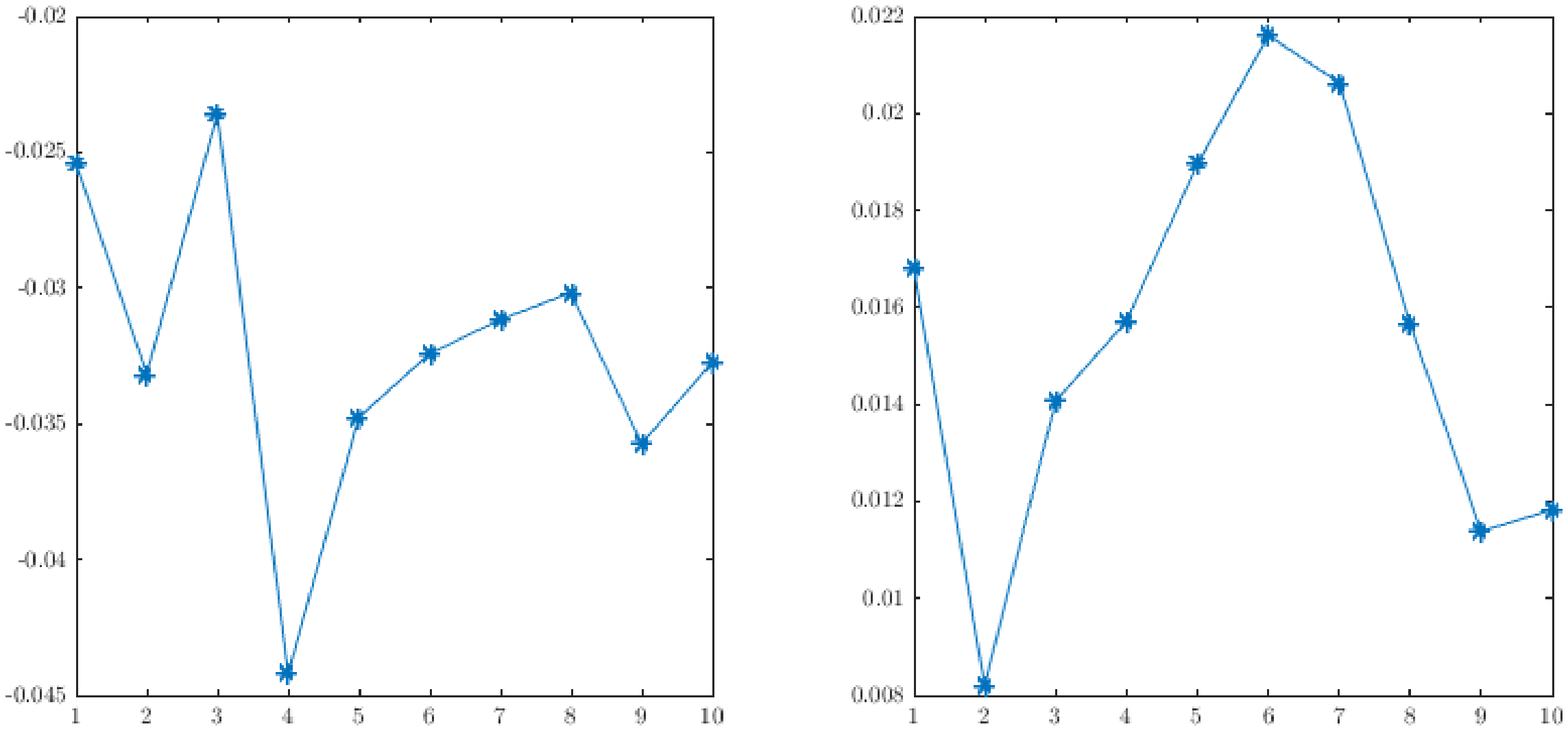}
\caption{Upper (\textit{left}) and lower (\textit{right}) tail dependence coefficients of the forecasted bivariate copula pdf $c_{T+h}(\cdot)$, for $h=1,\ldots,10$ (\textit{x-axis}), threshold $u=0.10$.}
\label{fig:lUL_avgH}
\end{figure}

Fig.~\ref{fig:lUL_avgH} shows the estimated tail coefficients for the forecasted copula pdfs, for each horizon $h=1,\ldots,10$, using  the threshold value $u=0.10$. The results are in line with the findings of the sample data: both estimated coefficients are different from zero and asymmetric between the upper and lower case, furthermore they change over time. Considered together, the findings in and out of sample estimated TDC points towards the rejection the use of copula families has either only one type of tail dependence or symmetric tail dependence, as show in Table~\ref{tab:tail_dependence_copula_families}.

\begin{remark}[Interpretation]
The proposed methodology, as opposed to standard (semi)parametric dynamic copula models allows to visualize and quantify the temporal evolution of both the upper and lower tail dependence between bivariate time series, as well as to estimate the associated TDC.
These findings suggest that the use of this methodology can improve the state-of-the-art on risk modelling due to its flexibility in modelling the dynamics of the dependence between random variables, which is the cornerstone for definition of adequate risk measures.
\end{remark}

\section{Conclusions}\label{sec:conclusions}
The time varying nature of the dependence pattern between financial variables is a challenging issue in statistics and econometrics. Common methods based on the specification of a dynamic copula model are not enough flexible to describe the temporal change, because each copula family has only a specific kind of tail dependence.

We contribute to this literature by proposing a nonparametric model for forecasting multivariate probability density functions with bounded or unbounded support. The methodology is used for studying the temporal evolution of the  copula probability density function encrypting the dependence structure between the S\&P500 and the NASDAQ indices.
We found evidence of time varying tail dependence which cannot be captured by commonly used econometric models based on dynamic copulas, whereas the model we propose is able to account for these changes. The forecasts highlight smooth but significant variation of the bivariate copula pdf.

The proposed methodology is quite general and can be applied also to other domains. An appealing framework deserving further research concerns the definition of time varying graphical models through dynamic vine copulas (\cite{Bedford02}, \cite{Joe11VineCopula_handbook}), which combine a tree-like graphical structure, for representing the conditional independence relationships among a set of variables, with bivariate copulas, which describe the pairwise conditional dependence.
Here, the method can be used for (separately) modelling the temporal evolution of each bivariate copula characterizing the edges of the network. Our methodology can be parallelized over the edges, for coupling with the issue of dimensionality.

Another stream of research worth further investigation regards the empirical analysis of multivariate (with dimension $d> 2$) pdfs with unbounded support, such as multivariate normal distributions, which are the building block of many well-known econometric models.

% BIBLIOGRAPHY - plain apalike siam chicago ecta
\bibliographystyle{ecta}
\bibliography{refsTVdensity.bib}

%%%%%%%%%%%%%%%%%%%%%%%%%%%%%%%%%%%%%%%%%%%%%%%%%%%%%%%%%%%%%%%%%%%%%%%%%%%%%%%%
\begin{appendix}
\section{Functional PCA}\label{sec:apdx_fPCA}
In this section, denote $\mathbf{f}(\cdot) = (f_1(\cdot),\ldots,f_T(\cdot))'$ a sequence of $T$ random functions $f_t:\mathbb{R}^n \rightarrow \mathbb{R}$ and let $V:\mathbb{R}^n\times \mathbb{R}^n \rightarrow \mathbb{R}$ be the covariance operator defined as:
\begin{equation}
V(f)(\cdot) = \int_{\mathbb{R}^n} v(\cdot,\mathbf{y}) f(\mathbf{y}) \: \mathrm{d}\mathbf{y} \, ,
\label{eq:apdx_fPCA_covariance_operator}
\end{equation}
where the kernel $v:\mathbb{R}^n \times \mathbb{R}^n \rightarrow \mathbb{R}$, expressed as $v(\mathbf{x},\mathbf{y})$, is the covariance function.

Functional principal component analysis (fPCA) is the infinite-dimensional analogue of multivariate principal component analysis (PCA), from which it borrows the terminology and interpretation (see~\cite[ch.8]{Ramsay05Functional_data_analysis_book} and~\cite{FerratyVieu06Nonparam_FDA_Book}).
It is possible to interpret fPCA as a truncated the Karhunen-Lo\'{e}ve decomposition (\cite{Karhunen47},~\cite{Loeve45}). The latter is used to represent a function $f:\mathbb{R}^n \rightarrow \mathbb{R}$ via an infinite linear combination of basis functions $\xi_j(\cdot)$ with coefficients $\beta_j$ given by:
\begin{equation}
f(\mathbf{x}) = \sum_{j=1}^{\infty} \beta_j \xi_j(\mathbf{x}) \, .
\label{eq:KL_multivariate}
\end{equation}
In fPCA, the infinite sum is truncated by keeping only $J$ components, thus reducing the infinite-dimensional problem into a finite-dimensional one, given by $(\xi_j(\cdot),\beta_j)$, $j=1,\ldots,J$.
In fact, the purpose of fPCA is to find out the linear combination of principal component functions (or factors) $\boldsymbol{\xi}(\cdot) = (\xi_1(\cdot),\ldots,\xi_J(\cdot))'$ and principal component scores (or loadings) $\boldsymbol{\beta} = (\beta_1,\ldots,\beta_J)'$, which best approximates a given function (or series of functions). The factors represent the main modes of variability and the scores specify the weight of each principal component function in the approximation of the observed function.
Let $\mathbf{\breve{f}} = (\breve{f}_1(\cdot),\ldots,\breve{f}_T(\cdot))'$ the set of functions approximating the series $\mathbf{f} = (f_1(\cdot),\ldots,f_T(\cdot))'$. Then each $\breve{f}_t(\cdot)$ is obtained as:
\begin{equation}
\breve{f}_t(\cdot) = \boldsymbol{\beta}_t' \boldsymbol{\xi}(\cdot) = \sum_{j=1}^J \beta_{t,j} \xi_j(\cdot) \, .
\end{equation}
For identification purposes, the principal component functions are often constrained to be orthonormal, that is $|| \xi_j(\cdot) ||_2 = 1$, $j=1,\ldots,J$ and $\langle \xi_k(\cdot), \xi_j(\cdot) \rangle = 0$, for $k\neq j$.

Different criteria are available for the choice of the number $J$ of principal components to take in the approximation\footnote{Notice that the number and shape of the factors necessary to approximate a function provide information about its complexity.} of eq.~\eqref{eq:apdx_fPCA2}. We interpret the estimation of the factors as an eigenproblem (see next paragraph) and, after having sorted the estimated eigenvalues in decreasing order, we keep the first $J$ eigenfunctions (corresponding to the factors) such that the proportion of variability explained is above a threshold $\bar{d}$. In the empirical analysis, regardless of the criterion used, the value of $J$ is generally very small, thus allowing to interpret and use fPCA as a dimensionality reduction technique for the original series $\mathbf{f}(\cdot)$.

There are several ways to estimate the principal component functions, according to the interpretation of the problem (see \cite{Ramsay05Functional_data_analysis_book} and \cite{FerratyVieu06Nonparam_FDA_Book} for a review). By interpreting them as the eigenfunctions of the covariance operator of the functions $(f_1(\cdot),\ldots,f_T(\cdot))$, we can estimate each pair $(\xi_j,\rho_j)$ of principal component function and score by solving the eiganproblem:
\begin{subnumcases}{}
\label{eq:multiKL_eigen_problem}
\displaystyle\int_{\mathbb{R}^n} V(\mathbf{x},\mathbf{y}) \xi_j(\mathbf{x}) \: \mathrm{d}\mathbf{x} = \rho_j \xi_j (\mathbf{y}) & \\[5pt]
%\displaystyle\int_{\mathbb{R}^n} \xi_j(\mathbf{x}) \xi_j(\mathbf{x}) \: \mathrm{d}\mathbf{x} = 1 & \\
\langle \xi_j(\cdot), \xi_j(\cdot) \rangle = 1 & 
\end{subnumcases}
subject to the additional constraint $\langle \xi_k(\cdot), \xi_j(\cdot) \rangle = 0$, for $k\neq j$. For $t=1,\ldots,T$, $j=1,\ldots,J$, the principal component scores, in the case of orthonormal eigenfunctions, satisfy:
\begin{equation}
\beta_{j,t} = \int_{\mathbb{R}^n} f_t(\mathbf{x}) \xi_j(\mathbf{x}) \: \mathrm{d}\mathbf{x} = \langle f_t(\cdot), \xi_j(\cdot) \rangle \, .
\label{eq:multiKL_PCscores}
\end{equation}
Following an alternative approach, each function $\xi_k(\cdot)$ is obtained by solving the optimization problem:
\begin{subnumcases}{}
\displaystyle\max_{\xi_k} \; \frac{1}{T} \sum_{t=1}^T \left( \int_{\mathbb{R}^n} f_t(\mathbf{x}) \xi_k(\mathbf{x}) \: \mathrm{d}\mathbf{x} \right)^2 & \\[5pt]
\text{s.t. } \norm{\xi_k(\mathbf{x})}_2 = 1 & 
\label{eq:apdx_fPCA1}
\end{subnumcases}
with the additional constraint that $\langle \xi_k, \xi_j \rangle = 0$, for $k\neq j$.
For $t=1,\ldots,T$, $j=1,\ldots,J$, the scores are obtained again from eq.~\eqref{eq:multiKL_PCscores}.
In both cases, the output is a sequence of estimated factors $\boldsymbol{\widehat{\xi}}(\cdot) = (\widehat{\xi}_1(\cdot),\ldots,\widehat{\xi}_J(\cdot))'$ and scores $\mathbf{\widehat{B}} = (\boldsymbol{\widehat{\beta}}_1,\ldots,\boldsymbol{\widehat{\beta}}_T)$, with $\boldsymbol{\widehat{\beta}}_t = (\widehat{\beta}_{t,1},\ldots,\widehat{\beta}_{t,J})'$ for $t=1,\ldots,T$. Then, we obtain:
\begin{equation}
\mathbf{f}(\cdot) \approx \mathbf{\breve{f}}(\cdot) = \mathbf{\widehat{B}}' \boldsymbol{\widehat{\xi}}(\cdot) \, ,   \qquad   \breve{f}_t(\cdot) = \boldsymbol{\widehat{\beta}}_t' \boldsymbol{\widehat{\xi}}(\cdot) \, .
\label{eq:apdx_fPCA2}
\end{equation}

In the paper we follow the first interpretation and estimate the principal component functions and scores by solving an eigenproblem. This poses the preliminary problem of estimating the covariance of the observed sample of functions $\mathbf{f}(\cdot)$. The standard sample covariance function estimator is given by:
\begin{equation}
\widehat{V}(\mathbf{x},\mathbf{y}) = \frac{1}{T} \sum_{t=1}^T f_t(\mathbf{x}) f_t(\mathbf{y}) = T^{-1}\langle f_t(\cdot), f_t(\cdot) \rangle \, .
\label{eq:apdx_fPCA_covariance_estimator}
\end{equation}
Alternative non-parametric estimators have been developed in the earlier contributions of \cite{Hall06PropertiesFPCA}, \cite{Li10ConvergenceRates_NonparametricRegression}, \cite{Yao05FunctionalDataAnalysis_SparseLongitudinal} and \cite{Staniswalis98NonparametricRegressionLongitudinal}.
In matrix notation, eq.~\eqref{eq:apdx_fPCA_covariance_estimator} is written as $v(\mathbf{x},\mathbf{y}) = T^{-1} \mathbf{f}'(\cdot) \mathbf{f}(\cdot)$. By exploiting eqs.~\eqref{eq:apdx_fPCA_covariance_estimator} and~\eqref{eq:apdx_fPCA2} we get:
\begin{equation}
T^{-1} \mathbf{f}'(\cdot) \mathbf{f}(\cdot) = T^{-1} \boldsymbol{\xi}(\cdot) \mathbf{B}' \mathbf{B} \boldsymbol{\xi}(\cdot) \, .
\end{equation}
Therefore, the $k$-th principal component function $\widehat{\xi}_k(\cdot)$ and the score $\boldsymbol{\widehat{\beta}}_t = (\widehat{\beta}_{t,1},\ldots,\widehat{\beta}_{t,J})' = (\langle f_t(\cdot), \xi_1(\cdot) \rangle, \ldots, \langle f_t(\cdot), \xi_J(\cdot) \rangle)'$, for $t=1,\ldots,T$, are obtained by solving the eigenproblem\footnote{This can also be interpreted as a $n$-dimensional Fredh\"{o}lm integral equation of the second type, see \cite{Atkinson09Numerical_IntegralEq2kind_book}, \cite{Atkinson05NumericalAnalysis_book}}:
\begin{align}
V \xi_k(\cdot) & = \rho_k \xi_k(\cdot) \, ,
\label{eq:apdx_eigen_compact}
\end{align}
under the constraints $\langle \xi_k(\cdot), \xi_j(\cdot) \rangle = 0$ for $k\neq j$ and $||\xi_k(\cdot)||_2 = 1$.
One way of solving the eigenproblem requires to discretize the functions on a specified grid of points $\lbrace \mathbf{x}_i \rbrace_{i=1}^N$, which permits to re-state eq.~\eqref{eq:apdx_eigen_compact} as a finite-dimensional eigenproblem in matrix form. Then, standard methods used in multivariate PCA are applied for obtaining the solution.

Alternatively, one may assume that both the original functions $f_t(\cdot)$ and the eigenfunctions $\xi_k(\cdot)$ can be expressed as a finite linear combination of some chosen basis functions $\boldsymbol{\psi}(\cdot) = (\psi_1(\cdot),\ldots,\psi_K(\cdot))'$, with different coefficients:
\begin{equation}
f_t(\cdot) = \sum_{k=1}^K d_{t,k} \psi_k(\cdot) = \mathbf{d}_t' \boldsymbol{\psi}(\cdot) \, ,  \qquad  \xi_j(\cdot) = \sum_{k=1}^K a_{j,k} \psi_k(\cdot) = \mathbf{a}_j' \boldsymbol{\psi}(\cdot) \, ,
\end{equation}
for $t=1,\ldots,T$ and $j=1,\ldots,J$. Given the choice of the basis functions, this reduces the infinite-dimensional problem for $\xi_j(\cdot)$ to a finite-dimensional one for the vector $\mathbf{a}_j = (a_{j,1},\ldots,a_{j,K})'$. From eqs.~\eqref{eq:apdx_fPCA_covariance_operator},~\eqref{eq:apdx_fPCA_covariance_estimator} and~\eqref{eq:apdx_eigen_compact} we obtain:
\begin{align}
T^{-1} \boldsymbol{\psi}(\cdot)' \mathbf{D}' \mathbf{D} \mathbf{M} \mathbf{a}_j & = \rho_j \boldsymbol{\psi}(\cdot)' \mathbf{a}_j \\
T^{-1} \mathbf{D}' \mathbf{D} \mathbf{M} \mathbf{a}_j & = \rho_j \mathbf{a}_j \, ,
\end{align}
with $\mathbf{D} = (\mathbf{d}_1,\ldots,\mathbf{d}_T)$ and $\mathbf{M} = (\langle \psi_k(\cdot), \psi_j(\cdot) \rangle)_{k,j}$, which is the identity matrix if the basis functions form an orthonormal system.

As common practice in multivariate PCA, the estimated eigenvalues $\widehat{\rho}_1,\widehat{\rho}_2,\ldots$ are then sorted in decreasing order and the number $J$ of principal component functions to take is decided on the basis of the proportion of total variation explained $J= \argmin_j \lbrace \sum_j \widehat{\rho}_j > \bar{d}\rbrace$.

\section{Computations}\label{sec:apdx_computations}
In this section we provide the details of the computations needed in Section~\ref{sec:methodology}.
We start by recalling a result from~\cite{LycheMorken08Spline_book} stating some useful properties of B-spline functions. A comprehensive discussion of spline functions and their properties can be found in~\cite{DeBoor01Spline_book} and~\cite{Schumaker07Spline_book}.

%\begin{remark}[{\cite[p.42]{LycheMorken08Spline_book}}]\label{rmk:spline_properties}
%The following properties hold:
%\begin{enumerate}[(i)]
%\item multiplicity:
%\begin{equation*}
%x=\lambda_{j+1} = \ldots = \lambda_{j+m} < \lambda_{j+m+1} \implies B_j^m(x) =1 \quad B_i^m(x) =0 \quad \forall \, i\neq j
%\end{equation*}
%\item local support I:
%\begin{equation*}
%x\notin [\lambda_i,\lambda_{i+m+1}) \implies B_i^{m+1}(x) = 0
%\end{equation*}
%\item local support II:
%\begin{equation*}
%x \in [\lambda_i, \lambda_{i+1}) \implies \begin{cases}
%B_j^{m+1}(x) =0 & j>i \\
%B_j^{m+1}(x) >0 & i-m \leq j \leq i \\
%B_j^{m+1}(x) =0 & j< i-m
%\end{cases}
%\end{equation*}
%\end{enumerate}
%\end{remark}

\subsection{Optimal smoothing problem}\label{sec:apdx_computations_opt_smooth}
In the following we show the procedure for solving the constrained optimal smoothing problem in eq.~\eqref{eq:constrained_smoothing_problem}. Let $\boldsymbol{\bar{\lambda}}^{x,y} = \boldsymbol{\bar{\lambda}}^x \otimes \boldsymbol{\bar{\lambda}}^y$ denote an extended knot sequence (see Section~\ref{sec:notation} for the notation).
We define the difference $\bar{\lambda}_{i,k}^{x,y} - \bar{\lambda}_{j,k}^{x,y}$ as the difference between the first coordinate, that is $\bar{\lambda}_{i,k}^{x,y} - \bar{\lambda}_{j,k}^{x,y} = \bar{\lambda}_i^x - \bar{\lambda}_j^x$ and $\bar{\lambda}_{k,i}^{x,y} - \bar{\lambda}_{k,j}^{x,y} = \bar{\lambda}_i^y - \bar{\lambda}_j^y$.
In this section, for ease of notation we omit the bar and the superscripts and we implicitly refer to augmented knot sequences, that is we use $\lambda_{i,j}$ instead of $\bar{\lambda}_{i,j}^{x,y}$.

The integral constraint in eq.~\eqref{eq:constrained_smoothing_problem} yields:
\begin{align}
\notag
\int_{a_1}^{b_1} \int_{a_2}^{b_2} s_m(u,v) \: \mathrm{d}v \: \mathrm{d}u & = \int_{a_1}^{b_1} \tilde{s}_{m}(u,b_2) - \tilde{s}_{m}(u,a_2) \: \mathrm{d}u \\ \notag
 & = s_{m+1}(b_1,b_2) - s_{m+1}(b_1,a_2) - s_{m+1}(a_1,b_2) + s_{m+1}(a_1,a_2) \\
 & = s_{m+1}(\lambda_{g+1,g+1}) - s_{m+1}(\lambda_{g+1,0}) - s_{m+1}(\lambda_{0,g+1}) + s_{m+1}(\lambda_{0,0}) = 0
\label{eq:integral_equality}
\end{align}
Starting from this result, we look for an equation allowing us to express the coefficient of a bivariate spline of order $k$ with those of a spline obtained after differentiating it with respect to both arguments (that is, we look for an analogue of eq.~\eqref{eq:univariate_spline_derivative}).

Now, we should derive the implication that the previous solution has on the coefficients $c_{i,j}$ of the spline $s_{m+1}(u,v)$ (e.g.: in the univariate case we end up with $0= s_{m+1}(\lambda_{g+1}) - s_{m+1}(\lambda_0) = c_{g} - c_{-m-1}$ thus implying $c_{-m-1} = c_{g}$)\footnote{They obtain the result by using the properties of B-splines in \cite{LycheMorken08Spline_book}. Some bases are exactly $1$, others $0$, reducing the spline function to the coefficient of the unique basis equal to $1$.}.

By exploiting known propertis of splines (see~\cite{LycheMorken08Spline_book}) we obtain:
\begin{align}
0 & = s_{m+1}(\lambda_{g+1,g+1}) - s_{k+1}(\lambda_{g+1,0}) - s_{m+1}(\lambda_{0,g+1}) + s_{m+1}(\lambda_{0,0}) \, , \\
0 & = \sum_i \sum_j c_{ij} \left[ B_i(\lambda_{g+1}) B_j(\lambda_{g+1}) -B_i(\lambda_{g+1}) B_j(\lambda_{0}) -B_i(\lambda_{0}) B_j(\lambda_{g+1}) +B_i(\lambda_{0}) B_j(\lambda_{0}) \right] \, .
\end{align}
By property (i) and (iii):
\begin{itemize}
\item for $i=j= g$ it holds:
\begin{equation}
B_g(\lambda_{g+1}) B_g(\lambda_{g+1}) -B_g(\lambda_{g+1}) B_g(\lambda_{0}) -B_g(\lambda_{0}) B_g(\lambda_{g+1}) +B_g(\lambda_{0}) B_g(\lambda_{0}) = 1 \, ,
\end{equation}
\item for $i=j= -m-1$ it holds:
\begin{align}
\notag
 & B_{-m-1}(\lambda_{g+1}) B_{-m-1}(\lambda_{g+1}) -B_{-m-1}(\lambda_{g+1}) B_{-m-1}(\lambda_{0}) \\
 & -B_{-m-1}(\lambda_{0}) B_{-m-1}(\lambda_{g+1}) +B_{-m-1}(\lambda_{0}) B_{-m-1}(\lambda_{0}) = 1 \, ,
\end{align}
\item for $i,j \notin \lbrace -m-1, g \rbrace$ the previous equation is always 0 since at least one of the terms of each product is 0.
\end{itemize}
Therefore we obtain:
\begin{equation}
s_{m+1}(\lambda_{g+1,g+1}) - s_{m+1}(\lambda_{g+1,0}) - s_{m+1}(\lambda_{0,g+1}) + s_{m+1}(\lambda_{0,0}) = 0 \, ,
\end{equation}
which implies:
\begin{equation}
c_{g,g} + c_{-m-1,-m-1} = 0 \iff c_{g,g} = - c_{-m-1,-m-1} \, .
\end{equation}
Now, by applying sequentially the recursion linking  spline function with its partial derivatives and using a knot sequence (or an extended knot sequence) with equal number of knots along both directions:
\begin{align}
\notag
\dfrac{\mathrm{d}}{\mathrm{d}u} \dfrac{\mathrm{d}}{\mathrm{d}v} s_{m+1}(u,v) & = \dfrac{\mathrm{d}}{\mathrm{d}u} \dfrac{\mathrm{d}}{\mathrm{d}v} \sum_{i=-m-1}^g \sum_{j=-m-1}^g c_{ij} B_i^{m+2}(u)B_j^{m+2}(v) \\ \notag
 & = \dfrac{\mathrm{d}}{\mathrm{d}v} \sum_{j=-m-1}^g B_j^{m+2}(v) \cdot \left( \dfrac{\mathrm{d}}{\mathrm{d}u}  \sum_{i=-m-1}^g c_{ij} B_i^{m+2}(u) \right) \\
 & = \dfrac{\mathrm{d}}{\mathrm{d}v} \sum_{j=-m-1}^g B_j^{m+2}(v) \cdot \left( \sum_{i=-m}^g c_{ij}^u B_i^{m+1}(u) \right) \\
 & = \sum_{i=-m}^g B_i^{m+1}(u) \left( \dfrac{\mathrm{d}}{\mathrm{d}v} \sum_{j=-m-1}^g c_{ij}^u B_j^{m+2}(v) \right) \\ \notag
 & = \sum_{i=-m}^g \sum_{j=-m}^g c_{ij}^{uy} B_i^{m+1}(u) B_j^{m+1}(v) \, .
\end{align}
Since by definition:
\begin{equation}
s_{m}(u,v) = \sum_{i=-m}^g \sum_{j=-m}^g b_{ij} B_i^{m+1}(u) B_j^{m+1}(v) \, ,
\end{equation}
by assuming $b_{ij}=c_{ij}^{uy}$ we get the equality:
\begin{equation}
\dfrac{\mathrm{d}}{\mathrm{d}u} \dfrac{\mathrm{d}}{\mathrm{d}v} s_{m+1}(u,v) = s_{m}(u,v) \, .
\end{equation}
It is now necessary to develop the above expression for the constraint on the coefficients in order to find out the precise relationship between the $c_{ij}$ (coefficients of $s_{m+1}$) and $b_{ij}$ (coefficients of $s_m$). This is required in order to derive the solution of eq.~\eqref{eq:constrained_smoothing_problem} by minimizing the first equation, thus solving an unconstrained optimization problem. First, recall that the coefficients of a univariate spline are related to those of its first order derivative via the relation:
\begin{equation}
\dfrac{\mathrm{d}}{\mathrm{d}u} s_m(u) = \sum_i \breve{c}_i B_i^{m}(u) = s_{m-1}(u) \qquad \breve{c}_i = m \dfrac{c_i - c_{i-1}}{\lambda_{i+m} - \lambda_i} \, .
\label{eq:coefficients_derivative_spline_univariate}
\end{equation}
In the bivariate case, first define $c_{ij}^u$, for fixed $j$ and $i=-m,\ldots,g$, as:
\begin{equation}
c_{ij}^u = (m+1) \dfrac{c_{i,j} - c_{i-1,j}}{\lambda_{i+m+1,j} - \lambda_{i,j}} \, .
\end{equation}
Then, iterated application eq.~\eqref{eq:coefficients_derivative_spline_univariate} along each dimension gives, for $j=-m,\ldots,g$:
\begin{align}
\notag
b_{i,j} = c_{i,j}^{uy} & = (m+1) \dfrac{c_{i,j}^u - c_{i,j-1}^u}{\lambda_{i,j+m+1} -\lambda_{i,j}} = \dfrac{(m+1)^2}{\lambda_{i,j+m+1} - \lambda_{i,j}} \left( \dfrac{c_{i,j} - c_{i-1,j}}{\lambda_{i+m+1,j} - \lambda_{i,j}} - \dfrac{c_{i,j-1} - c_{i-1,j-1}}{\lambda_{i+m+1,j-1} - \lambda_{i,j-1}} \right) \\
 & = \dfrac{(m+1)^2}{\lambda_{i,j+m+1} - \lambda_{i,j}} \left( \dfrac{c_{i,j} - c_{i-1,j}}{\lambda_{i+m+1,j} - \lambda_{i,j}} - \dfrac{c_{i,j-1} - c_{i-1,j-1}}{\lambda_{i+m+1,j-1} - \lambda_{i,j-1}} \right) \, .
\label{eq:bivariate_coefficients_link}
\end{align}
This implies that the matrix $\mathbf{B}$ has the following top-left (i.e. $b_{-m,-m}$) and bottom-right (i.e. $b_{g,g}$) entries:
\begin{align}
b_{-m,-m} & = \dfrac{(m+1)^2}{\lambda_{-m,1}-\lambda_{-m,-m}} \left( \dfrac{c_{-m,-m} -c_{-m-1,-m}}{\lambda_{1,-m}-\lambda_{-m,-m}} -\dfrac{c_{-m,-m-1} -c_{-m-1,-m-1}}{\lambda_{1,-m-1}-\lambda_{-m,-m-1}} \right) \, , \\[8pt]
b_{g,g} & = \dfrac{(m+1)^2}{\lambda_{g,g+m+1}-\lambda_{g,g}} \left( \dfrac{c_{g,g} -c_{g-1,g}}{\lambda_{g+m+1,g}-\lambda_{g,g}} -\dfrac{c_{g,g-1} -c_{g-1,g-1}}{\lambda_{g+m+1,g-1}-\lambda_{g,g-1}} \right) \, .
\end{align}
We need conditions for linking the $(g+m+1)\times(g+m+1)$ coefficient matrix $\mathbf{B} = (b_{i,j})_{i,j}$ of the spline function $s_m(u,v)$ and the $(g+m+2)\times(g+m+2)$ coefficient matrix $\mathbf{C} = (c_{i,j})_{i,j}$ of the spline function $s_{m+1}(u,v)$. In the univariate case they are two vectors whose lengths differ by one, and the condition to be imposed consists in the equality of the first and last entry of the coefficient vector of the spline with higher degree.
In the bivariate case, instead, $2(g+m+1)+1$ constraints are required:
\begin{equation}
\mathbf{C} = \left[ \begin{array}{c|c}
\textcolor{blue}{c_{-m-1,-m-1}} & \begin{matrix} \textcolor{red}{c_{-m-1,-m}} & \textcolor{red}{\ldots} & \textcolor{red}{c_{-m-1,g}} \end{matrix} \\
\hline
\begin{matrix} \textcolor{purple}{c_{-m,-m-1}} \\ \textcolor{purple}{\vdots} \\ \textcolor{purple}{c_{g,-m-1}} \end{matrix} & \mathbf{\bar{C}} 
\end{array} \right]
\end{equation}
From the previous computations, we obtain the constraint:
\begin{equation}
\textcolor{blue}{c_{-m-1,-m-1}} = -c_{g,g} \, .
\label{eq:condition_from_integral_constraint}
\end{equation}
%In the univariate case, the $N\times 1$ vector $\mathbf{b}$ and the $(N+1)\times 1$ vector $\mathbf{c}$ are linked through the $N\times N$ matrix $\mathbf{D}$ and the $N\times (N+1)$ (first difference) matrix $\mathbf{K}$ by:
%\begin{equation}
%\mathbf{b} = \mathbf{D} \mathbf{K} \mathbf{c} \, .
%\end{equation}
%Then, in order to include in the matrix representation the constraint stemming from the integral equation (that is, $c_{-m-1} = c_g$), the matrix has been transformed into a square one, with the first column removed and a $-1$ inserted in top-left and top-right position.
We need to incorporate this result obtained from the integral constraint. From eq.~\eqref{eq:bivariate_coefficients_link} we have that\footnote{We used the notation $\mathbf{A}_{j,j}^i$ to mean the $(j,j)$-th entry of the diagonal matrix $\mathbf{A}^i$.}:
\begin{align}
\notag
b_{i,j} & = \dfrac{(m+1)^2}{\lambda_{i,j+(m+1)} -\lambda_{i,j}} \left[ \dfrac{1}{\lambda_{i+(m+1),j} -\lambda_{i,j}} (c_{i,j}-c_{i-1,j}) - \dfrac{1}{\lambda_{i+(m+1),j-1} -\lambda_{i,j-1}} (c_{i,j-1}-c_{i-1,j-1}) \right] \\[5pt]
 & = \mathbf{D}_{j,j}^i \left[ \mathbf{E}_{i,i}^j (\mathbf{K} \mathbf{c}_{:,j})_i - \mathbf{E}_{i,i}^{j-1} (\mathbf{K} \mathbf{c}_{:,j-1})_i \right] \, ,
\label{eq:entry}
\end{align}
which, using the shorthand $N=(g+m+1)$ and letting $\lbrace \mathbf{e}_1,\ldots,\mathbf{e}_N \rbrace$ be the canonical basis of the space of $N \times N$ matrices, gives the following expression for the column vector $\mathbf{b}_{:,j}$:
%\begin{align}
%\notag
%\mathbf{b}_{:,j} & = \left[ \sum_{k=1}^N \mathbf{e}_k \mathbf{e}_k' \otimes (\mathbf{D}_{j,j}^k \mathbf{E}_{k,k}^j) \right] \mathbf{K}\mathbf{c}_{:,j} - \left[ \sum_{k=1}^N \mathbf{e}_k \mathbf{e}_k' \otimes (\mathbf{D}_{j-1,j-1}^k \mathbf{E}_{k,k}^{j-1}) \right] \mathbf{K}\mathbf{c}_{:,j-1} \\ \notag
% & = \begin{bmatrix}
%\mathbf{D}_{j,j}^1 \mathbf{E}_{1,1}^j & & & & \\
% & \mathbf{D}_{j,j}^2 \mathbf{E}_{2,2}^j & & & \\
% & & \ddots & & \\
% & & & \ddots & \\
% & & & & \mathbf{D}_{j,j}^N \mathbf{E}_{N,N}^j
%\end{bmatrix}
%\begin{bmatrix}
%(\mathbf{K} \mathbf{c}_{:,j})_1 \\
%(\mathbf{K} \mathbf{c}_{:,j})_2 \\
%\vdots \\
%(\mathbf{K} \mathbf{c}_{:,j})_N
%\end{bmatrix} \\
%& - \begin{bmatrix}
%\mathbf{D}_{j-1,j-1}^1 \mathbf{E}_{1,1}^{j-1} & & & & \\
% & \mathbf{D}_{j-1,j-1}^2 \mathbf{E}_{2,2}^{j-1} & & & \\
% & & \ddots & & \\
% & & & \ddots & \\
% & & & & \mathbf{D}_{j-1,j-1}^N \mathbf{E}_{N,N}^{j-1}
%\end{bmatrix}
%\begin{bmatrix}
%(\mathbf{K} \mathbf{c}_{:,j-1})_1 \\
%(\mathbf{K} \mathbf{c}_{:,j-1})_2 \\
%\vdots \\
%(\mathbf{K} \mathbf{c}_{:,j-1})_N
%\end{bmatrix} \, .
%\end{align}
\begin{align}
\notag
\mathbf{b}_{:,j} & = \left[ \sum_{k=1}^N \mathbf{e}_k \mathbf{e}_k' \otimes (\mathbf{D}_{k,k}^j \mathbf{E}_{k,k}^j) \right] \mathbf{K}\mathbf{c}_{:,j} - \left[ \sum_{k=1}^N \mathbf{e}_k \mathbf{e}_k' \otimes (\mathbf{D}_{k,k}^{j-1} \mathbf{E}_{k,k}^{j-1}) \right] \mathbf{K}\mathbf{c}_{:,j-1} \\ \notag
 & = \begin{bmatrix}
\mathbf{D}_{1,1}^j \mathbf{E}_{1,1}^j & & & & \\
 & \mathbf{D}_{2,2}^j \mathbf{E}_{2,2}^j & & & \\
 & & \ddots & & \\
 & & & \ddots & \\
 & & & & \mathbf{D}_{N,N}^j \mathbf{E}_{N,N}^j
\end{bmatrix}
\begin{bmatrix}
(\mathbf{K} \mathbf{c}_{:,j})_1 \\
(\mathbf{K} \mathbf{c}_{:,j})_2 \\
\vdots \\
(\mathbf{K} \mathbf{c}_{:,j})_N
\end{bmatrix} \\
& \quad - \begin{bmatrix}
\mathbf{D}_{1,1}^j \mathbf{E}_{1,1}^{j-1} & & & & \\
 & \mathbf{D}_{2,2}^j \mathbf{E}_{2,2}^{j-1} & & & \\
 & & \ddots & & \\
 & & & \ddots & \\
 & & & & \mathbf{D}_{N,N}^j \mathbf{E}_{N,N}^{j-1}
\end{bmatrix}
\begin{bmatrix}
(\mathbf{K} \mathbf{c}_{:,j-1})_1 \\
(\mathbf{K} \mathbf{c}_{:,j-1})_2 \\
\vdots \\
(\mathbf{K} \mathbf{c}_{:,j-1})_N
\end{bmatrix} \, .
\end{align}
For $j = \textcolor{red}{-m,\ldots,g}$, we define the $(g+m+1) \times (g+m+1)$ diagonal matrix $\mathbf{D}^i$ by:
\begin{equation}
\mathbf{D}^j = \diag{\dfrac{(m+1)^2}{\lambda_{-m,j+m+1} -\lambda_{-m,j}}, \, \dfrac{(m+1)^2}{\lambda_{-m+1,j+m+1} -\lambda_{-m+1,j}}, \, \ldots, \, \dfrac{(m+1)^2}{\lambda_{g,j+m+1} -\lambda_{g,j}}} \, .
%\mathbf{D}^i = \begin{bmatrix}
%\dfrac{(m+1)^2}{\lambda_{-m,i+m+1} -\lambda_{-m,i}} & & & & & \\
% & \dfrac{(m+1)^2}{\lambda_{-m+1,i+m+1} -\lambda_{-m+1,i}} & & & & \\
% & & & \ddots & & \\
% & & & & \ddots & \\
% & & & & & \dfrac{(m+1)^2}{\lambda_{g,i+m+1} -\lambda_{g,i}}
%\end{bmatrix} \, ,
\end{equation}
%Notice that the matrices $\mathbf{D}^i$ are indexed by $i = \textcolor{red}{-m,\ldots,g}$, by contrast, the matrices $\mathbf{E}^j$ are indexed by $j= \textcolor{blue}{-m-1,}\textcolor{red}{-m,\ldots,g}$. This is due to the fact each matrix $\mathbf{E}^j$ must include an additional value for the index $j$, in order to account for the ``lagged'' initial value.
and, for $j= \textcolor{blue}{-m-1,}\textcolor{red}{-m,\ldots,g}$, we define the $(g+m+1) \times (g+m+1)$ diagonal matrix $\mathbf{E}^j$ by:
\begin{equation}
\mathbf{E}^j = \diag{\dfrac{1}{\lambda_{1,j} -\lambda_{-m,j}}, \, \dfrac{1}{\lambda_{2,j} -\lambda_{-m+1,j}}, \, \ldots, \, \dfrac{1}{\lambda_{g+m+1,j} -\lambda_{g,j}}} \, .
%\mathbf{E}^j = \begin{bmatrix}
%\dfrac{1}{\lambda_{1,j} -\lambda_{-m,j}} & & & & & \\
% & \dfrac{1}{\lambda_{2,j} -\lambda_{-m+1,j}} & & & & \\
% & & & \ddots & & \\
% & & & & \ddots & \\
% & & & & & \dfrac{1}{\lambda_{g+m+1,j} -\lambda_{g,j}}
%\end{bmatrix} \, .
\end{equation}
The matrix $\mathbf{K}$ coincides with the matrix representation of the linear operator $L$ which performs first differences $\mathbf{L}_{g+m+1}$ given by the $(g+m+1)\times (g+m+2)$:
\begin{equation}
\mathbf{K} = \mathbf{L}_{g+m+1} = \begin{bmatrix}
-1 & 1 & & & \\
 & -1 & 1 & & \\
 & & \ddots & \ddots & \\
 & & & -1 & 1
\end{bmatrix} \, .
\label{eq:K_unconstrained}
\end{equation}
Therefore, for $j=-m,\ldots,g$, the $(g+m+1) \times 1$ vector of first differences $\mathbf{K} \mathbf{c}_{:,j}$ is given by:
\begin{equation}
\mathbf{K} \mathbf{c}_{:,j} = \begin{bmatrix}
c_{-m,j} -c_{-m-1,j} \\
c_{-m+1,j} -c_{-m,j} \\
\vdots \\
c_{g,j} -c_{g-1,j} \\
\end{bmatrix} \, .
\end{equation}
In order to rewrite eq.~\eqref{eq:entry} in more compact form, we introduce the following $N \times (N+1)$ matrices, which allow to select the first (or last, respectively) $N$ columns from another one by post-multiplication. Let $\lbrace \mathbf{e}_1^N,\ldots,\mathbf{e}_N^N \rbrace$ be the canonical basis of the space of square matrices of size $N\times N$ and define the $N\times (N+1)$ matrix $P_{N,m} = [\mathbf{e}_1^N,\ldots,\mathbf{e}_{m-1}^N,\mathbf{0}_N,\mathbf{e}_m^N,\ldots,\mathbf{e}_N^N]$ with $m \in \lbrace 1,\ldots,N+1 \rbrace$, which, by pre-multiplying a vector of length $(N+1)$, selects all but the $m$-th entry. The $(N+1)\times N$ matrices $\mathbf{S}_c^f,\mathbf{S}_c^\ell$ defined as follows, instead, when pre-multiplied by a $(N+1)\times(N+1)$ matrix $\mathbf{A}$ select the sub-matrix made with the first (last, respectively) $N$ columns and rows $\mathbf{A}$:
\begin{equation}
\mathbf{S}_c^f = P_{N,N+1}' = \left[ \begin{array}{c c c c}
1 & & \\
  & \ddots & \\
  & & 1 \\ \hline
0 & \ldots & 0
\end{array} \right] \, ,    \qquad    
\mathbf{S}_c^\ell = P_{N,1}' = \left[ \begin{array}{c c c c}
0 & \ldots & 0 \\ \hline
1 & & \\
  & \ddots & \\
  & & 1
\end{array} \right] \, .
\end{equation}
For example, if $\mathbf{A}=[\mathbf{A}^f|\mathbf{a}_{N+1}]=[\mathbf{a}_1|\mathbf{A}^\ell]$ then $\mathbf{A} \mathbf{S}_c^f = \mathbf{A}^f$ and $\mathbf{A}\mathbf{S}_c^\ell = \mathbf{A}^\ell$. Notice also that the transposed versions, that is $\mathbf{S}_r^f = (\mathbf{S}_c^f)'$ and $\mathbf{S}_r^\ell = (\mathbf{S}_c^\ell)'$, allow to select rows instead of columns.
It is now possible to rewrite eq.~\eqref{eq:entry} as:
\begin{align}
\notag
\vecc{\mathbf{B}} & = \begin{bmatrix}
\mathbf{D}^{-m} \mathbf{E}^{-m} &  &  \\
 & \ddots &  \\
 &  & \mathbf{D}^{g} \mathbf{E}^{g}
\end{bmatrix} \vecc{\mathbf{S}_r^\ell \mathbf{\tilde{C}} \mathbf{S}_c^\ell} -\begin{bmatrix}
\mathbf{D}^{-m} \mathbf{E}^{-m-1} &  &  \\
 & \ddots &  \\
 &  & \mathbf{D}^{g} \mathbf{E}^{g-1}
\end{bmatrix} \vecc{\mathbf{S}_r^f \mathbf{\tilde{C}} \mathbf{S}_c^f} \\
& = \mathbf{D} \mathbf{E} \vecc{\mathbf{S}_r^\ell \mathbf{\tilde{C}} \mathbf{S}_c^\ell} -\mathbf{D} \mathbf{F} \vecc{\mathbf{S}_r^f \mathbf{\tilde{C}} \mathbf{S}_c^f} \, ,
\label{eq:vectorised}
\end{align}
where, letting $\lbrace \mathbf{e}_{-m},\ldots,\mathbf{e}_g \rbrace$ be the canonical basis for the space of square matrices of size $(g+m+1)\times(g+m+1)$, we defined:
\begin{align*}
\mathbf{D} & = \sum_{i=-m}^g \mathbf{e}_i \mathbf{e}_i' \otimes \mathbf{D}^i \quad \text{ size } (g+m+1)^2 \times (g+m+1)^2 \\[6pt]
\mathbf{E} & = \sum_{i=-m}^g \mathbf{e}_i \mathbf{e}_i' \otimes \mathbf{E}^i \quad \text{ size } (g+m+1)^2 \times (g+m+1)^2 \\[6pt]
\mathbf{F} & = \sum_{i=-m-1}^{g-1} \mathbf{e}_i \mathbf{e}_i' \otimes \mathbf{E}^i \quad \text{ size } (g+m+1)^2 \times (g+m+1)^2 \\[6pt]
\mathbf{\tilde{C}} & = \left[ \mathbf{K} \mathbf{c}_{:,-m} | \ldots | \mathbf{K} \mathbf{c}_{:,g+1} \right] \quad \text{ size } (g+m+1+1) \times (g+m+1+1) \, ,
\end{align*}
and $\mathbf{S}_r^\ell \mathbf{\tilde{C}} \mathbf{S}_c^\ell$ (respectively, $\mathbf{S}_r^f \mathbf{\tilde{C}} \mathbf{S}_c^f$) select the bottom-right (respectively, top-left) square submatrix of $\mathbf{\tilde{C}}$ of size $(g+m+1) \times (g+m+1)$, denoted $(\mathbf{\tilde{C}})_{-m}^g$ (respectively, $(\mathbf{\tilde{C}})_{-m-1}^{g-1}$).

Now, we need to include in eq.~\eqref{eq:vectorised} the information obtained in eq.~\eqref{eq:condition_from_integral_constraint} from the integral constraint in eq.~\eqref{eq:constrained_smoothing_problem}, that is:
\begin{equation}
c_{-m-1,-m-1} = -c_{g,g} \, .
\end{equation}
To this end, notice that there is no possibility of defining a unique matrix $\mathbf{K}$ which is able to give the desired result by a suitable choice of its entries (as opposed to the univariate case). This is mainly due to the fact that in the linear system representation of $\mathbf{\tilde{C}}$ the terms $c_{-m-1,-m-1}$ and $c_{g,g}$ are never in the same equation and the constraints on the other coefficients of $\mathbf{K}$ prevent to obtain the result. We propose to solve this issue as follows. Instead of transforming the matrix $\mathbf{C}$ and then vectorize it, reverse the order, that is, vectorize $\mathbf{C}$ then apply a suitable transformation. Then, define the $((g+m+1+1)^2-1) \times 1$ vector:
\begin{equation}
\mathbf{\tilde{c}} = P_{(g+m+1+1)^2-1,1} \cdot \vecc{\mathbf{C}} \, ,
\end{equation}
which corresponds to the vectorization of the matrix $\mathbf{C}$ without the element\footnote{We remove from $\vecc{\mathbf{C}}$ all the entries equal to $c_{-m-1,-m-1}$, thus obtaining a vector $\mathbf{c}^*$ of length equal to the length of $\vecc{\mathbf{C}}$ minus the number of occurrences of $c_{-m-1,-m-1}$.} $c_{-m-1,-m-1}$.
Consider the $N\times (N+1)$ matrix $\mathbf{L}_N$ representing the first difference operator $L$:
\begin{equation}
\mathbf{L}_N = \begin{bmatrix}
-1 & 1 & & & \\
& -1 & 1 & & \\
& & \ddots & \ddots & \\
& & & -1 & 1
\end{bmatrix} \, .
\label{eq:difference_operator_matrix}
\end{equation}
Since we are dealing with the vectorisation of a matrix, $\vecc{\mathbf{C}}$, we must keep in mind that taking first differences of the whole vector implies taking differences also between the first entry of a column and the last of the previous one, which is undesired. Therefore we need to modify the structure of the difference operator matrix accordingly: we can do it by ``shifting'' to the right the blocks of non-zero entries every $N-1$ rows, where $N$ is the number of rows of the original matrix $\mathbf{C}$.
Moreover, taking into account the integral constraint\footnote{The constraint here has the opposite sign as compared to the univariate case, but it is coherent. In fact, it stems from the integral constraint and in the univariate case the integral is obtained by taking the difference between the value at the extrema of integration, while in the bivariate case the values at the top-right and bottom-left corners are added while those at the other two vertices of the rectangle are subtracted.} in eq.~\eqref{eq:condition_from_integral_constraint} we get in top right corner $1$ instead of $0$ whereas the first column of the difference operator matrix is removed.
Consequently, we define $\mathbf{K}^*$ to be the matrix with number of columns equal to the size of $\mathbf{\tilde{c}}$ (that is, $(g+m+1+1)^2-1$) and number of rows equal to the number of entries of $\mathbf{C}$ minus a row (that is, $(g+m+1)(g+m+2)$), which is lost by taking differences. We obtain the $(g+m+1)(g+m+2) \times ((g+m+2)^2-1)$ block diagonal matrix:
\begin{equation}
\mathbf{K}^* = \left[ \begin{array}{c | c | c | c | c}
\mathbf{K}^{11} & & & & \mathbf{K}^{1N} \\ \hline
 & \mathbf{K}^{-m} & & & \\ \hline
 & & \mathbf{K}^{-m+1} & & \\ \hline
 & & & \ddots & \\ \hline
 & & & & \mathbf{K}^{g}
\end{array} \right] \, ,
\label{eq:Kstar}
\end{equation}
where each $(g+m+1)\times(g+m+2)$ matrix $\mathbf{K}^{i} = \mathbf{L}_{g+m+1}$, for $i=-m,\ldots,g$, whereas $\mathbf{K}^{1N}$ is a $(g+m+1)\times(g+m+2)$ with all zeros but the top-right entry and the $(g+m+1)\times(g+m+1)$ square matrix $\mathbf{K}^{11}$ is given by:
\begin{equation}
\mathbf{K}^{11} = \left[ \begin{array}{c c c c c}
 1 & & & &  \\
-1 & 1 & & & \\
 & -1 & 1 & & \\
 & & \ddots & \ddots & \\
 & & & -1 & 1
\end{array} \right] \, .
\end{equation}
%\begin{equation}
%\mathbf{K}^* = \left[ \begin{array}{c c|c c c|c c c|c c c|c c c}
%1 & & & & & & & & & & & & & 1 \\
%-1 & 1 & & & & & & & & & & \\ \hline
% & & -1 & 1 & & & & & & & & & & \\
% & & & -1 & 1 & & & & & & & & & \\ \hline
% & & & & & -1 & 1 & & & & & & & \\
% & & & & & & -1 & 1 & & & & & & \\ \hline
% & & & & & & & & \ddots & \ddots & & & \\
% & & & & & & & & & \ddots & \ddots & & \\ \hline
% & & & & & & & & & & & -1 & 1 & \\
% & & & & & & & & & & & & -1 & 1
%\end{array} \right] \, .
%\label{eq:new_K}
%\end{equation}
For example, let $I_R,I_C$ be the number of rows and columns in $\mathbf{C}$, respectively. Then the size of the matrix $\mathbf{K}^*$ is $(I_C (I_R-1)) \times (I_C I_R -1)$.

By exploiting the previously defined diagonal matrices $\mathbf{D}^j, \mathbf{E}^j$, we define the following $(g+m+1)^2 \times (g+m+1)^2$ block diagonal matrices:
\begin{align}
\mathbf{D} = \sum_{i=-m}^g \mathbf{e}_i \mathbf{e}_i' \otimes \mathbf{D}^i & = \left[ \begin{array}{c|c|c|c}
\mathbf{D}^{-m} & & & \\ \hline
& \mathbf{D}^{-m-1} & & \\ \hline
& & \ddots & \\ \hline
& & & \mathbf{D}^{g}
\end{array} \right] \, , \\
\mathbf{E} = \sum_{i=-m}^g \mathbf{e}_i \mathbf{e}_i' \otimes \mathbf{E}^i & = \left[ \begin{array}{c|c|c|c}
\mathbf{E}^{-m} & & & \\ \hline
& \mathbf{E}^{-m-1} & & \\ \hline
& & \ddots & \\ \hline
& & & \mathbf{E}^{g}
\end{array} \right] \, , \\
\mathbf{F} = \sum_{i=-m}^g \mathbf{e}_i \mathbf{e}_i' \otimes \mathbf{E}^{i-1} & = \left[ \begin{array}{c|c|c|c}
\mathbf{E}^{-m-1} & & & \\ \hline
& \mathbf{E}^{-m} & & \\ \hline
& & \ddots & \\ \hline
& & & \mathbf{E}^{g-1}
\end{array} \right] \, .
\end{align}
Finally, we define $\mathbf{T}^f, \mathbf{T}^l$ to be two selection matrices of size $(g+m+1)^2 \times (g+m+1)(g+m+2)$ which select entries from a vector of length $(g+m+1)(g+m+2)$ by pre-multiplication:
\begin{equation}
\mathbf{T}^f = \left[ \begin{array}{c c c |c c c}
1 & & & 0 & \ldots & 0 \\
& \ddots & & \vdots & & \vdots \\
& & 1 & 0 & \ldots & 0
\end{array} \right] \, ,  \qquad  
\mathbf{T}^l = \left[ \begin{array}{c c c|c c c}
0 & \ldots & 0 & 1 & & \\
\vdots & & \vdots & & \ddots & \\
0 & \ldots & 0 & & & 1
\end{array} \right] \, .
\end{equation}
Finally, define $\mathbf{A} = [\mathbf{E} \mathbf{T}^f - \mathbf{F} \mathbf{T}^l] \mathbf{K}^*$. We obtain the following equation relating the vectorized matrices of spline coefficients $\mathbf{B}$ and $\mathbf{C}$:
\begin{equation}
\overline{\mathbf{b}} = \vecc{\mathbf{B}} = \mathbf{D} \left[ \mathbf{E} \mathbf{T}^f \mathbf{K}^* \mathbf{\tilde{c}} - \mathbf{F} \mathbf{T}^l \mathbf{K}^* \mathbf{\tilde{c}} \right] = \mathbf{D} \left[ \mathbf{E} \mathbf{T}^f - \mathbf{F} \mathbf{T}^l \right] \mathbf{K}^* \mathbf{\tilde{c}} = \mathbf{D} \mathbf{A} \mathbf{\tilde{c}} \, ,
\label{eq:relation_coefficients_b_c}
\end{equation}
The next step consists in re-writing the objective function of the optimization problem~\eqref{eq:constrained_smoothing_problem} using matrix notation. First of all, since the B-spline basis for the bivariate spline is the product of two univariate B-splines, given a sample $(\mathbf{z},\mathbf{u},\mathbf{v}) = \lbrace z_i, (u_i,v_i) \rbrace_{i=1}^n$ we define the modified version of the matrix $\mathbf{C}^{m+1}(\mathbf{u})$ used in the univariate case as follows:
\begin{equation}
\mathbf{C}^{m+1}(\mathbf{u}, \mathbf{v}) = \begin{bmatrix}
B_{-m}^{m+1}(u_1)B_{-m}^{m+1}(v_1) & \ldots & B_{g}^{m+1}(u_1)B_{-m}^{m+1}(v_1) & \ldots & B_{g}^{m+1}(u_1)B_{g}^{m+1}(v_1) \\
\vdots & & & & \vdots \\
B_{-m}^{m+1}(u_n)B_{-m}^{m+1}(v_n) & \ldots & B_{g}^{m+1}(u_n)B_{-m}^{m+1}(v_n) & \ldots & B_{g}^{m+1}(u_n)B_{g}^{m+1}(v_n)
\end{bmatrix} \, ,
\end{equation}
whose size is $n \times (g+m+1)^2$ and generic entry $\mathbf{C}_{i,j}^{m+1}(\mathbf{u},\mathbf{v}) = B_{j_1}^{m+1}(u_i)B_{j_2}^{m+1}(v_i)$, with $j_1,j_2$ obtained by inverting\footnote{The inversion is obtained by solving a linear system with two equations and two unknowns, $j_1,j_2$, which has a unique solution: $j_1= j-(j_2-1)(g+m+1)$ and $j_2= \lfloor j/(g+m+1) \rfloor$, where $\lfloor x \rfloor$ denote the integer part of $x$.} the linear indexing $j= j_1 + (j_2-1)(g+m+1)$. Each row is constructed by first fixing the index for the B-spline along the direction of $v$, that is $B_j^{m+1}(v_i)$, then considering all the combinations with the B-spline along the direction of $u$, that is $B_h^{m+1}(u_i)$. Then the index $j$ is incremented and the process is iterated until the exhaustion of the basis. Notice that each row of the matrix corresponds to the same observation point. This construction is necessary to rewrite a bivariate spline function given the observation points $\lbrace (u_i,v_i)\rbrace_{i=1}^n$ in matrix form, obtaining a vector of size $n \times 1$:
\begin{equation}
s_m(\mathbf{u},\mathbf{v}) = \mathbf{C}^{m+1}(\mathbf{u},\mathbf{v}) \overline{\mathbf{b}} \, .
\label{eq:spline_bivariate}
\end{equation}
In order to write the matrix form of the integral of the squared derivative (of order $\ell \leq m-1$) of the bivariate spline function, start by defining the $(g+m+1-\ell)^2 \times (g+m+1-\ell)^2$ matrix of inner products of the B-spline basis functions of order $m-\ell$ as follows:
\begin{equation}
\mathbf{M}_{m,\ell} = \begin{bmatrix}
\langle B_{-m+\ell}^{m+1-\ell}B_{-m+\ell}^{m+1-\ell}, \: B_{-m+\ell}^{m+1-\ell}B_{-m+\ell}^{m+1-\ell} \rangle & \ldots & \langle B_{g}^{m+1-\ell}B_{g}^{m+1-\ell}, \: B_{-m+\ell}^{m+1-\ell}B_{-m+\ell}^{m+1-\ell} \rangle \\
\vdots & & \vdots \\
\langle B_{-m+\ell}^{m+1-\ell}B_{-m+\ell}^{m+1-\ell}, \: B_{g}^{m+1-\ell}B_{g}^{m+1-\ell} \rangle & \ldots & \langle B_{g}^{m+1-\ell}B_{g}^{m+1-\ell}, \: B_{g}^{m+1-\ell}B_{g}^{m+1-\ell} \rangle
\end{bmatrix} \, ,
\end{equation}
where the generic entry is $\mathbf{M}_{m,\ell; i,j} = \langle B_{j_1}^{m+1-\ell}B_{j_2}^{m+1-\ell}, B_{i_1}^{m+1-\ell}B_{i_2}^{m+1-\ell} \rangle$, where $i_1,i_2,j_1,j_2$ are obtained, as for $\mathbf{C}_{i,j}^{m+1}(\mathbf{u},\mathbf{v})$, by inverting the linear indexing $i=i_1+(i_2-1)(g+m+1-\ell)$ and $j=j_1+(j_2-1)(g+m+1-\ell)$.
% The first component of each inner product is the B-spline along the direction of $u$, while the second is that along the direction of $v$.
%Each row is defined for a fixed value of the indices of the second term of the inner products and the ordering within each row corresponds to that chosen for the matrix $\mathbf{C}^{m+1}(\mathbf{u},\mathbf{v})$.
The inner product is defined in the usual way (see Algorithm 5.22 in \cite{Schumaker07Spline_book} for numerical computation) as:
\begin{align}
\notag
 & \langle B_{i}^{m+1-\ell}B_{j}^{m+1-\ell}, \: B_{h}^{m+1-\ell}B_{l}^{m+1-\ell} \rangle \\[8pt]
 & = \int_{\lambda_0}^{\lambda_g} \int_{\lambda_0}^{\lambda_g} B_{i}^{m+1-\ell}(u) \: B_{j}^{m+1-\ell}(v) \: B_{h}^{m+1-\ell}(u) \: B_{l}^{m+1-\ell}(v) \; \; \mathrm{d}u \: \mathrm{d}v \geq 0 \, .
\end{align}
Under the assumptions made in the text, that is $(a_1,b_1) = (\lambda_{0,0},\lambda_{g,g})$, $(a_2,b_2) = (\lambda_{u,0},\lambda_{u,g})$, $\ell_1 = \ell_2 = \ell$, $n_1 = n_2 = n$, the objective function in eq.~\eqref{eq:constrained_smoothing_problem} can be re-written as the sum of two terms:
\begin{align}
\notag
J_\ell(s_m) & = \int_{a_1}^{b_1} \int_{a_2}^{b_2} \left[ s_m^{(\ell_1,\ell_2)}(u,v) \right]^2 \: \mathrm{d}v \: \mathrm{d}u + \alpha \left[ \sum_{i=1}^{n_1}\sum_{j=1}^{n_2} w_{ij} \left( z_{ij} -s_m(u_i,v_j) \right)^2 \right] \\[6pt] \notag
 & = \int_{\lambda_{0,0}}^{\lambda_{g,g}} \int_{\lambda_{u,0}}^{\lambda_{u,g}} \left[ s_m^{(\ell,\ell)}(u,v) \right]^2 \: \mathrm{d}v \: \mathrm{d}u + \alpha \left[ \sum_{i=1}^{n} \sum_{j=1}^{n} w_{ij} \left( z_{ij} -s_m(u_{i},v_{j}) \right)^2 \right] \\[6pt]
 & = \int_{\lambda_{0,0}}^{\lambda_{g,g}} \int_{\lambda_{u,0}}^{\lambda_{u,g}} \left[ s_m^{(\ell,\ell)}(u,v) \right]^2 \: \mathrm{d}v \: \mathrm{d}u + \alpha \left[ \sum_{i'=1}^{n'} w_{i'} \left( z_{i'} -s_m(u_{i'},v_{i'}) \right)^2 \right] \\[6pt]
 & = J_\ell^1(s_m) + J_\ell^2(s_m) \, .
\label{eq:objective_function_integral_bivariate}
\end{align}
The double sum has been reduced to a single sum under the hypothesis that the sample consists of a value $z_i$ and a point $(u_i,v_i)$. The third line has been obtained after vectorization.
%, which implies the change of index $i' = n(i-1)+j$.
The extrema of integration are the same as in the univariate case, but now it is necessary to stress formally that when integrating with respect to $v$, the extrema of integration in principle may depend on $u$. The idea is nonetheless simple: the area of integration consists of the points included in the square with vertices $(\lambda_{0,0}, \lambda_{0,g}, \lambda_{g,0}, \lambda_{g,g})$. Concerning the derivative, by choosing $\ell_1 = \ell_2 = \ell = 2$ we are performing second-order derivative, thus obtaining a solution in the class of cubic spline functions.

Let $\mathbf{Z}=(z_{ij})_{ij}$, $\mathbf{W}=(w_{ij})_{ij}$ be $n \times n$ matrices and $\mathbf{z}=\vecc{\mathbf{Z}}$, $\mathbf{w}=\vecc{\mathbf{W}}$ be their vectorization. By exploiting eq.~\eqref{eq:spline_bivariate}, the second component of eq.~\eqref{eq:objective_function_integral_bivariate} can be written in matrix form as:
\begin{equation}
J_\ell^2(\overline{\mathbf{b}}) = \alpha \left( \mathbf{z} - \mathbf{C}^{m+1}(\mathbf{u},\mathbf{v}) \overline{\mathbf{b}} \right)' \mathbf{W} \left( \mathbf{z} - \mathbf{C}^{m+1}(\mathbf{u},\mathbf{v}) \overline{\mathbf{b}} \right) \, .
\label{eq:objective_fucntion_part2}
\end{equation}
As for the first addendum, since the derivative of spline is another spline of lower degree, it can be represented in matrix form. In particular, it is given by the product of the vectorised coefficient matrix $\overline{\mathbf{b}}$ and a vector of B-spline basis functions $\mathbf{g}(u,v)$ defined as:
\begin{equation}
\mathbf{g}(u,v) = \left[ B_{-m}^{m+1}B_{-m}^{m+1}(v), \ldots, B_{g}^{m+1}(u)B_{-m}^{m+1}(v), \ldots, B_{g}^{m+1}(u)B_{g}^{m+1}(v) \right] \, .
\end{equation}
This allows to write:
\begin{align}
\notag
J_\ell^1(s_m) & = \int_{\lambda_{0,0}}^{\lambda_{g,g}} \int_{\lambda_{u,0}}^{\lambda_{u,g}} [s_m^{(\ell,\ell)}(u,v) ]^2 \; \mathrm{d}v \: \mathrm{d}u = \int_{\lambda_{0,0}}^{\lambda_{g,g}} \int_{\lambda_{u,0}}^{\lambda_{u,g}}  \overline{\mathbf{b}}^{(\ell)'} \mathbf{g}(u,v)' \mathbf{g}(u,v) \overline{\mathbf{b}}^{(\ell)}  \; \mathrm{d}v \: \mathrm{d}u \\[8pt] \notag
 & = \overline{\mathbf{b}}^{(\ell)'} \left[ \int_{\lambda_{0,0}}^{\lambda_{g,g}} \int_{\lambda_{u,0}}^{\lambda_{u,g}}   \mathbf{g}(u,v)' \mathbf{g}(u,v)  \; \mathrm{d}v \: \mathrm{d}u \right] \overline{\mathbf{b}}^{(\ell)} \\
 & = \overline{\mathbf{b}}^{(\ell)'} \mathbf{M}_{m,\ell} \overline{\mathbf{b}}^{(\ell)} = J_\ell^1(\overline{\mathbf{b}}) \, ,
\end{align}
where the last line follows from the definition of the matrix $\mathbf{M}_{m,\ell}$.
We are left to find an explicit form for the vectorised coefficient matrix of the original spline of degree $m$, $\overline{\mathbf{b}}$, and that of its $\ell$-th derivative, $\overline{\mathbf{b}}^{(\ell)}$. Recall the previous manipulation of the integral constraint gave a linear relation between $\overline{\mathbf{b}}$ and $\mathbf{\tilde{c}}$, with a restriction was accounted for in the construction of the matrix $\mathbf{K}^*$.
In the problem at hand there no constraints, therefore we use the matrix $\mathbf{K}$ defined in eq.~\eqref{eq:K_unconstrained}.
Finally, notice that we can compute the $\ell$-th derivative of $s_m(u,v)$ by simply iterating $\ell$ times the procedure previous used for the first order derivative, with a slight modification of the matrices involved. In fact, the size of the matrices need to shrink at each derivation step (in fact, the degree of a spline determines the length of its coefficient vector). Therefore, by indexing each matrix with a subscript corresponding to the order of the derivative, we obtain for the $\ell$-th order derivative (similar to eq.~\eqref{eq:relation_coefficients_b_c}):
\begin{equation}
\overline{\mathbf{b}}^{(\ell)} = \mathbf{S}_\ell \mathbf{\overline{b}} = \left[ \prod_{h=1}^\ell \mathbf{D}_h \left[ \mathbf{E}_h \mathbf{T}_h^f - \mathbf{F}_h \mathbf{T}_h^l \right] \mathbf{K}_h \right] \mathbf{\overline{b}} \, .
\label{eq:coeff_derivative_ell}
\end{equation}
All the definitions are provided below in eq.~\eqref{eq:Dh},~\eqref{eq:Kh},~\eqref{eq:Thfl},~\eqref{eq:Eh},~\eqref{eq:Fh}. Notice however that they are just simple generalisations of the matrices used when we dealing with the integral constraint: in fact in that case we were considering a first order derivative, while here we are considering a $\ell$-th order derivative. The only significant difference consists in the substitution of the matrix $\mathbf{K}$ with the difference operator matrix $\mathbf{K}_h$ defined in eq.~\eqref{eq:Kh}.
Let $\lbrace \mathbf{e}_{-m+h},\ldots,\mathbf{e}_g \rbrace$ be the canonical basis of the space of $(g+m+1-h)\times(g+m+1-h)$ matrices, with $h=1,\ldots,\ell$. The block diagonal $(g+m+1-h)^2 \times (g+m+1-h)^2$ matrix $\mathbf{D}_h$ is given by:
\begin{equation}
\mathbf{D}_h = \sum_{i=-m+h}^g \mathbf{e}_i \mathbf{e}_i' \otimes \mathbf{D}_h^i = \left[ \begin{array}{c|c|c|c}
\mathbf{D}_h^{-m+h} & & & \\ \hline
& \mathbf{D}_h^{-m+h+1} & & \\ \hline
& & \ddots & \\ \hline
& & & \mathbf{D}_h^g
\end{array} \right] \, ,
\label{eq:Dh}
\end{equation}
where each block $\mathbf{D}_h^j$, for $j= -m+h,\ldots,g$, $h=1,\ldots,\ell$, is a $(g+m+1-h) \times (g+m+1-h)$ diagonal matrix:
\begin{equation}
\mathbf{D}_h^j = \diag{\dfrac{(m+1-h)^2}{\lambda_{-m+h,j+m+1-h} -\lambda_{-m+h,j}}, \, \dfrac{(m+1-h)^2}{\lambda_{-m+h+1,j+m+1-h} -\lambda_{-m+h+1,j}}, \, \ldots, \, \dfrac{(m+1-h)^2}{\lambda_{g,j+m+1-h} -\lambda_{g,j}}} \, .
%\mathbf{D}_h^j = \begin{bmatrix}
%\dfrac{(m+1-h)^2}{\lambda_{-m+h,j+m+1-h} -\lambda_{-m+h,j}} & & & & \\
% & \dfrac{(m+1-h)^2}{\lambda_{-m+h+1,j+m+1-h} -\lambda_{-m+h+1,j}} & & & \\
% & & & \ddots & \\
% & & & & \dfrac{(m+1-h)^2}{\lambda_{g,j+m+1-h} -\lambda_{g,j}}
%\end{bmatrix} \, .
\end{equation}
Notice that when $\ell=1$ and the original spline has degree $m+1$ we are back in the previous case. As previously noted, in dealing with derivatives without constraints, the matrix $\mathbf{K}$ can be substituted by the difference operator matrix defined in eq.~\eqref{eq:difference_operator_matrix}. It has the same structure and entries for each $h=1,\ldots,\ell$, but with different size. Since we are dealing with the iterative vectorisation of a matrix, we must account that at each derivative the last row of the original matrix is lost due to differentiation. This is reflected in a reduction of the number of rows $\mathbf{K}_h$ by a factor of $hN$ at each step, where $N$ is the number of columns of the original matrix.
Finally, since differencing is performed iteratively, the vector to be differenced is the outcome of the previous iteration, hence its length (which is equal to the number of columns of $\mathbf{K}_h$) corresponds to the number of rows of $\mathbf{K}_h$ plus $hN$. Summarizing, for $h=1,\ldots,\ell$ we define the $(g+m+1)(g+m+1-h) \times (g+m+1)(g+m+2-h)$ matrix:
\begin{equation}
\mathbf{K}_h = \begin{bmatrix}
-1 & 1 & & \\
& \ddots & \ddots & \\
& & -1 & 1
\end{bmatrix} \, .
\label{eq:Kh}
\end{equation}
Similarly, she selection matrices $\mathbf{T}_h^f,\mathbf{T}_h^l$, with $h=1,\ldots,\ell$, have the structure as $\mathbf{T}^f,\mathbf{T}^l$, but their size is $(g+m+1-h)^2 \times (g+m+1)(g+m+1-h)$. They are defined as follows:
\begin{equation}
\mathbf{T}_h^f = \left[ \begin{array}{c c c |c c c}
1 & & & 0 & \ldots & 0 \\
& \ddots & & \vdots & & \vdots \\
& & 1 & 0 & \ldots & 0
\end{array} \right] \, \qquad  
\mathbf{T}_h^l = \left[ \begin{array}{c c c|c c c}
0 & \ldots & 0 & 1 & & \\
\vdots & & \vdots & & \ddots & \\
0 & \ldots & 0 & & & 1
\end{array} \right] \, .
\label{eq:Thfl}
\end{equation}
Finally, the matrices $\mathbf{E}$ and $\mathbf{F}$ are generalized to obtain the $(g+m+1-h)^2 \times (g+m+1-h)^2$ block diagonal matrices:
\begin{equation}
\mathbf{E}_h = \sum_{i=-m+h}^g \mathbf{e}_i \mathbf{e}_i' \otimes \mathbf{E}_h^i = \left[ \begin{array}{c|c|c|c}
\mathbf{E}_h^{-m} & & & \\ \hline
& \mathbf{E}_h^{-m-1} & & \\ \hline
& & \ddots & \\ \hline
& & & \mathbf{E}_h^{g}
\end{array} \right] \, ,
\label{eq:Eh}
\end{equation}
\begin{equation}
\mathbf{F}_h = \sum_{i=-m+h}^g \mathbf{e}_i \mathbf{e}_i' \otimes \mathbf{E}_h^{i-1} = \left[ \begin{array}{c|c|c|c}
\mathbf{E}_h^{-m-1} & & & \\ \hline
& \mathbf{E}_h^{-m} & & \\ \hline
& & \ddots & \\ \hline
& & & \mathbf{E}_h^{g-1}
\end{array} \right] \, ,
\label{eq:Fh}
\end{equation}
where each $\mathbf{E}_h^j$, with $j = -m-1+h,\ldots,g$, $h=1,\ldots,\ell$, is the $(g+m+1-h) \times (g+m+1-h)$ diagonal matrix:
\begin{equation}
\mathbf{E}_h^j = \diag{\dfrac{1}{\lambda_{1,j} -\lambda_{-m+h,j}}, \, \dfrac{1}{\lambda_{2,j} -\lambda_{-m+h+1,j}}, \, \ldots, \, \dfrac{1}{\lambda_{g+m+1-h,j} -\lambda_{g,j}}} \, .
%\mathbf{E}_h^j = \begin{bmatrix}
%\dfrac{1}{\lambda_{1,j} -\lambda_{-m+h,j}} & & & & & \\
% & \dfrac{1}{\lambda_{2,j} -\lambda_{-m+h+1,j}} & & & & \\
% & & & \ddots & & \\
% & & & & \ddots & \\
% & & & & & \dfrac{1}{\lambda_{g+m+1-h,j} -\lambda_{g,j}}
%\end{bmatrix} \, .
\end{equation}
To sum up, we can re-write the first addendum of the objective function of the optimization problem in eq.~\eqref{eq:objective_function_integral_bivariate} in compact form as follows:
\begin{equation}
J_\ell^1 (\overline{\mathbf{b}}) = \overline{\mathbf{b}}^{(\ell)'} \mathbf{M}_{m,\ell} \overline{\mathbf{b}}^{(\ell)} = \overline{\mathbf{b}}' \mathbf{S}_\ell' \mathbf{M}_{m,\ell} \mathbf{S}_\ell \overline{\mathbf{b}} \, .
\label{eq:objective_function_part1}
\end{equation}
Putting together eq.~\eqref{eq:objective_function_part1} and~\eqref{eq:objective_fucntion_part2} we obtain the following matrix representation of the objective function of the optimisation problem in eq.~\eqref{eq:constrained_smoothing_problem}:
\begin{equation}
J_\ell(\overline{\mathbf{b}}) = \overline{\mathbf{b}}' \mathbf{S}_\ell' \mathbf{M}_{m,\ell} \mathbf{S}_\ell \overline{\mathbf{b}} + \alpha \left( \mathbf{z} - \mathbf{C}^{m+1}(\mathbf{u},\mathbf{v}) \overline{\mathbf{b}} \right)' \mathbf{W} \left( \mathbf{z} - \mathbf{C}^{m+1}(\mathbf{u},\mathbf{v}) \overline{\mathbf{b}} \right) \, .
\label{eq:objetive_function_vectorised}
\end{equation}
We can now exploit the linear relation obtained in eq.~\eqref{eq:relation_coefficients_b_c} by working out the integral constraint and substitute it in eq.~\eqref{eq:objetive_function_vectorised}. This transforms the constrained optimization problem in eq.~\eqref{eq:constrained_smoothing_problem} into an unconstrained optimisation problem for $\mathbf{\tilde{c}}$, with objective function:
\begin{equation}
J_\ell (\mathbf{\tilde{c}}) = \mathbf{\tilde{c}}' \mathbf{A}' \mathbf{D}' \mathbf{S}_\ell' \mathbf{M}_{m,\ell} \mathbf{S}_\ell \mathbf{D} \mathbf{A} \mathbf{\tilde{c}} + \alpha \left( \mathbf{z} - \mathbf{C}^{m+1}(\mathbf{u},\mathbf{v}) \mathbf{D} \mathbf{A} \mathbf{\tilde{c}} \right)' \mathbf{W} \left( \mathbf{z} - \mathbf{C}^{m+1}(\mathbf{u},\mathbf{v}) \mathbf{D} \mathbf{A} \mathbf{\tilde{c}} \right) \, .
\label{eq:uncontrained_opt_c}
\end{equation}
The system of first order necessary conditions for an optimum is obtained from:
\begin{align}
\notag
\frac{\mathrm{d}J_\ell(\mathbf{\tilde{c}})}{\mathrm{d} \mathbf{\tilde{c}}'} & = 2 \mathbf{A}' \mathbf{D}' \mathbf{S}_\ell' \mathbf{M}_{m,\ell} \mathbf{S}_\ell \mathbf{D} \mathbf{A} \mathbf{\tilde{c}} -2 \alpha \mathbf{A}' \mathbf{D}' \mathbf{C}^{m+1}(\mathbf{u},\mathbf{v})' \mathbf{W} \mathbf{z} \\
 & +2 \alpha \mathbf{A}' \mathbf{D}' \mathbf{C}^{m+1}(\mathbf{u},\mathbf{v})' \mathbf{W} \mathbf{C}^{m+1}(\mathbf{u},\mathbf{v}) \mathbf{D} \mathbf{A} \mathbf{\tilde{c}} = 0 \, .
\label{eq:system_FOCs_1}
\end{align}
Define the following variables for easing the notation:
\begin{align*}
\mathbf{N}_{m,\ell} & = \mathbf{A}' \mathbf{D}' \mathbf{S}_\ell' \mathbf{M}_{m,\ell} \mathbf{S}_\ell \mathbf{D} \mathbf{A} \, , \\
\mathbf{H}(\mathbf{u},\mathbf{v}) & = \mathbf{C}^{m+1}(\mathbf{u},\mathbf{v}) \mathbf{D} \mathbf{A} \, .
\end{align*}
Therefore, one gets:
\begin{align}
\notag
\mathbf{N}_{m,\ell} \mathbf{\tilde{c}} +\alpha \mathbf{H}(\mathbf{u},\mathbf{v})' \mathbf{W} \mathbf{H}(\mathbf{u},\mathbf{v}) \mathbf{\tilde{c}} & = \alpha \mathbf{H}(\mathbf{u},\mathbf{v})' \mathbf{W} \mathbf{z} \\
\left[ \mathbf{N}_{m,\ell} +\alpha \mathbf{H}(\mathbf{u},\mathbf{v})' \mathbf{W} \mathbf{H}(\mathbf{u},\mathbf{v}) \right] \mathbf{\tilde{c}} & = \alpha \mathbf{H}(\mathbf{u},\mathbf{v})' \mathbf{W} \mathbf{z} \, .
\label{eq:system_FOCs_2}
\end{align}
If the condition of the Rouch\'{e}-Capelli theorem for the system to admit solution is satisfied and the matrix $\left[ \mathbf{N}_{m,\ell} +\alpha \mathbf{H}(\mathbf{u},\mathbf{v})' \mathbf{W} \mathbf{H}(\mathbf{u},\mathbf{v}) \right]$ has full rank, then the system has a unique solution $\mathbf{\tilde{c}}^*$. By contrast, if the matrix is singular the problem admits an infinite number of solutions which can be obtained by computing the Moore-Penrose pseudo-inverse (denoted by $\dag$):
\begin{equation}
\mathbf{\tilde{c}}^* = \alpha \left[ \mathbf{N}_{m,\ell} +\alpha \mathbf{H}(\mathbf{u},\mathbf{v})' \mathbf{W} \mathbf{H}(\mathbf{u},\mathbf{v}) \right]^\dag \mathbf{H}(\mathbf{u},\mathbf{v})' \mathbf{W} \mathbf{z} \, ,
\label{eq:solution_c*}
\end{equation}
with a slight abuse of notation. Among this set of solutions, we choose the one with smallest norm. As a final step, we use eq.~\eqref{eq:relation_coefficients_b_c} and plug-in the optimal value of $\mathbf{\tilde{c}}$ for obtaining the optimal value of the coefficients $\overline{\mathbf{b}}^*$:
\begin{equation}
\overline{\mathbf{b}}^* = \mathbf{D} \mathbf{A} \mathbf{\tilde{c}}^* \, .
\label{eq:solution_b*}
\end{equation}

\subsection{Eigenproblem}\label{sec:apdx_computations_eigenproblem}
In the following we show the computations required for obtaining eq.~\eqref{eq:eigenproblem_final}. In this section, differently from the previous ones, we explicitly denote all the arguments of a function for making notation clearer. Thus, for example, we have that  $\breve{f}_t(\cdot) = \breve{f}_t(\cdot,\cdot)$, where the first is follows the notation of the previous sections, while the second is according to the notation of this section.
We use the standard estimator for the sample covariance (alternative non-parametric estimatiors have been proposed by \cite{Hall06PropertiesFPCA}, \cite{Li10ConvergenceRates_NonparametricRegression}, \cite{Yao05FunctionalDataAnalysis_SparseLongitudinal} and \cite{Staniswalis98NonparametricRegressionLongitudinal}), that is:
\begin{equation}
v(l_1,m_1,l_2,m_2) = \dfrac{1}{T} \sum_{t=1}^T \breve{f}_t(l_1,m_1) \breve{f}_t(l_2,m_2) \, .
\end{equation}
Then, the eigenproblem can be formulated as follows, for $j=1,2,\ldots$:
\begin{equation}
\int_{a_1}^{b_1} \int_{a_2}^{b_2} v(\cdot,\cdot,l_2,m_2) \xi_j(l_2,m_2) \; \; \mathrm{d}l_2 \: \mathrm{d}m_2 = \rho_j \xi_j(\cdot,\cdot) \, .
\label{eq:eigenproblem_bivariate}
\end{equation}
In order to solve this problem, we choose to express the eigenfunctions as finite linear combinations of the same set of basis functions used for the functions $\breve{f}_t(\cdot,\cdot)$, that is the basis B-spline functions $\boldsymbol{\psi}(\cdot,\cdot) = (\psi_1(\cdot,\cdot),\ldots,\psi_K(\cdot,\cdot))'$ in eq.~\eqref{eq:spline_solution}. Define the coefficient vectors $\mathbf{a}_j = (a_{j,1},\ldots,a_{j,K})'$. To summarize, we have:
\begin{align}
\label{eq:eigenproblem_basis_expansion_f}
\breve{f}_t(\cdot,\cdot) & = \mathbf{d}_t' \boldsymbol{\psi}(\cdot,\cdot) = \sum_{k=1}^K d_{t,k} \psi_k(\cdot,\cdot) \, , \\
\label{eq:eigenproblem_basis_expansion_xi}
\xi_j(\cdot,\cdot) & = \mathbf{a}_j' \boldsymbol{\psi}(\cdot,\cdot) = \sum_{k=1}^K a_{j,k} \psi_k(\cdot,\cdot) \, .
\end{align}
By stacking all data together in $\mathbf{\breve{f}}(\cdot,\cdot) = (\breve{f}_1(\cdot,\cdot),\ldots,\breve{f}_T(\cdot,\cdot))'$ and $\mathbf{D} = (\mathbf{d}_1,\ldots,\mathbf{d}_T)$, we obtain:
\begin{equation}
\mathbf{f}(\cdot,\cdot) = \mathbf{D} \boldsymbol{\psi}(\cdot,\cdot) \, .
\label{eq:eigenproblem_f_stacked}
\end{equation}
We can thus rewrite eq.~\eqref{eq:eigenproblem_bivariate} in matrix notation:
\begin{align}
\frac{1}{T} \int_a^b \int_a^b  \boldsymbol{\psi}(\cdot,\cdot)' \mathbf{D}' \mathbf{D} \boldsymbol{\psi}(l_2,m_2) \boldsymbol{\psi}(l_2,m_2)' \mathbf{a}_j \; \; \mathrm{d}l_2 \: \mathrm{d}m_2 & = \rho_j \boldsymbol{\psi}(\cdot,\cdot)' \mathbf{a}_j \\
\frac{1}{T} \boldsymbol{\psi}(\cdot,\cdot)' \mathbf{D}' \mathbf{D} \left[ \int_a^b \int_a^b \boldsymbol{\psi}(l_2,m_2) \boldsymbol{\psi}(l_2,m_2)' \; \; \mathrm{d}l_2 \: \mathrm{d}m_2 \right] \mathbf{a}_j & = \rho_j \boldsymbol{\psi}(\cdot,\cdot)' \mathbf{a}_j \, ,
\end{align}
then define the matrix of inner products:
\begin{align}
& \mathbf{M} = \int_a^b \int_a^b \boldsymbol{\psi}(l_2,m_2) \boldsymbol{\psi}(l_2,m_2)' \; \; \mathrm{d}l_2 \: \mathrm{d}m_2 = \begin{bmatrix}
\langle \psi_1(\cdot,\cdot), \psi_1(\cdot,\cdot) \rangle & \ldots & \langle \psi_1(\cdot,\cdot), \psi_K(\cdot,\cdot) \rangle \\
\vdots & & \vdots \\
\langle \psi_K(\cdot,\cdot), \psi_1(\cdot,\cdot) \rangle & \ldots & \langle \psi_K(\cdot,\cdot), \psi_K(\cdot,\cdot) \rangle
\end{bmatrix} \\
& \langle \psi_i(\cdot,\cdot), \psi_j(\cdot,\cdot) \rangle = \int_a^b \int_a^b \psi_i(l,m) \psi_j(l,m) \; \; \mathrm{d}l \: \mathrm{d}m \, ,
\end{align}
thus obtaining:
\begin{equation}
T^{-1} \mathbf{D}' \mathbf{D} \mathbf{M} \mathbf{a}_j = \rho_j \mathbf{a}_j \, .
\label{eq:eigenproblem_bivariate_matrix}
\end{equation}
In order to obtain a positive semi-definite matrix, we apply the linear transformation $\mathbf{u}_j = \mathbf{M}^{1/2} \mathbf{a}_j$, where $\mathbf{A}^{1/2}$ is the principal square root of the positive definite matrix $\mathbf{A}$. Then, re-write the previous equation as an eigenproblem for $\mathbf{u}_j$ as follows:
\begin{equation}
T^{-1} \mathbf{M}^{1/2} \mathbf{D}' \mathbf{D} \mathbf{M}^{1/2} \mathbf{u}_j = \rho_j \mathbf{u}_j \, ,
\label{eq:eigenproblem_bivariate_matrix_u}
\end{equation}
which is a standard multivariate eigenproblem for the matrix $T^{-1} \mathbf{M}^{1/2} \mathbf{D}' \mathbf{D} \mathbf{M}^{1/2}$. The number of components to take, $J$, is determined by the fraction of variability explained: we sort the estimated eigenvalues $\widehat{\rho}_j$, for $j=1,2,\ldots$, in decreasing order. Then, we fix a threshold $\bar{d}$ and retain all the pairs of eigenvalues and eigenvectors until the corresponding cumulated proportion of explained variability reaches, that is $J = \argmin_j \lbrace \sum_j \widehat{\rho}_j \geq \bar{d} \rbrace$.

\section{Additional plots}\label{sec:apdx_application_plots}

\begin{figure}[H]
\centering
\captionsetup{width=0.9\linewidth}
\includegraphics[trim= 35mm 0mm   0mm 0mm, clip, height= 3.2cm, width= 8.0cm]{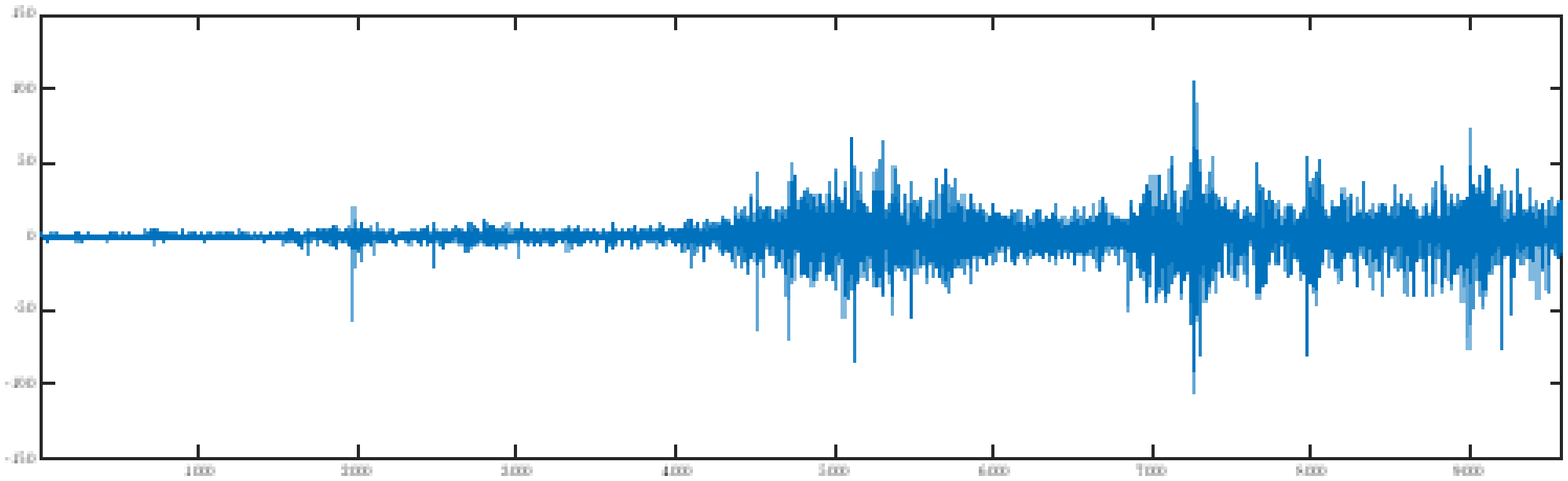} \hspace*{-4ex}
\includegraphics[trim= 35mm 0mm   0mm 0mm, clip, height= 3.2cm, width= 8.0cm]{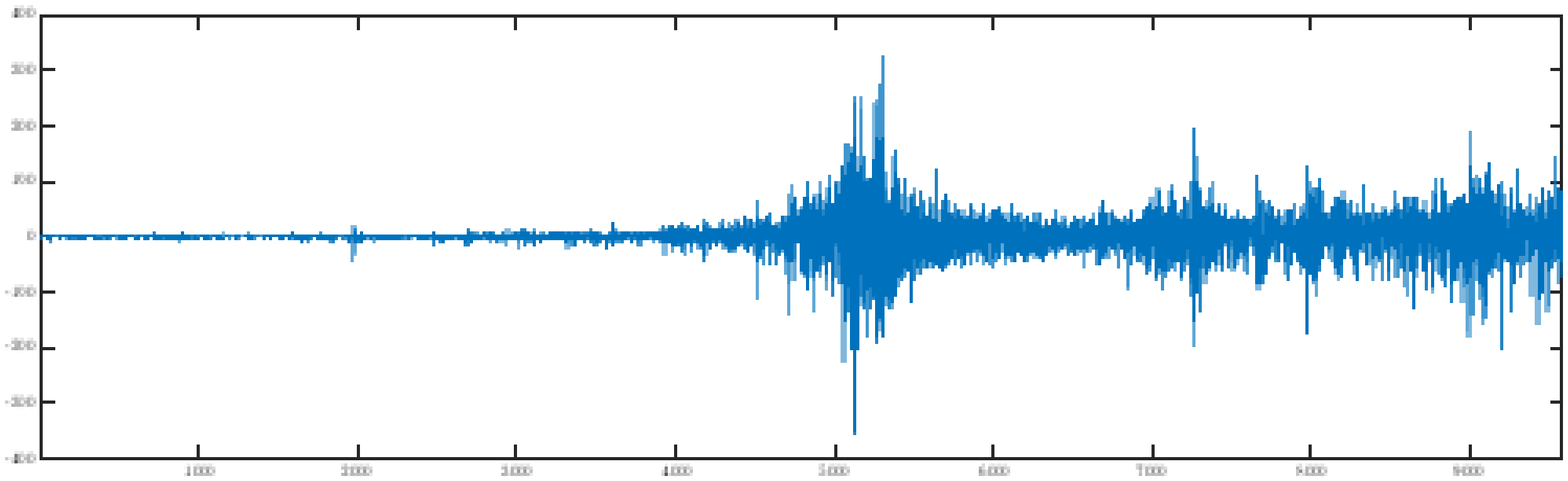}
\caption{\label{fig:plot_raw_data} First differenced series of S\&P500 (\textit{left}) and NASDAQ (\textit{right}).}
\end{figure}

\begin{figure}[H]
\centering
\captionsetup{width=0.9\linewidth}
\includegraphics[trim= 25mm 0mm   0mm 0mm, clip, height= 3.0cm, width= 8.0cm]{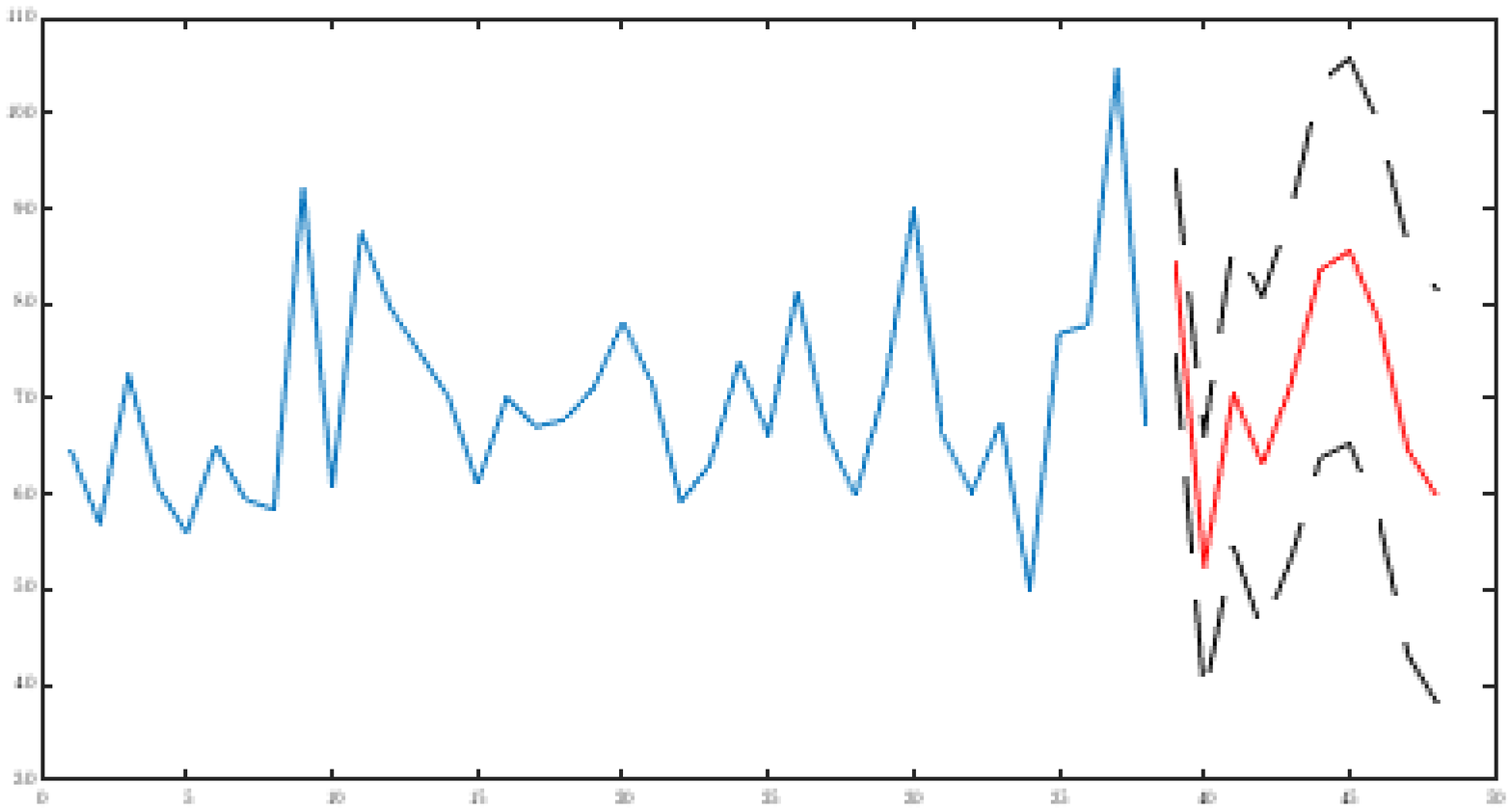} \hspace*{-4ex}
\includegraphics[trim= 25mm 0mm   0mm 0mm, clip, height= 3.0cm, width= 8.0cm]{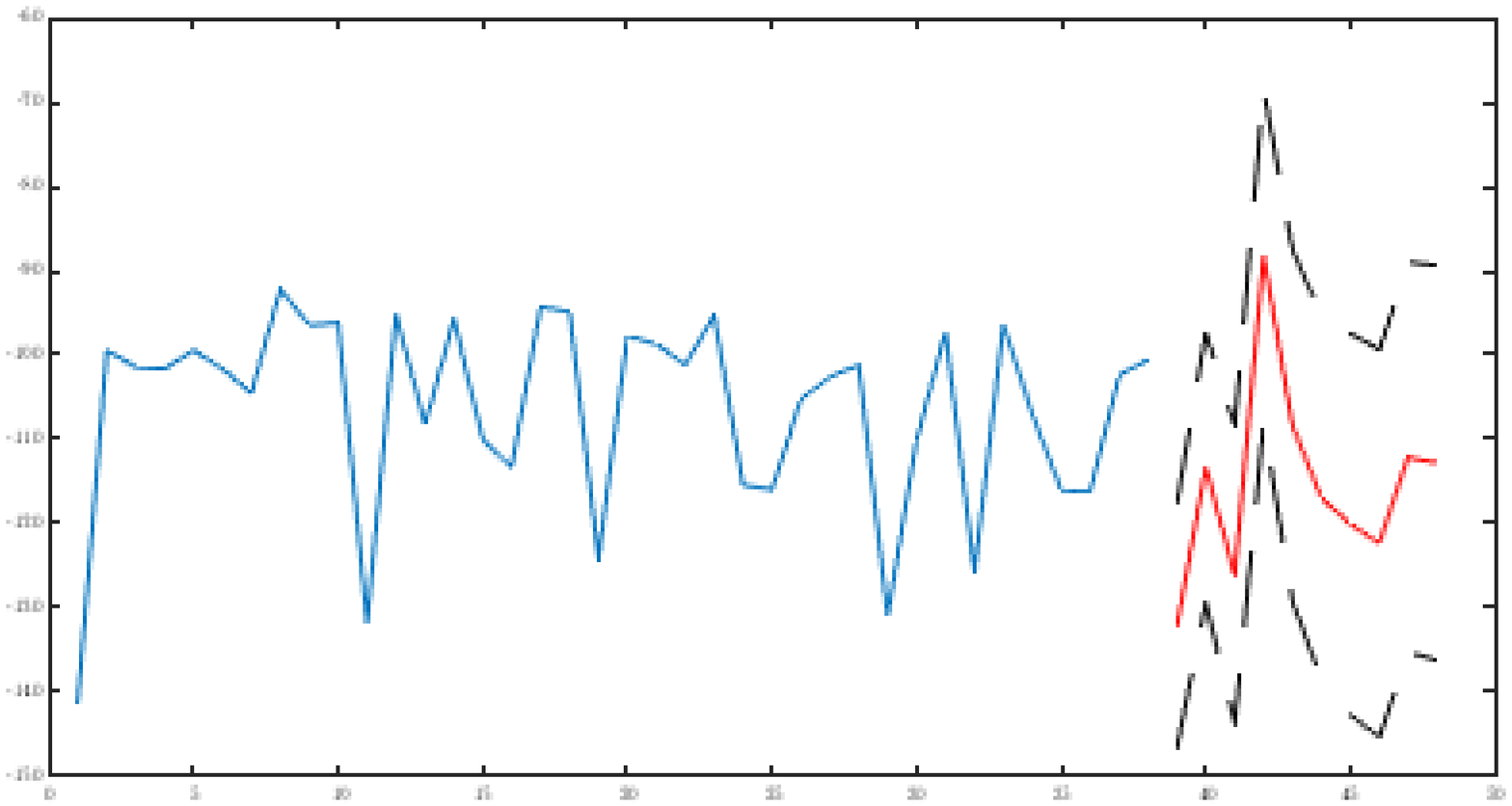}
\includegraphics[trim= 25mm 0mm   0mm 0mm, clip, height= 3.0cm, width= 8.0cm]{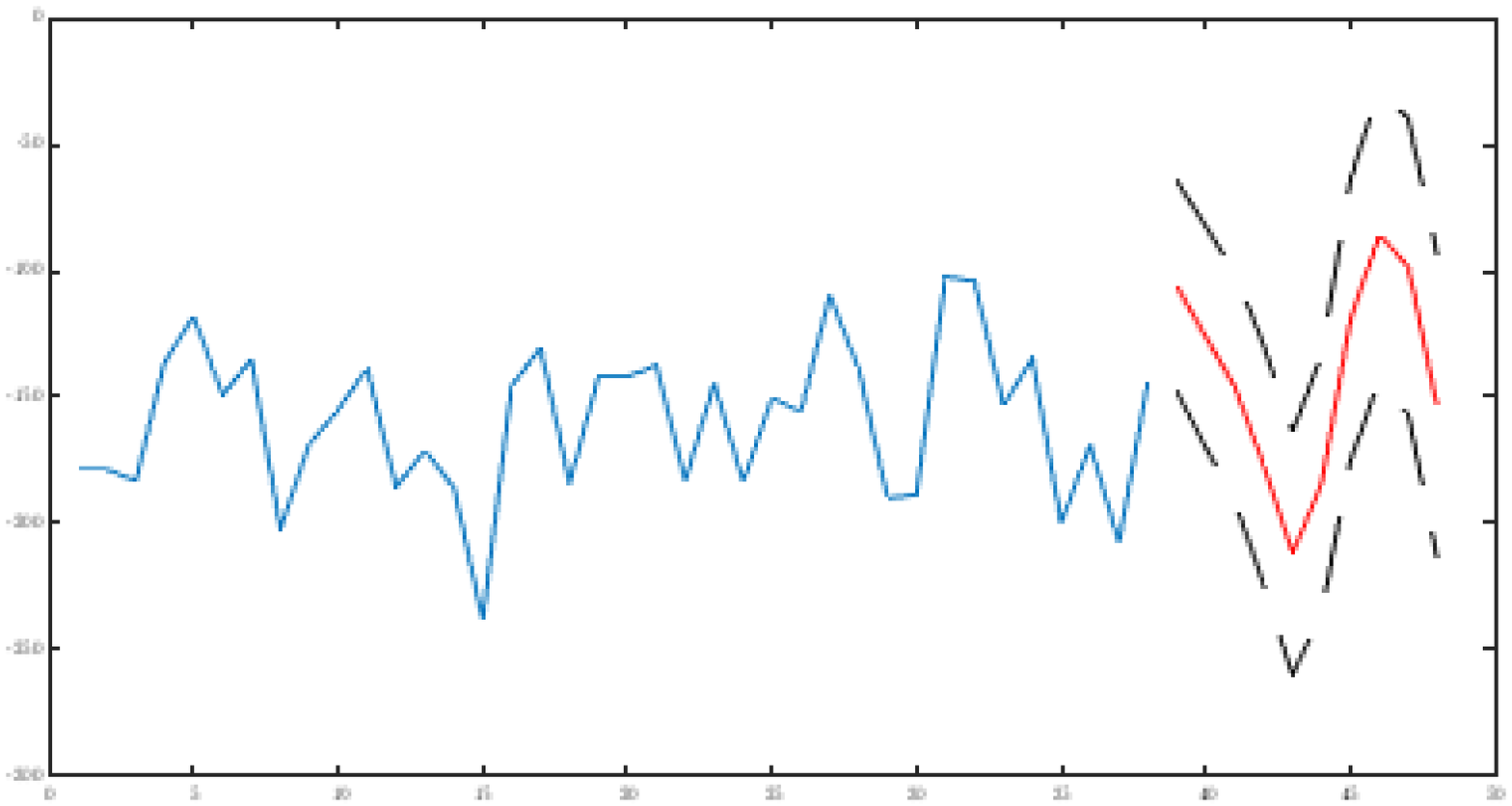} \hspace*{-4ex}
\includegraphics[trim= 25mm 0mm   0mm 0mm, clip, height= 3.0cm, width= 8.0cm]{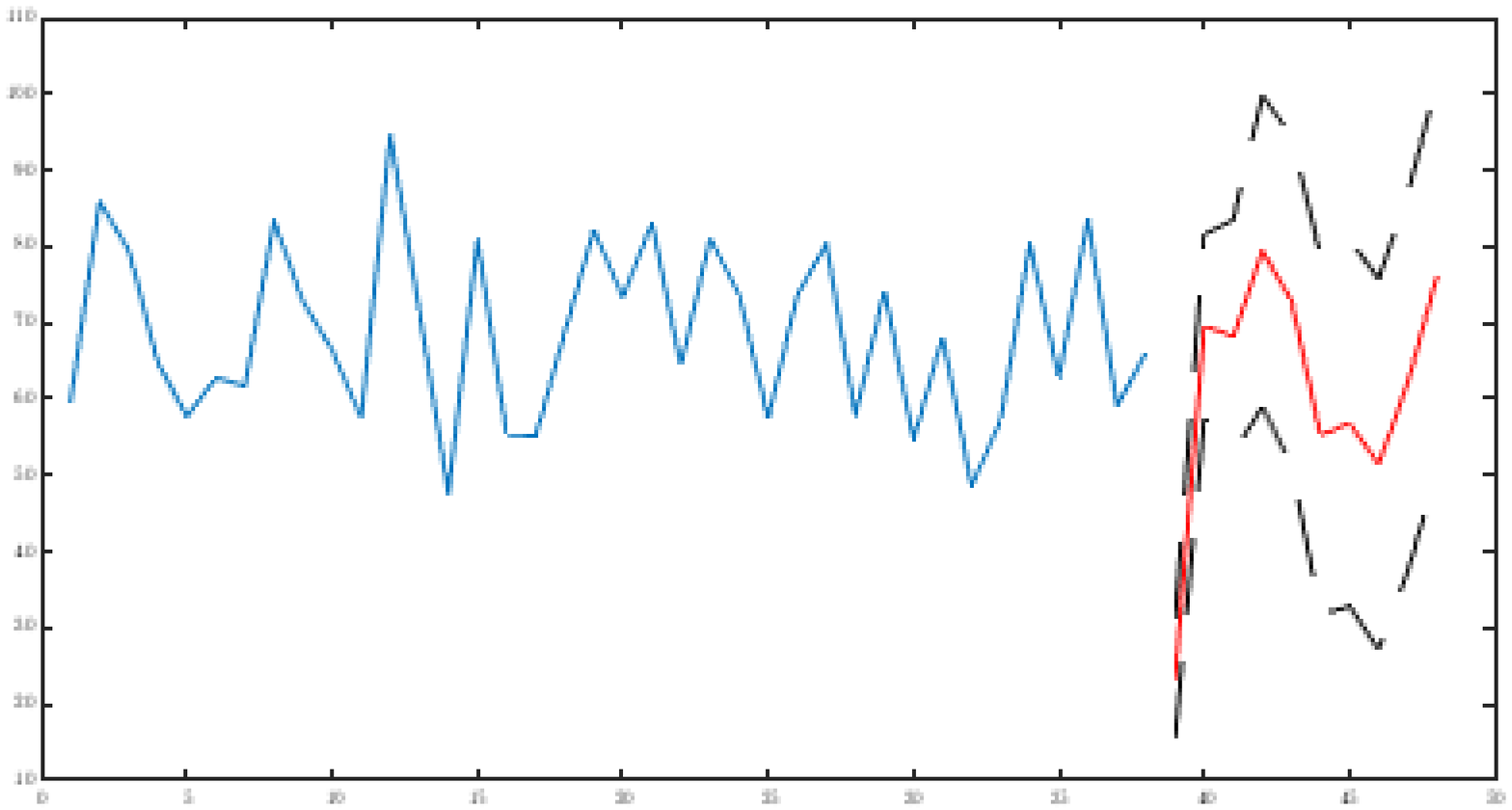}
\caption{\label{fig:ts_forec_beta} Estimated time series (\textit{solid, blue}) and forecast (\textit{solid, red}) with 95\% confidence intervals (\textit{dashed, black}) of each entry of the vector of fPCA scores $\lbrace \boldsymbol{\widehat{\beta}}_t \rbrace_t$,from $j=1$ (\textit{top left}) to $j=J$ (\textit{bottom}).}
\end{figure}

\begin{sidewaysfigure}
%\centering
\captionsetup{width=0.9\linewidth}
\includegraphics[trim=40mm 10mm  0mm 10mm, clip, height= 15.0cm, width= 25.0cm]{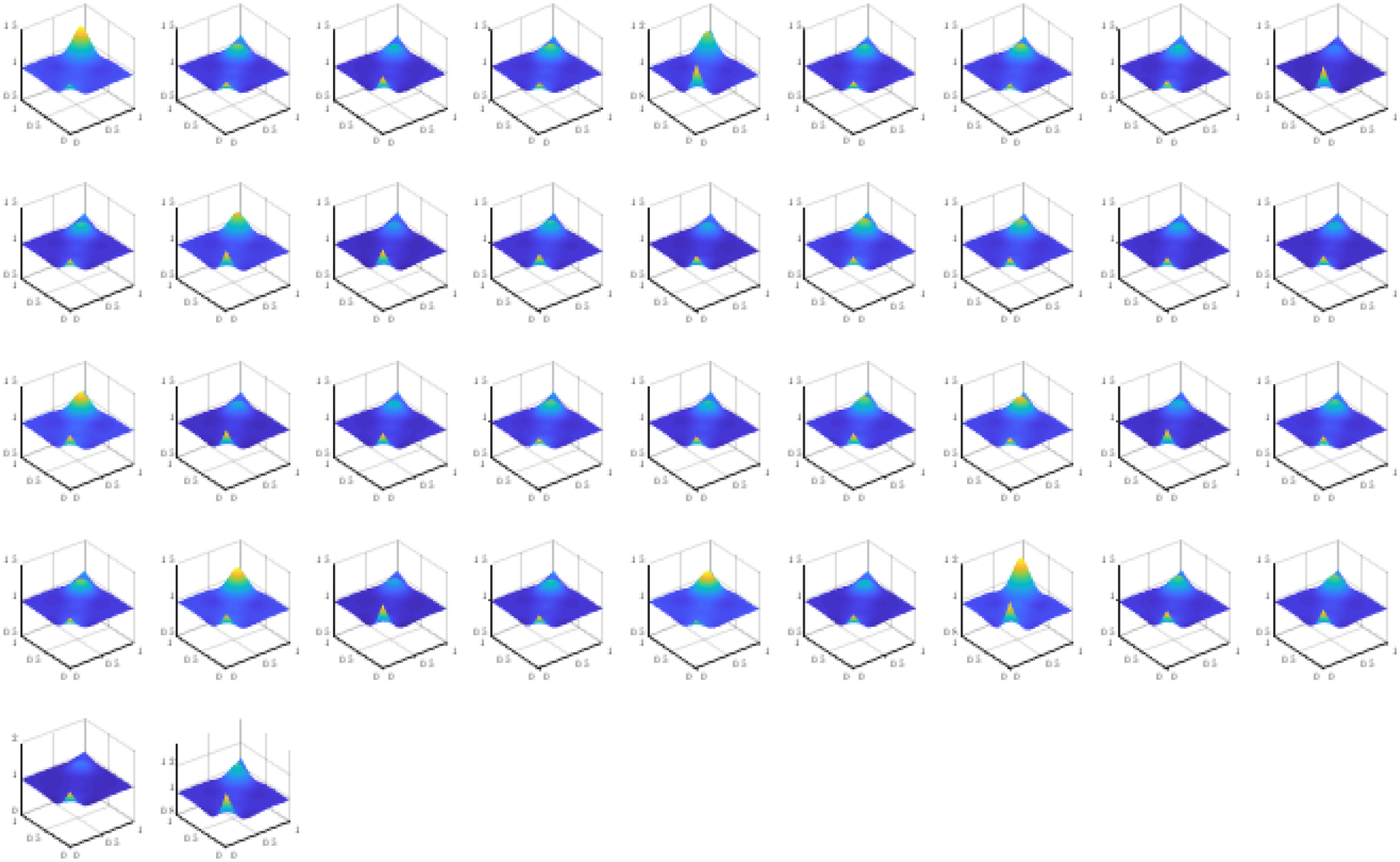}
\caption{\label{fig:ts_3D} 3D density plot of time series of bivariate copula pdfs, approximated via fPCA, for each year $t=1,\ldots,T$, starting from $t=1$ in the top-left panel.}
\end{sidewaysfigure}

\end{appendix}

\end{document}